\documentclass[<preprint>]{elsarticle}

\biboptions{longnamesfirst,semicolon}

\usepackage[margin=2.5cm]{geometry}

\usepackage{natbib}            
\usepackage{setspace}
\usepackage{psfrag}
\usepackage{graphicx}
\usepackage{amsfonts,amssymb,amsmath}
\usepackage{setspace,subfigure}
\usepackage{latexsym}
\usepackage{pifont}
\usepackage{mathtools}
\usepackage{mathdots}
\usepackage{bm}

\usepackage{multirow}
\usepackage{multicol}
\usepackage{color}
\usepackage{url}
\usepackage{algorithm}
\usepackage{algorithmic}

\usepackage{rotating}
\usepackage{calc}
\usepackage{relsize}
\usepackage{array}

\newcommand{\enma}[1]   {\ensuremath{#1}}

\newcommand{\beq}{\begin{equation}}
\newcommand{\eeq}{\end{equation}}
\newcommand{\bseq}{\begin{subequations}}
\newcommand{\eseq}{\end{subequations}}
\newcommand{\beqn}{\begin{eqnarray}}
\newcommand{\eeqn}{\end{eqnarray}}
\newcommand{\ba}{\begin{array}}
\newcommand{\ea}{\end{array}}
\newcommand{\bct}{\begin{center}}
\newcommand{\ect}{\end{center}}
\newcommand{\btmz}{\begin{itemize}}
\newcommand{\etmz}{\end{itemize}}
\newcommand{\benum}{\begin{enumerate}}
\newcommand{\eenum}{\end{enumerate}}

\newcommand{\cH}{\enma{\mathcal H}}

\newcommand{\inner}[2]{\left\langle #1,#2 \right\rangle}

\newcommand{\bv}{{\bf v}}

\newcommand{\matbegin}{
        \left[
}
\newcommand{\matend}{
        \right]
}

\newcommand{\tbo}[2]{
  \matbegin \begin{array}{c}
       #1 \\ #2
       \end{array} \matend }

\newcommand{\obt}[2]{
  \matbegin \begin{array}{cc}
       #1 & #2
       \end{array} \matend }

\newcommand{\tbt}[4]{
  \matbegin \begin{array}{cc}
       #1 & #2 \\ #3 & #4
       \end{array} \matend }

\newcommand{\ca}{{\cal A}}

\newcommand{\ce}{{\cal E}}

\newcommand{\be}{\begin{equation}}
\newcommand{\ee}{\end{equation}}

\newcommand{\cplxs}{ C\kern -.35em \rule{0.03 em}{.7 ex}~   }

\def\complex{\hbox{C\kern -.45em \rule{0.03 em}{1.5 ex}}~}

\newcommand{\bi}{\begin{itemize}}
\newcommand{\ei}{\end{itemize}}

\newcommand{\bA}{\mathbf{A}}
\newcommand{\bB}{\mathbf{B}}
\newcommand{\bC}{\mathbf{C}}
\newcommand{\bE}{\mathbf{E}}

\newcommand{\bH}{\mathbf{H}}
\newcommand{\bL}{\mathbf{L}}
\newcommand{\bM}{\mathbf{M}}
\newcommand{\bN}{\mathbf{N}}

\newcommand{\bQ}{\mathbf{Q}}

\newcommand{\bJ}{\mathbf{J}}
\newcommand{\bK}{\mathbf{K}}

\newcommand{\cc}{{\cal C}}
\newcommand{\cb}{{\cal B}}

\newcommand{\cf}{{\cal F}}

\newcommand{\ct}{{\cal T}}

\newcommand{\bbC}{\mathbb{C}}

\newcommand{\non}{\nonumber}
\newcommand{\ds}{\displaystyle}

\newcommand{\mrd}{\mathrm{d}}
\newcommand{\mre}{\mathrm{e}}
\newcommand{\mri}{\mathrm{i}}

\newcommand{\bd}{\mathbf{d}}
\newcommand{\bD}{{\bf D}}

\newcommand{\bx}{{\bf x}}
\newcommand{\bu}{{\bf u}}

\newcommand{\bW}{{\bf W}}
\newcommand{\bYY}{{\bf Y}}
\newcommand{\bxi}{\mbox{\boldmath$\xi$}}
\newcommand{\bphi}{\mbox{\boldmath$\phi$}}
\newcommand{\balpha}{\mbox{\boldmath$\alpha$}}

\newcommand{\bnu}{\mbox{\boldmath$\nu$}}

\newcommand{\bbeta}{\mbox{\boldmath$\beta$}}
\newcommand{\bgamma}{\mbox{\boldmath$\gamma$}}
\newcommand{\bpsi}{\mbox{\boldmath$\psi$}}

\newcommand{\bPhi}{\mbox{\boldmath$\Phi$}}

\newcommand{\bnabla}{\mbox{\boldmath$\nabla$}}

\newcommand{\py}{\partial_y}

\newcommand{\bI}{{\bf I}}

\newcommand{\btau}{\mbox{\boldmath$\tau$}}

\newcommand{\We}{W\!e}

\newcommand{\bk}{\mathbf{k}}

\newcommand{\bg}{\mathbf{g}}
\newcommand{\bbf}{\mathbf{f}}

\newcommand{\bz}{\mathbf{z}}
\newcommand{\bq}{\mathbf{q}}

\newcommand{\cg}{{\cal G}}
\newcommand{\bvphi}{\mbox{\boldmath$\varphi$}}

\journal{Journal of Computational Physics}

\begin{document}
\begin{frontmatter}
	\title{\LARGE \bf
		Computation of frequency responses for linear time-invariant PDEs on a compact interval
	}

	\author[umn]{Binh K. Lieu}
	\ead{lieux006@umn.edu}
	\ead[url]{www.ece.umn.edu/$\sim$blieu}
	
   	\author[umn]{Mihailo R.\ Jovanovi\'c}
	\ead{mihailo@umn.edu}
	\ead[url]{www.ece.umn.edu/$\sim$mihailo}

	\address[umn]{Department of Electrical and Computer Engineering,~University of Minnesota,~Minneapolis,~MN~55455,~USA}
	
	\begin{abstract}
		We develop mathematical framework and computational tools for calculating frequency responses of linear time-invariant PDEs in which an independent spatial variable belongs to a compact interval. In conventional studies this computation is done numerically using spatial discretization of differential operators in the evolution equation. In this paper, we introduce an alternative method that avoids the need for finite-dimensional approximation of the underlying operators in the evolution model. This method recasts the frequency response operator as a two point boundary value problem and uses state-of-the-art automatic spectral collocation techniques for solving integral representations of the resulting boundary value problems with accuracy comparable to machine precision. Our approach has two advantages over currently available schemes: first, it avoids numerical instabilities encountered in systems with differential operators of high order and, second, it alleviates difficulty in implementing boundary conditions. We provide examples from Newtonian and viscoelastic fluid dynamics to illustrate utility of the proposed method.		
	\end{abstract}
	

\end{frontmatter}

\section{Introduction}
\label{sec.intro}

	In many physical systems there is a need to examine the effects of exogenous disturbances on the variables of interest. Frequency response analysis represents an effective means for quantifying the system's  performance in the presence of a stimulus, and it characterizes the steady-state response of a stable system to persistent harmonic forcing. At each temporal frequency, the frequency response of finite dimensional linear time-invariant systems with scalar input and output is a complex number that determines the magnitude and phase of the output relative to the input. In systems with many inputs and outputs (multi-variable systems), the frequency response is a complex matrix whose dimension is determined by the number of inputs and outputs. In systems with infinite dimensional input and output spaces that are considered in this paper, the frequency response is an operator. It is well-known that the singular values of the frequency response matrix (in multi-variable systems) or the frequency response operator (in infinite dimensional systems) represent proper generalization of the magnitude characteristics for single-input single-output systems. At a specific frequency, the largest singular value determines the largest amplification from the input forcing to the desired output~\cite{schhen01}. Furthermore, the associated left and right principal singular functions identify the spatial distributions of the output (that exhibits this largest amplification) and the input (that has the strongest influence on the system's dynamics), respectively.
	
	\vspace*{0.15cm}
		
	In this paper, we study the frequency responses of linear time-invariant partial differential equations (PDEs) in which an independent spatial variable belongs to a compact interval. We are interested in computing the largest singular value of the frequency response operator and its corresponding singular functions. Computation of frequency responses for PDEs is typically done numerically using finite-dimensional approximations of the operators in the evolution equation. Pseudo-spectral methods represent a powerful tool for discretization of spatial differential operators because they possess superior numerical accuracy compared to approximation schemes based on finite differences~\cite{canhusquazan88,boy89,treFinite96,treSpectral00}. In spite of their advantages, pseudo-spectral methods may produce unreliable results and even fail to converge upon grid refinement when dealing with systems that contain differential operators of high order; this lack of convergence is attributed to the loss of accuracy arising from ill-conditioning of the discretized differentiation matrices~\cite{hei89}. Furthermore, implementation of general boundary conditions may be challenging.
	
	\vspace*{0.15cm}
		
	To alleviate these difficulties, we introduce a method that avoids the need for finite dimensional approximations of differential operators in the evolution equation. This is accomplished by recasting the frequency response operator as a two point boundary value problem (TPBVP) that is given by either an input-output differential equation of high order or by an equivalent system of first order differential equations (i.e., spatial state-space representation). Furthermore, we present a procedure for converting these differential representations into the corresponding systems of integral equations. This transformation facilitates the use of recently developed computing environment, {\sf Chebfun}~\cite{chebfunv4}, that is capable of solving boundary value problems and eigenvalue problems with superior accuracy. Our mathematical framework in conjunction with {\sf Chebfun}'s state-of-the-art numerical algorithms has two main advantages over standard methods: first, it alleviates numerical ill-conditioning encountered in systems with differential operators of high order; and second, it enables easy implementation of a wide range of boundary conditions.
			
	\vspace*{0.15cm}
		
	{\sf Chebfun} is a collection of powerful algorithms for numerical computations that involve continuous and piecewise-continuous functions. Instead of working in a finite dimensional setting, {\sf Chebfun} allows users to symbolically represent functions and operators in their infinite dimensional form with simple and compact {\sc Matlab} syntaxes. This provides an elegant high-level language for solving linear and nonlinear boundary value and eigenvalue problems with few lines of code. Internally, functions are computed numerically using automatic Chebyshev polynomial interpolation techniques, and the operators are approximated using automatic spectral collocation methods. Finite dimensional approximations of functions and operators are automatically refined in order to obtain accurate and convergent representations. Furthermore, once the boundary conditions are specified {\sf Chebfun} makes sure that they are automatically satisfied internally when solving differential or integral equations.
	
	\vspace*{0.15cm}
	
	The proposed method has many potential applications in numerical analysis, physics, and engineering, especially in systems with generators that do not commute with their adjoints~\cite{treemb05}. In these systems, standard modal analysis may fail to capture amplification of exogenous disturbances, low stability margins, and large transient responses. In contrast, singular value decomposition of the frequency response operator represents an effective tool for identifying these non-modal aspects of the system's dynamics. In particular, wall-bounded shear flows of both Newtonian and viscoelastic fluids have non-normal dynamical generators of high spatial order and the ability to accurately compute frequency responses for these systems is of paramount importance; additional examples of systems with non-normal generators, for which the tools developed in this paper are particularly well-suited, can be found in the outstanding book by Trefethen and Embree~\cite{treemb05} and the references therein. The utility of non-modal analysis in understanding the dynamics of infinitesimal fluctuations around laminar flow conditions has been well-documented; see~\cite{,schhen01,tretrereddri93,gro00,jovbamJFM05,sch07,niclel11,garlesschcho12,garlesschhue13} for Newtonian fluids and~\cite{hodjovkumJFM08,hodjovkumJFM09,jovkumPOF10,jovkumJNNFM11,liejovkumJFM13} for viscoelastic fluids. In viscoelastic fluids with large polymer relaxation times, analysis is additionally complicated by the fact that pseudo-spectral methods exhibit spurious numerical instabilities~\cite{kup05,gra98}. We use examples from fluid mechanics to demonstrate the ease of incorporating boundary conditions and superior accuracy of our method compared to conventional finite dimensional approximation schemes.
	
	\vspace*{0.15cm}
		
	Our presentation is organized as follows. In Section~\ref{sec.preliminaries}, we formulate the problem and discuss the notion of a frequency response for PDEs. In Section~\ref{sec.io-ss}, we present the method for converting the frequency response operator into a TPBVP that can be posed as an input-output differential equation or as a spatial state-space representation. In Section~\ref{sec.computation}, we show how to transform a family of differential equations into equivalent integral equations and describe the use of {\sf Chebfun}'s indefinite integration operator for determining the eigenvalues and corresponding eigenfunctions of the resulting integral equations. In Section~\ref{sec.example}, we demonstrate the utility of our developments by providing two examples from Newtonian and viscoelastic fluid dynamics. We conclude with a brief summary of the paper in Section~\ref{sec.conclusion}, and relegate the mathematical developments to the appendices.
	
\section{Motivating examples and problem formulation}
\label{sec.preliminaries}

    	In this section, we provide two examples that are used to motivate our developments and to illustrate the classes of systems that we consider. These examples are used throughout the paper to explain the problem setup and utility of the proposed method. We then describe the class of PDEs that we study and briefly review the notion of a frequency response operator.
		
\subsection{Motivating examples}
    \label{sec.motivating-examples}

    	We next present two physical examples: the one-dimensional (1D) diffusion equation, and the system of PDEs that governs the dynamics of the flow fluctuations in an inertialess channel flow of viscoelastic fluids. The 1D diffusion equation has simple dynamics and it is used to illustrate mathematical framework developed in the paper. The example from viscoelastic fluid mechanics is used to demonstrate utility of our approach on a system that is known to produce spurious numerical instabilities. We show how numerical difficulties encountered in the computation of the frequency responses can be overcome using the developed framework in conjunction with state-of-the-art automatic spectral collocation techniques.
	
    	\subsubsection{One-dimensional diffusion equation}
	\label{sec.motivating-ex-heat}

Let a one-dimensional diffusion equation with homogenous Dirichlet boundary conditions and zero initial conditions be subject to spatially and temporally distributed forcing $d (y,t)$,
		\begin{equation}
		\label{eq.heat-eq}
			\begin{array}{rcl}
				\phi_t (y,t)
				& \!\! = \!\! &
				\phi_{yy} (y,t)
				\; + \;
				d(y,t),
				\\[0.1cm]
				\phi(\pm 1, t)
				& \!\! = \!\! &
				0,
				\\[0.1cm]
				\phi(y,0)
				& \!\! = \!\! &
				0,
				\;\;
				y \in \left[ -1, \, 1 \right].
			\end{array}
		\end{equation}
		Throughout the paper, the spatially independent variable is denoted by $y$, the time is denoted by $t$, and the subscripts denote differentiation with respect to time/space. Considering $\phi$ as the field of interest, the frequency response operator for this system (from input $d$ to output $\phi$) is obtained by evaluating the resolvent on the $\mri \omega$-axis
	\begin{equation}
		\label{eq.heat-eq-T}
		\ct(\omega)
        \; = \;
        \left( \mri \omega I \, - \, D^{(2)} \right)^{-1},
	\end{equation}
	where $D^{(2)}$ is the second derivative operator with homogenous Dirichlet boundary conditions, $I$ is the identity operator, $\omega$ is the temporal frequency, and $\mri$ is the imaginary unit.
	
	\vspace*{0.15cm}

It is well known that the second derivative operator with Dirichlet boundary conditions is self-adjoint with a complete set of orthonormal eigenfunctions, $v_n (y) = \sin \left( (n \pi/2) (y + 1) \right)$, $n = \{ 1, 2, \ldots \}$. This information can be used to diagonalize operator $D^{(2)}$ in $\ct (\omega)$ which facilitates straightforward determination of the frequency response. For systems with spatially varying coefficients and non-normal generators the frequency response analysis is typically done numerically using finite dimensional approximations of the differential operators. For example, the pseudospectral method~\cite{weired00} with $N$ collocation points can be used to transform the frequency response operator~(\ref{eq.heat-eq-T}) of system~(\ref{eq.heat-eq}) into an $N \times N$ matrix. However, for systems with differential operators of high order, spectral differentiation matrices may be poorly conditioned and implementation of boundary conditions may be challenging. 

	\vspace*{0.15cm}

Alternatively, by applying the temporal Fourier transform to system~(\ref{eq.heat-eq}) we obtain the following input-output differential equation
	\begin{subequations}
			\label{eq.heat-eq-tpbvp}
			\begin{align}
            \label{eq.heat-eq-tpbvp-a}
		\hat{\phi}''(y,\omega)
        \, - \,
        \mri \omega \hat{\phi}(y,\omega)
        \; = & \;\;
        - \hat{d}(y,\omega),
        \\
            \label{eq.heat-eq-tpbvp-b}
		\hat{\phi}(\pm 1,\omega)
        \; = & \;\;
        0,
		\end{align}
		\end{subequations}
	where $\hat{d}$ and $\hat{\phi}$ are the Fourier transformed input and output fields, and $\hat{\phi}' = \mrd \hat{\phi} / \mrd y$. At each $\omega$,~(\ref{eq.heat-eq-tpbvp-a}) is a second-order ordinary differential equation (in $y$) subject to the boundary conditions~(\ref{eq.heat-eq-tpbvp-b}). Equivalently, by defining $x_{1} = \hat{\phi}$ and $x_{2} = \hat{\phi}'$,~(\ref{eq.heat-eq-tpbvp}) can be brought into a system of first order differential equations
			\begin{equation}
			\label{eq.TPBVSR-heat}
				\ct(\omega):
				\left\{
				\begin{array}{rcl}
					\left[
					\begin{array}{c}
					x_1'(y) \\[0.1cm]
					x_2'(y)
					\end{array}
					\right]
					& \!\! = \!\! &
					\left[
					\begin{array}{cc}
					0 & 1 \\[0.1cm]
					\mri \omega & 0
					\end{array}
					\right] \,
					\left[
					\begin{array}{c}
					x_1(y) \\[0.1cm]
					x_2(y)
					\end{array}
					\right]
					\, + \,
					\left[
					\begin{array}{r}
					0 \\[0.1cm]
					-1
					\end{array}
					\right] \,
					\, d(y),
					\\[0.5cm]
					\phi(y)
					& \!\! = \!\! &
					\left[
					\begin{array}{cc}
						1 & 0
					\end{array}
					\right]
					\left[
					\begin{array}{c}
					x_1(y) \\[0.1cm]
					x_2(y)
					\end{array}
					\right],
					\\[0.5cm]
					\left[
					\begin{array}{c}
						0 \\[0.1cm]
						0
					\end{array}
					\right]
					& \!\! = \!\! &
					\left[
					\begin{array}{cc}
						1 & 0 \\[0.1cm]
						0 & 0
					\end{array}
					\right]
					\left[
					\begin{array}{c}
					x_1(-1) \\[0.1cm]
					x_2(-1)
					\end{array}
					\right]
					\, + \,
					\left[
					\begin{array}{cc}
						0 & 0 \\[0.1cm]
						1 & 0
					\end{array}
					\right]
					\left[
					\begin{array}{c}
					x_1(1) \\[0.1cm]
					x_2(1)
					\end{array}
					\right].
				\end{array}
				\right.
			\end{equation}

We will utilize structures of the TPBVPs~(\ref{eq.heat-eq-tpbvp}) and~(\ref{eq.TPBVSR-heat}) in conjunction with recently developed automatic spectral collocation techniques to study the frequency response across $\omega$.
			
		\subsubsection{Inertialess channel flow of viscoelastic fluids}
		\label{sec.motivating-ex-SOB}
		
\begin{figure}
				\begin{center}
			            	{
			            	\includegraphics[width=0.75\columnwidth]
					{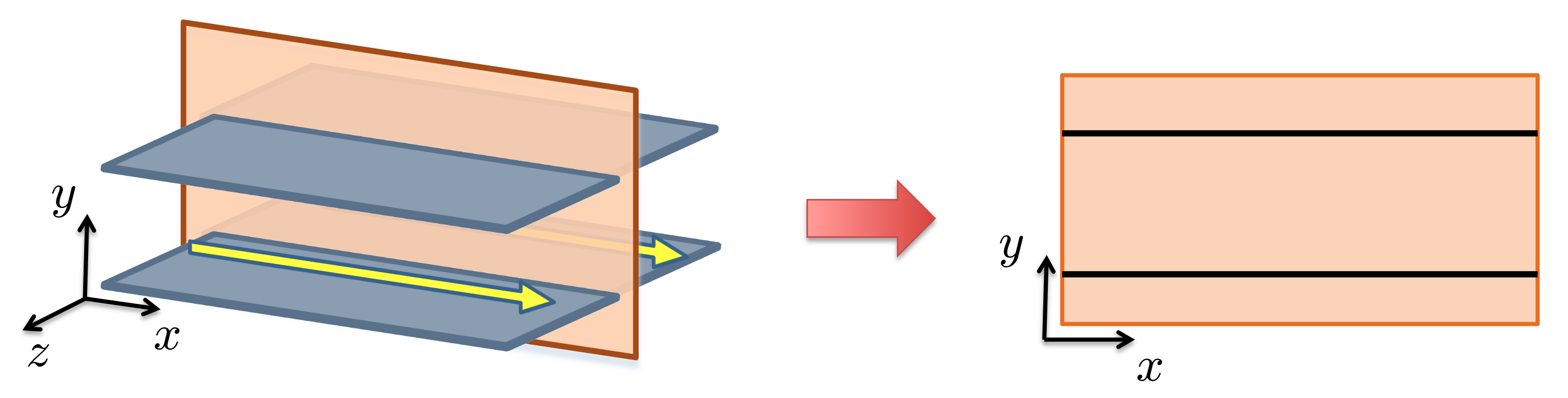}
					}
			  	\end{center}
				\vspace{-0.5cm}
				\caption{We consider the dynamics of flow fluctuations in the ($x,y$)-plane of the channel.}
				\label{fig.channel}
		 	\end{figure}

			We next consider a system that describes the dynamics of two-dimensional velocity and polymer stress fluctuations in an inertialess channel flow of viscoelastic fluids; see figure~\ref{fig.channel} for geometry.
			The dynamics of infinitesimal fluctuations around the mean flow ($\bar{\bv}$, $\bar{\btau}$) are given by
			\begin{subequations}
			\label{eq.stokes-governing}
				\begin{align}
					0 \; = \; &
					\, - \,
					\bnabla p \, + \, \left( 1 - \beta \right) \bnabla \cdot \btau \, + \, \beta \, \bnabla^2 \bv \, + \, \bd,
					\label{eq.stokes}\\			
					0 \; = \; & \, \bnabla \cdot \bv,
					\label{eq.vcontinuity} \\ \nonumber
					\btau_t \; = \; & \,
					\bnabla \bv \, + \, \left( \bnabla \bv \right)^{T} \, - \, \btau \, +  \, \We \left( \btau \cdot \bnabla \bar{\bv} \, + \, \bar{\btau} \cdot \bnabla \bv \right. \\
					& \, + \, \left( \bar{\btau} \cdot \bnabla \bv \right)^{T} \, + \, \left( \btau \cdot \bnabla \bar{\bv} \right)^{T} \, - \, \bv \cdot \bnabla \bar{\btau} \, - \, \left .\bar{\bv} \cdot \bnabla \btau \right).
					\label{eq.tconstitutive}
				\end{align}
			\end{subequations}
			In shear driven flow, $\bar{\bv}$ and $\bar{\btau}$ are
			\begin{equation*}
				\begin{array}{c}
					\bar{\bv}
					\; = \;
					\left[
					\begin{array}{c}
						y \\
						0
					\end{array}
					\right],
					\,\,\,\,
					\bar{\btau}
					\; = \;
					\left[
					\begin{array}{cc}
						\bar{\tau}_{11} & \bar{\tau}_{12} \\
						\bar{\tau}_{12} & \bar{\tau}_{22}
					\end{array}
					\right]
					\; = \;
					\left[
					\begin{array}{cc}
						2 \We & 1 \\
						1 & 0
					\end{array}
					\right],
				\end{array}
			\end{equation*}
			$\bv = \left[ \begin{array}{cc} u & v \end{array} \right]^{T}$, $p$, and $\btau$ are the velocity, pressure, and stress fluctuations; $u$ and $v$ are velocities in $x$ and $y$ directions; $\bnabla$ is the gradient; and $\bnabla^2 = \bnabla \cdot \bnabla$ is the Laplacian. System~(\ref{eq.stokes-governing}) is driven by spatially distributed and temporally varying body force fluctuations $\bd = \left[ \begin{array}{cc} d_{1} & d_{2} \end{array}\right]^{T}$ with $d_1$ and $d_2$ representing the forcing in $x$ and $y$. The non-dimensional parameters in~(\ref{eq.stokes-governing}) are the ratio of the solvent to the total viscosity $\beta \in (0,1)$, and the ratio of the fluid relaxation time to the characteristic flow time $\We$ (the Weissenberg number).
			
			\vspace*{0.15cm}
			
			Static-in-time momentum~(\ref{eq.stokes}) and continuity~(\ref{eq.vcontinuity}) equations describe the motion of incompressible fluids in the Stokes flow, i.e., at zero Reynolds number. The constitutive equation~(\ref{eq.tconstitutive}) captures the influence of the velocity gradients on the dynamics of stress fluctuations in dilute polymer solutions~\cite{lar99}. For background material on the use of frequency response analysis in understanding the dynamics of viscoelastic fluids, we refer the reader to~\cite{hodjovkumJFM08,hodjovkumJFM09,jovkumPOF10,jovkumJNNFM11,liejovkumJFM13}.
			
			\vspace*{0.15cm}
			
By expressing the velocity fluctuations in terms of the stream function $\psi$,
			\[
				u \; = \; \py \psi, \,\,\,\, v \; = \; -\partial_x \psi,
			\]
pressure can be removed form the equations~(\ref{eq.stokes-governing}). Furthermore, by applying the Fourier transform in $x$ and $t$ on~(\ref{eq.tconstitutive}) and by substituting the resulting expression for stresses into the equation for $\psi$, we arrive at the following ordinary differential equation (ODE) in $y$ for the stream function,
			\begin{equation}
			\label{eq.ex-psi-ode}
				\ct(\omega):
				\left\{
				\begin{array}{l}
					\left( D^{(4)} + a_{3}(y) D^{(3)} + a_{2}(y) D^{(2)} + a_{1}(y) D^{(1)} + a_{0}(y) \right) \hat{\psi}(y) ~ = \\[0.1cm]
					\left( \mathbf{b}_{1}(y) \, \bD^{(1)} + \mathbf{b}_{0}(y) \right) \hat{\bd}(y), \\[0.2cm]
					\left[
					\begin{array}{c}
						\hat{u}(y) \\[0.1cm]
						\hat{v}(y)
					\end{array}
					\right]
					\; = \;
					\left[
					\begin{array}{c}
						D^{(1)} \\[0.1cm]
						-\mri k_x
					\end{array}
					\right] \hat{\psi}(y), \\[0.5cm]
					0 \; = \; \hat{\psi}(y = \pm 1) \; = \; \hat{\psi}'(y = \pm 1),
				\end{array}
				\right.
			\end{equation}
			where $D^{(k)} = \partial^{k} / \partial y^{k}$, $k_x$ is the horizontal wavenumber, and
			\begin{equation*}
				\bD^{(1)}
				\; = \;
				\left[
				\begin{array}{cc}
					D^{(1)} & 0 \\[0.1cm]
					0 & D^{(1)}
				\end{array}
				\right].
			\end{equation*}
			The coefficients $\{ a_{i}(y), \mathbf{b}_{j}(y)\}$ in~(\ref{eq.ex-psi-ode}) are reported in~\ref{sec.app-SOB-kz0-TPBVSR}. The system of equations~(\ref{eq.ex-psi-ode}) is parameterized by $\omega$, $k_x$, $\beta$, and $\We$. For notational convenience, we have suppressed the dependence of $\hat{\psi}$, $\hat{\bd}$, $\hat{u}$, and $\hat{v}$ on these four parameters.
			
			\vspace*{0.15cm}
			
			In Section~\ref{sec.example}, we show that spatial discretization of the operators in~(\ref{eq.stokes-governing}) using the pseudo-spectral method~\cite{weired00} can produce erroneous frequency responses. In contrast, transformation of the system into a TPBVP
(which is then recast into an equivalent integral form) followed by the use of the Chebfun environment~\cite{chebfunv4} yields reliable results.		
		
	\subsection{Problem formulation}
	\label{sec.problem-formulation}

		We now formulate the problem for PDEs with the evolution equation
		\begin{subequations}
			\label{eq.system}
			\begin{align}
				\label{eq.system-state}
				\ce \, \bphi_t (y, t)
                \; = & \;\;
                \cf \, \bphi (y, t)
                \; + \; \cg \, \bd(y, t),
                \\
				\label{eq.system-output}
				\bvphi(y, t)
                \; = & \;\;
                \cH \, \bphi(y, t),
			\end{align}
		\end{subequations}
		where $t \in [0, \infty)$ and $y \in [a, b]$ denote the temporal and spatial variables. The spatially distributed and temporally varying state, input, and output fields  are represented by $\bphi$, $\bd$, and $\bvphi$, respectively. At each $t$, $\bd(\cdot, t)$ and $\bvphi(\cdot, t)$ denote the square-integrable vector-valued functions, and $\{ \ce$, $\cf$, $\cg$, $\cH \}$ are matrices of differential operators with, in general, spatially varying coefficients. For example, the $ij$th entry of the operator $\cf$ can be expressed as
		\begin{equation*}
			\cf_{i j}
        \; = \;
        \sum_{k \, = \, 0}^{n_{i j}} f_{i j, k}(y) \, D^{(k)},
		\end{equation*}
		where each $f_{i j, k}$ is a function that is at least $k$ times continuously differentiable on the interval $[a, b]$~\cite{renrog04}, $D^{(k)} = \partial^{k}/\partial y^{k}$, and $n_{i j}$ is the order of the highest derivative in $\cf_{ij}$.
		
		\vspace*{0.15cm}
		
		The application of the temporal Fourier transform yields the frequency response operator of system~(\ref{eq.system})
		\begin{equation}
			\label{eq.freq-response}
			\ct (\omega)
            \; = \;
            \cH
            \left(
            \mri \omega \ce \, - \, \cf
            \right)^{-1}
            \cg,
		\end{equation}
		For an exponentially stable system~(\ref{eq.system}), $\ct (\omega)$ describes the steady-state response to harmonic input signals across the temporal frequency $\omega$. Namely, if the input is harmonic in $t$, i.e.,
    \[
    \bd(y, t)
    \; = \;
    \hat{\bd}(y, \omega)
    \,
    {\mre}^{\mri \omega t},
    \]
with $\hat{\bd}(\cdot, \omega)$ denoting a square-integrable spatial distribution in $y$, then the output is also harmonic in $t$ with the same frequency but with a modified amplitude and phase
		\beq
            \ba{rcl}
			\bvphi(y, t)
			& \!\! = \!\! &
	    		\left(
	    			\left[
	    			\ct(\omega) \, \hat{\bd}(\cdot,\omega)
	    			\right] (y)
	    		\right)
	    		\mre^{ \mri \omega t}
			\; = \;
            \hat{\bvphi}(y, \omega) \, \mre^{ \mri \omega t}
            \\[0.25cm]
            & \!\! = \!\! &
	    		{\ds \left(
	    		\int_{a}^{b} \ct_{\mathrm{ker}}(y,\xi; \omega) \, \hat{\bd} (\xi, \omega)
               \,
               \mrd \xi
	    		\right)
	    		\mre^{ \mri \omega t}}.
                \ea
	    		\non
	   	\eeq
		The amplitude and phase of the output at the frequency $\omega$ are precisely determined by the frequency response operator $\ct (\omega)$, with $\ct_{\mathrm{ker}}$ denoting the kernel representation of the operator $\ct$.

	\vspace*{0.15cm}
		
		For the class of systems that we consider, the kernel representation of the frequency response operator $\ct_{\mathrm{ker}}(\, \cdot \, , \, \cdot \; ; \omega)$ is a bounded matrix valued function on $[a, b] \times [a, b]$. This implies that the operator $\ct (\omega)$ can be represented using the singular value (i.e., Schmidt) decomposition~\cite{naysel00},
		\begin{equation}
		\label{eq.schmidt}
			 \hat{\bvphi}(y,\omega)
			 \; = \;
			 \left[ \ct (\omega) \, \hat{\bd} (\cdot, \omega) \right] (y)
			\; = \;
			\ds{\sum_{n \, = \, 1}^{\infty}}
			\,
			\sigma_n (\omega) \, \hat{\bu}_n (y,\omega) \inner{\hat{\bv}_n}{\hat{\bd}},
		\end{equation}
where $\inner{\cdot}{\cdot}$ is the standard $L_{2}\left[ a, b\right]$ inner product,
		\begin{equation*}
			 \inner{\hat{\bv}_{1}}{\hat{\bv}_{2}}
                \; = \;
                \int_{a}^{b} \hat{\bv}_{1}^{*}(y) \, \hat{\bv}_{2}(y) \, \mrd y,
		\end{equation*}
and $\hat{\bv}_{1}^{*} (y)$ is the complex-conjugate-transpose of the vector $\hat{\bv}_1 (y)$. In~(\ref{eq.schmidt}), $\{ \hat{\bu}_n \}$ and $\{ \hat{\bv}_n \}$ denote the left and the right singular functions of the operator $\ct$ associated with the singular value $\sigma_n$. These are obtained from the eigenvalue decomposition of the operators $\ct \, \ct^\star$ and $\ct^\star \, \ct$,
		\begin{equation*}
			\ba{rcl}
			\left[ \ct (\omega) \, \ct^\star (\omega) \, \hat{\bu}_n (\cdot,\omega) \right] (y)
			& \!\! = \!\! &
			\sigma_n^2 (\omega) \, \hat{\bu}_n (y,\omega),
			\\[0.15cm]
			\left[ \ct^\star (\omega) \, \ct (\omega) \, \hat{\bv}_n (\cdot,\omega) \right] (y)
			& \!\! = \!\! &
			\sigma_n^2 (\omega) \, \hat{\bv}_n (y,\omega),
			\ea
		\end{equation*}
		where $\ct^{\star}$ is the adjoint of the operator $\ct$. The singular values are positive numbers arranged in descending order,
    \[
    \sigma_1
    ~ \geq ~
    \sigma_2
    ~ \geq ~
    \cdots
    ~ > ~
    0,
    \]
and they are determined by the square root of the non-zero eigenvalues of $\ct \, \ct^\star$ (or $\ct^\star \, \ct$). On the other hand, the singular functions $\{ \hat{\bu}_n \}$ and $\{ \hat{\bv}_n \}$ form the orthonormal bases for the spaces of square integrable functions to which the output $\hat{\bvphi}$ and the input $\hat{\bd}$ belong.

	\vspace*{0.15cm}

From~(\ref{eq.schmidt}) we see that the action of the operator $\ct (\omega)$ on $\hat{\bd} (y,\omega)$ is determined by the linear combination of the left singular functions $\{ \hat{\bu}_n \}$. The product between the singular values, $\sigma_n$, and the inner product of the input $\hat{\bd}$ and the right singular function $\hat{\bv}_n$, $\left< \hat{\bv}_n,\hat{\bd} \right>$, yields the corresponding weights. Thus, for $\hat{\bd} = \hat{\bv}_m$, the output is in the direction of $\hat{\bu}_m$ and its energy is determined by $\sigma_m^2$,
    \[
    \hat{\bd} (y,\omega)
    ~ = ~
    \hat{\bv}_m (y,\omega)
    ~~ \Rightarrow ~~
    \hat{\bvphi} (y,\omega)
    ~ = ~
    \sigma_m (\omega) \, \hat{\bu}_m (y,\omega),
    \]
implying that at any frequency $\omega$ the largest singular value $\sigma_1 (\omega)$ quantifies the largest energy of the output for unit energy inputs. This largest energy can be achieved by selecting
    $
    \hat{\bd} (y,\omega)
    =
    \hat{\bv}_1 (y,\omega),
    $
and the most energetic spatial output profile resulting from the action of $\ct (\omega)$ is given by
    $
    \hat{\bvphi} (y,\omega)
    =
    \sigma_1 (\omega) \, \hat{\bu}_1 (y,\omega).
    $
    
    \vspace*{0.15cm}

     In linear dynamical systems, spectral decomposition of the dynamical generators is typically used to identify instability. Appearance of the eigenvalues with positive real part implies exponential temporal growth of infinitesimal fluctuations and the associated eigenfunctions characterize spatial patterns of these growing modes. For systems with normal dynamical generators (i.e., operators that {\em commute\/} with their adjoints) the eigenfunctions are mutually orthogonal and the eigenvalues provide complete information about system's response. However, for systems with non-normal generators eigenvalues may give misleading information about system's responses. Even in the stable regime, non-normality can cause (i) substantial transient growth of fluctuations before their asymptotic decay; (ii) significant amplification of ambient disturbances; and (iii) substantial decrease of stability margins. We note that singular value decomposition of the frequency response operator represents an effective tool for capturing these non-modal aspects of the system's response.
     
     \vspace*{0.15cm}
		
		In what follows, we describe the procedure for reformulating the frequency response operator~(\ref{eq.freq-response}) into corresponding two point boundary value problems that are given by either an input-output differential equation or by a spatial state-space representation. These can be solved with superior accuracy using recently developed computational tools~\cite{chebfunv4}. We illustrate the utility of our developments on an example from viscoelastic fluid dynamics, where standard finite dimensional approximation techniques fail to produce reliable results.

\section{Two point boundary value representations of $\ct$, $\ct^{\star}$, and $\ct \ct^{\star}$}
\label{sec.io-ss}

In this section, we first describe the procedure for determining the two point boundary value representations of the frequency response operator~(\ref{eq.freq-response}). These are given by either a high-order input-output differential equation or by a system of first-order differential equations in spatial variable $y$. We then discuss the procedure for obtaining corresponding representations of the adjoint operator $\ct^{\star}$ and the operator $\ct \ct^{\star}$.
		
	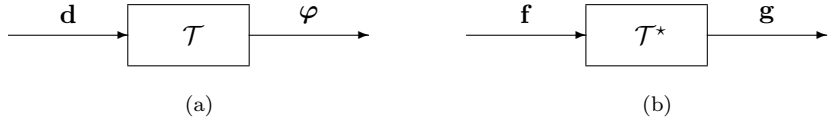
\begin{figure}
		\begin{center}
		\begin{tabular}{cc}
	            	\subfigure[]
	            	{
	            	\setlength{\unitlength}{0.8cm}
		    	\begin{picture}(6.5,2)

	\put(0,1){\vector(1,0){2}}
	\put(1,1.2){\makebox(0,0)[b]{$ \bd $}}
	\put(2,0.5){\framebox(2,1){
	$
	\ct
	$}}
	\put(4,1){\vector(1,0){2}}
	\put(5,1.2){\makebox(0,0)[b]{$ \bvphi $}}

\end{picture}
			\label{fig.blkdiag-t}
			}
			&
	            	\subfigure[]
	            	{
	            	\setlength{\unitlength}{0.8cm}
		    	\begin{picture}(6.5,2)

	\put(0,1){\vector(1,0){2}}
	\put(1,1.2){\makebox(0,0)[b]{$ \bbf $}}
	\put(2,0.5){\framebox(2,1){
	$
	\ct^{\star}
	$}}
	\put(4,1){\vector(1,0){2}}
	\put(5,1.2){\makebox(0,0)[b]{$ \bg $}}

\end{picture}
			\label{fig.blkdiag-ts}
			}
			\end{tabular}
			\vspace{-0.7cm}
	  	\end{center}
		\caption{Block diagrams of (a) the frequency response operator $\ct$: $\bd \mapsto \bvphi$; and (b) the adjoint operator $\ct^{\star}$: $\bbf \mapsto \bg$.}
		\label{fig.blkdiag-t-ts}
 	\end{figure}
	
	\subsection{Representations of the frequency response operator $\ct$}
	\label{sec.representations-ct}
		The application of the temporal Fourier transform to~(\ref{eq.system}) yields
		\begin{subequations}
			\label{eq.system-ode1}
			\begin{align}
				\label{eq.system-ode1-state}
				(\mri \omega \ce \, - \, \cf) \, \bphi(y,\omega)
                \; = & \;\;
                \cg \, \bd(y,\omega),
                \\[0.cm]
				\label{eq.system-ode1-output}
				\bvphi(y,\omega)
                \; = & \;\;
                \cH \, \bphi(y, \omega),
			\end{align}
		\end{subequations}
        where we have omitted hats from the Fourier transformed fields for notational convenience (a convention that we adopt from now on).
		System~(\ref{eq.system-ode1}) represents an $\omega$-parameterized family of ordinary differential equations (ODEs) in $y$, with boundary conditions at $a$ and $b$. From the definitions of the operators $\{ \ce$, $\cf$, $\cg$, $\cH \}$ described in Section~\ref{sec.problem-formulation},~(\ref{eq.system-ode1}) can be represented by the following system of differential equations
		\begin{equation}
			\label{eq.system-ode2}
			\ct:
			\left\{
			\begin{array}{rcl}
				\left[ \, \ca_{0} \, \bphi \, \right] (y)
    & \!\! = \!\! &
    \left[ \, \cb_{0} \, \bd \, \right] (y),
    \\[0.1cm]
				\bvphi(y)
    & \!\! = \!\! &
    \left[ \, \cc_{0} \, \bphi \, \right] (y),
    \\[0.1cm]
				0
    & \!\! = \!\! &
    {\cal N}_{0} \, \bphi \, (y),
			\end{array}
			\right.
		\end{equation}
		where
		\begin{equation*}
			\begin{array}{rcl}
				\ca_{0}
                & \!\! = \!\! &
                {\ds \sum_{i \, = \, 0}^{n} \balpha_{i}(y) \, \bD^{(i)}},
				\,\,
				\cb_{0}
                \; = \,
                {\ds \sum_{i \, = \, 0}^{m} \bbeta_{i}(y) \, \bD^{(i)}},
				\,\,
				\cc_{0} \; = \;
                {\ds \sum_{i \, = \, 0}^{k} \bgamma_{i}(y) \, \bD^{(i)}}, \\[0.5cm]
				{\cal N}_{0}
                & \!\! = \!\! &
                {\ds \sum_{i \, = \, 0}^{\ell}
                \left( \bW_{a, i} \, \bE_{a} \,  +  \, \bW_{b, i} \, \bE_{b} \right) \bD^{(i)}},
				\\[0.5cm]
				\bD^{(i)}
                & \!\! = \!\! &
				\left[
				\begin{array}{ccc}
				D^{(i)} & &
                \\[-0.15cm]
				& \ddots &
                \\[-0.1cm]
				& & D^{(i)}
			\end{array}
				\right],
				\,\,
				\bphi \; = \,
				\left[
				\begin{array}{c}
					\phi_{1} \\
					\vdots \\
					\phi_{s}
				\end{array}
				\right],
                \,\,
				\bd \; = \,
				\left[
				\begin{array}{c}
					d_{1} \\
					\vdots \\
					d_{r}
				\end{array}
				\right],
                \,\,
				\bvphi \; = \,
				\left[
				\begin{array}{c}
					\varphi_{1} \\
					\vdots \\
					\varphi_{p}
				\end{array}
				\right].
				\end{array}
		\end{equation*}
		Here, $D^{(i)} \phi_j = \mrd^{i} \phi_j / \mrd y^{i}$, $\bE_{a}$ and $\bE_{b}$ denote the point evaluation functionals at the boundaries, e.g.,
		    \[
		    \bE_{a} \, \bphi
		    \,
		    (y)
		    \; = \;
		    \bphi(a),
		    \]
		and $\{ \bW_{a, i}$, $\bW_{b, i}\}$ are constant matrices that specify the boundary conditions on $\bphi$. For notational convenience we have omitted the dependence on $\omega$ in~(\ref{eq.system-ode2}), which is a convention that we adopt from now on. Here, $n$, $m$, $k$, and $\ell$ denote the highest differential orders of the operators $\ca_0$, $\cb_0$, $\cc_0$, and ${\cal N}_{0}$, respectively. If the number of components in $\bphi$, $\bd$, and $\bvphi$ is given by $s$, $r$, and $p$, then $\{ \balpha_{i}(y) \}$ are matrices of size $s \times s$ with entries determined by the coefficients of the operator $(\mri \omega \ce - \cf)$; $\{ \bbeta_{i}(y) \}$ are matrices of size $s \times r$ with entries determined by the coefficients of the operator $\cg$; and $\{ \bgamma_{i}(y) \}$ are matrices of size $p \times s$ with entries determined by the coefficients of the operator $\cH$. We also normalize the coefficient of the highest derivative of each $\phi_{i}$ to one, i.e.,
		\begin{equation*}
			\alpha_{n_{i}, i i}
            \; = \; 1,
            \,\,\,
            i
            \; = \;
            1, \ldots, s,
		\end{equation*}
		where $\alpha_{n_{i}, i i}$ is the $ii$th component of the matrix $\balpha_{n_i}$, and $n_{i}$ identifies the highest derivative of $\phi_{i}$. In order to make sure that the input field $\bd$ in~(\ref{eq.system-ode2}) does not directly influence the boundary conditions and the output field $\bvphi$, we impose the following technical assumptions on system~(\ref{eq.system-ode2}),
		\[
	    \ell \, < \, n,
        ~~
		m \, < \, n \, - \, \ell,
        ~~
		k \, < \, n \, - \, m.
		\]		
This assumption is satisfied in most physical problems of interest.

	\vspace*{0.15cm}

		Alternatively we can bring~(\ref{eq.system-ode2}) into a system of first-order differential equations (in $y$). This can be done by introducing state variables, $\{ \bx_i(y) \}$, where each of the states represents a linear combination of $\bphi$ and $\bd$, and their derivatives up to a certain order. A procedure for converting a high-order two point boundary value realization~(\ref{eq.system-ode2}) with spatially varying coefficients to a system of first-order ODEs is described in~\ref{sec.app-realization}. This transformation yields the {\em spatial\/} state-space representation of the frequency response operator~$\ct$
		\begin{equation}
	 		\label{eq.T-ssfirst}
			\ct:
			\left\{
			\begin{array}{rcl}
				\bx'(y)
				& \!\! = \!\! &
				\bA_0(y) \, \bx(y) ~+~ \bB_0(y) \, \bd(y),
				\\[.1cm]
				\bvphi(y)
                & \!\! = \!\! &
                \bC_0(y) \, \bx(y),
				\\[.1cm]
				0
                & \!\! = \!\! &
                \bN_a \, \bx (a) ~+~ \bN_b \, \bx (b),
			\end{array}
			\right.
		\end{equation}
		where $\bx$ is the state vector, $\bA_{0}$, $\bB_{0}$, and $\bC_{0}$ are matrices with, in general, spatially varying entries, $\bN_a$ and $\bN_b$ are constant matrices that specify the boundary conditions, and $\bx' = \mrd \bx/\mrd y$. To avoid redundancy in boundary conditions, $\bN_a$ and $\bN_b$ are chosen so that the matrix $\obt{\bN_a}{\bN_b}$ has a full row rank. We note that~(\ref{eq.T-ssfirst}) is well-posed (that is, it has a unique solution for any input $\bd$) if and only if~\cite{gohkaa84}
	\[
	\det
    \left(
    \bN_a \,+\, \bN_b \, \bPhi_0 (b,a)
    \right)
    \; \neq \;
    0,
	\]
	where $\bPhi_0 (y,\eta)$ is the state transition matrix of $\bA_0(y)$,
    \beq
        \dfrac{\mrd \bPhi_0 (y, \eta)}{\mrd y}
        \; = \;
        \bA_0 (y) \, \bPhi_0 (y, \eta),
        ~~
        \bPhi_0 (\eta, \eta)
        \; = \;
        I,
        \non
        \eeq
and $\det \left( \cdot \right)$ is the determinant of a given matrix.

	\vspace*{0.15cm}
	
For the 1D diffusion equation of Section~\ref{sec.motivating-ex-heat}, the input-output differential equation and the corresponding spatial state-space representation of the frequency response operator are given by~(\ref{eq.heat-eq-tpbvp}) and~(\ref{eq.TPBVSR-heat}), respectively. Note that the boundary conditions~(\ref{eq.heat-eq-tpbvp-b}) can be rewritten into the form required by~(\ref{eq.system-ode2}),
		\begin{equation}
		\label{eq.heat-BC}
			\left(
			\left[
			\begin{array}{cc}
				1 \\
				0
			\end{array}
			\right]
			E_{-1}
			\, + \,
			\left[
			\begin{array}{cc}
				0 \\
				1
			\end{array}
			\right]
			E_{1} \right) \phi(y)
			\; = \,
        \tbo{0}{0}.
        \non
		\end{equation}
				
	\subsection{Representations of the adjoint operator $\ct^{\star}$}
	\label{sec.representations-ctstar}

		We next describe the procedure for obtaining the two point boundary value representations of the adjoint of the frequency response operator, $\ct$, $\ct^{\star}$: $\bbf \mapsto \bg$; see figure~\ref{fig.blkdiag-ts}. As shown above, the operator $\ct$ can be recast into the input-output differential equation~(\ref{eq.system-ode2}), and the corresponding representation of $\ct^{\star}$ is given by
		\begin{equation}
			\label{eq.system-ode2-ad}
			\ct^{\star}:
			\left\{
			\begin{array}{rcl}
				\left[ \, \ca_0^{\star} \, \bpsi \, \right](y)
                & \!\! = \!\! &
                \left[ \, \cc_0^{\star} \, \bbf \, \right](y),
                \\[0.1cm]
				\bg(y)
                & \!\! = \!\! &
                \left[ \, \cb_0^{\star} \, \bpsi \, \right](y),
                \\[0.1cm]
				0
                & \!\! = \!\! &
                {\cal N}_0^{\star} \, \bpsi \, (y).
			\end{array}
			\right.
		\end{equation}
		Here, the adjoint operators are~\cite{renrog04,crolagsch95}
		\begin{equation*}
			\begin{array}{rclrcl}
				\left[ \, \ca_0^{\star} \, \bpsi \, \right] (y)
				& \!\! = \!\! &
				\ds{\sum_{i \, = \, 0}^{n} \left( -1 \right)^{i}
			       	\left[ \bD^{(i)} \left( \balpha^*_{i} \, \bpsi \right) \right] (y)},
				&
				\left[ \, \cc_0^{\star} \, \bbf \, \right] (y)
                & \!\! = \!\! &
                {\ds \sum_{i \, = \, 0}^{k} \left( -1 \right)^{i}  \left[ \bD^{(i)} \left( \bgamma^*_{i} \, \bbf \right) \right] (y)},
				\\[0.1cm]
				\left[ \, \cb_0^{\star} \, \bpsi \, \right] (y)
                & \!\! = \!\! &
                {\ds \sum_{i \, = \, 0}^{m} \left( -1 \right)^{i}  \left[ \bD^{(i)} \left( \bbeta^{*}_{i} \, \bpsi \right) \right] (y)},
                			&
				\left[ \, {\cal N}_{0}^{\star} \, \bpsi \, \right] (y)
		                & \!\! = \!\! &
		                {\ds \sum_{i \, = \, 0}^{\ell}
		                \left( \bW^{\star}_{a, i} \, \bE_{a} \,  +  \, \bW^{\star}_{b, i} \, \bE_{b} \right) \left[ \bD^{(i)} \, \bpsi \right] (y)},
			\end{array}
		\end{equation*}
		where $\balpha_{i}^*$, $\bbeta_{i}^*$, and $\bgamma_{i}^*$ are the complex-conjugate-transposes of the matrices $\balpha_{i}$, $\bbeta_{i}$, and $\bgamma_{i}$.
		The boundary conditions on the adjoint variable $\bpsi$ are determined so that the boundary terms vanish when determining the adjoint of the operator $\ca_{0}$. A procedure describing how to determine the boundary conditions of the adjoint system is given in~\cite[][Section~5.5]{renrog04}.
		
		\vspace*{0.15cm}
					
		On the other hand, the state-space representation of the adjoint of the operator $\ct$ is given in~\cite{gohkaa84}
		\beq
		 	\label{eq.T-sssecond}
			\ct^{\star}:
			\left\{
			\begin{array}{rcl}
				\bz'(y)
				& \!\! = \!\! &
				-\bA^{*}_0(y) \, \bz(y)
                \; - \;
                \bC^{*}_0(y) \, \bbf(y),
				\\[.1cm]
				\bg(y)
                & \!\! = \!\! &
                \bB^{*}_0(y) \, \bz(y),
				\\[.1cm]
				0
                & \!\! = \!\! &
                \bM_a \, \bz (a) \; + \; \bM_b \, \bz (b),
			\end{array}
			\right.
		\eeq
		where $\bA^{*}_{0}$, $\bB^{*}_{0}$, and $\bC^{*}_{0}$ denote the complex-conjugate-transposes of the matrices $\bA_{0}$, $\bB_{0}$, and $\bC_{0}$. The boundary condition matrices $\bM_a$ and $\bM_b$ are determined so that $\obt{\bM_a}{\bM_b}$ has a full row rank and
		 \beq
		 	\label{eq.M1M2}
			\obt{\bM_a}{\bM_b}
			\left[
            \ba{r}
            {\bN_a^*}
            \\[0.1cm]
            {- \bN_b^*}
            \ea
            \right]
			\; = \;
			0.
		\eeq
		A procedure for selecting $\bM_a$ and $\bM_b$ that satisfy these two requirements is described in~\cite[][Section~3.1]{jovbamSCL06}. Furthermore, we note that the well-posedness of the adjoint representation~(\ref{eq.T-sssecond}) is guaranteed by the well-posedness of $\ct$.
		
		\begin{figure}
			\begin{center}
			    	{ \setlength{\unitlength}{0.8cm}
			    	\begin{picture}(9,2.5)

	\put(0,1){\vector(1,0){2}}
	\put(1,1.2){\makebox(0,0)[b]{$ \bbf $}}
	\put(2,0.5){\framebox(2,1){
	$
	\ct^{\star}
	$}}
	\put(4,1){\vector(1,0){2}}
	\put(5,1.2){\makebox(0,0)[b]{$ \bg = \bd$}}

	\put(6,0.5){\framebox(2,1){
	$
	\ct
	$}}
	\put(8,1){\vector(1,0){2}}
	\put(9,1.2){\makebox(0,0)[b]{$ \bvphi $}}

\end{picture}
			    	}
				\vspace{-0.7cm}
			\end{center}
			\caption{A cascade connection of $\ct^{\star}$ and $\ct$ with $\ct \ct^{\star}$: $\bbf \mapsto \bvphi$.}
			\label{fig.blkdiag-tts}
		\end{figure}
		
		\vspace*{0.15cm}

		For the 1D diffusion equation of Section~\ref{sec.motivating-ex-heat}, the adjoint of the operator $\ct (\omega)$ described by~(\ref{eq.heat-eq-tpbvp}) has the following input-output representation
		\begin{equation}
		\label{eq.IO-heat-ad}
			\ct^{\star}(\omega):
			\left\{
			\begin{array}{rcl}
				\left( D^{(2)} + \mri \omega I \right) \psi(y) & \!\! = \!\! & f(y),
                \\[0.15cm]
				g(y) & \!\! = \!\! & -\psi(y),
                \\[0.15cm]
				\left(
				\left[
				\begin{array}{cc}
					1 \\
					0
				\end{array}
				\right]
				E_{-1}
				\, + \,
				\left[
				\begin{array}{cc}
					0 \\
					1
				\end{array}
				\right]
				E_{1} \right) \psi(y)
				& \!\! = \!\! &
                \tbo{0}{0}.
			\end{array}
			\right.
		\end{equation}
		As specified in~(\ref{eq.T-sssecond}), the state-space representation of $\ct^{\star}(\omega)$ is determined by taking the appropriate complex-conjugate-transposes of the corresponding matrices in~(\ref{eq.TPBVSR-heat}) with the following boundary condition matrices
		\begin{equation}
		\label{eq.TPBVSR-heat-ad-bc}
			M_{1} \, = \,
			\left[
			\begin{array}{cc}
				0 & 1 \\
				0 & 0
			\end{array}
			\right],
			\,\,\,
			M_{2} \, = \,
			\left[
			\begin{array}{cc}
				0 & 0 \\
				0 & 1
			\end{array}
			\right].
        \non
		\end{equation}

	\subsection{Representations of $\ct \ct^{\star}$}
		
From the above described representations of $\ct$ and $\ct^{\star}$, we can determine corresponding representations of the operator $\ct \ct^{\star}$: $\bbf \mapsto \bvphi$. As illustrated in figure~\ref{fig.blkdiag-tts}, this operator represents a cascade connection of the frequency response operator $\ct$ and its adjoint $\ct^{\star}$. The input-output differential equation for $\ct \ct^{\star}$ is obtained by equating the output of $\ct^{\star}$ in~(\ref{eq.system-ode2-ad}) with the input of $\ct$ in~(\ref{eq.system-ode2}), i.e., $\bd = \bg$, yielding
		\begin{equation}
			\label{eq.system-ode2-tts}
			\ct \ct^{\star}:
			\left\{
			\begin{array}{rcl}
				\left[ \, \ca \, \bxi \, \right] (y)
                & \!\! = \!\! &
                \left[ \, \cb \, \bbf \, \right] (y),
                \\[0.1cm]
				\bvphi(y)
                & \!\! = \!\! &
                \left[ \, \cc \, \bxi \, \right] (y), \\[0.1cm]
				0
                & \!\! = \!\! &
                {\cal N} \, \bxi \, (y),
			\end{array}
			\right.
		\end{equation}
		where
		\beq
			\non
			\ba{c}
				\bxi(y) \; = \; \tbo{\bphi(y)}{\bpsi(y)},
				\,\,\,
				\ca
                \; = \;
				\left[
				\ba{cc}
					\ca_0 & -\cb_0 \, \cb_0^{\star} \\[.15cm]
					0 & \ca_0^{\star}
				\ea
				\right],
				\\[0.5cm]
				{\cal N}
				\; = \;
				\tbt{{\cal N}_0}{0}{0}{{\cal N}_0^{\star}},
				\,\,
				\cb
				\; = \;
				\tbo{0\\[-.3cm]}{\cc_0^{\star}},
				\,\,
				\cc =  \obt{\cc_0}{0}.
			\ea
		\eeq
		Similarly, the spatial state-space representation of $\ct \ct^{\star}$ is obtained by equating the input $\bd$ in~(\ref{eq.T-ssfirst}) to the output $\bg$ in~(\ref{eq.T-sssecond}), which yields
		\beq
			\label{eq.HHstar-ss}
			{\cal T} {\cal T}^\star:
			\left\{
			\begin{array}{rcl}
				\bq'(y)
				& \!\! = \!\! &
				\bA(y) \, \bq(y)
                \; + \;
                \bB(y) \, \bbf(y),
				\\[.1cm]
				\bvphi(y)
                & \!\! = \!\! &
                \bC(y) \, \bq(y),
				\\[.1cm]
				0
                & \!\! = \!\! &
                \bL_a \, \bq(a) \; + \; \bL_b \, \bq(b),
			\end{array}
			\right.
		\eeq
		with
		\beq
			\non
			\ba{rclrcl}
				\bq(y) & \!\! = \!\! & \tbo{\bx(y)}{\bz(y)},
				&
				\bA(y) & \!\! = \!\! &
				\left[
				\ba{cc}
					\bA_0(y) & \bB_0(y) \, \bB_0^*(y) \\[.15cm]
					0 & - \bA_0^*(y)
				\ea
				\right],
				\\[0.5cm]
				\bB(y)
				& \!\! = \!\! &
				\tbo{0\\[-.3cm]}{-\bC_0^*(y)},
				&
				\bC(y) & \!\! = \!\! & \obt{\bC_0(y)}{0},
				\\[0.5cm]
				\bL_a
				& \!\! = \!\! &
				\tbt{\bN_a}{0}{0}{\bM_a},
				&
				\bL_b
				& \!\! = \!\! &
				\tbt{\bN_b}{0}{0}{\bM_b}.
			\ea
		\eeq
		Since a cascade connection of two well-posed systems is well-posed, the existence and uniqueness of solutions of~(\ref{eq.system-ode2-tts}) and~(\ref{eq.HHstar-ss}) is guaranteed by the well-posedness of the corresponding two point boundary value representations of $\ct$ and $\ct^{\star}$.
		
		\vspace*{0.15cm}
		
	We next present a procedure for computing the largest singular value of $\ct$ using the above representations of the operator $\ct \ct^\star$.

\section{Computation of the largest singular value of $\ct$}
\label{sec.computation}
	
	In this section, we utilize the structure of the two point boundary value representations~(\ref{eq.system-ode2-tts}) and~(\ref{eq.HHstar-ss}) of $\ct \ct^{\star}$ to develop a method for computing the largest singular value of the frequency response operator $\ct (\omega)$,
	\begin{equation*}
		\sigma_{\max}^{2} \left( \ct (\omega) \right)
		\; = \;
		\lambda_{\max} \left( \ct (\omega) \, \ct^{\star} (\omega) \right),
	\end{equation*}
	where $\lambda_{\max} (\cdot)$ denotes the largest eigenvalue of a given operator. In what follows, we present the procedure for computing the eigenvalues of $\ct \ct^{\star}$ using both input-output~(\ref{eq.system-ode2-tts}) and state-space~(\ref{eq.HHstar-ss}) representations of $\ct \ct^{\star}$. This is done by first recasting the system of differential equations into a corresponding integral formulation; we then employ the recently developed automatic Chebyshev spectral collocation method~\cite{chebfunv4} to solve the eigenvalue problem for the resulting integral equation. Note that the eigenfunction corresponding to the largest singular value identifies the output of the system that is most amplified in the presence of disturbances. Similar procedure can be used to determine the principal eigenfunction of the operator $\ct^{\star} \ct$, thereby yielding the input that has the largest influence on the system's output.
	
	\vspace*{0.15cm}
	
    		The solution to a two point boundary value problem~(\ref{eq.system-ode2-tts}) can be obtained numerically by approximating the differential operators using, e.g., a pseudo-spectral collocation technique~\cite{canhusquazan88,boy89,treFinite96,treSpectral00}. For differential equations of a high-order, the resulting finite-dimensional approximations may be poorly conditioned. This difficulty can be overcome by converting a high-order differential equation into a corresponding integral equation~\cite{gre91}. This conversion utilizes indefinite integration operators that are characterized by condition numbers that remain bounded upon discretization refinement, thereby alleviating ill-conditioning associated with finite dimensional approximation of high-order differential operators. The procedure for achieving this conversion, described in Section~\ref{sec.inteigs}, extends the result of~\cite{dri10} from a scalar case to a system of high-order differential equations. Furthermore, in Section~\ref{sec.ss-solve} we show how a spatial state-space representation~(\ref{eq.HHstar-ss}) can be transformed to an equivalent integral form. Finally, we employ {\sf Chebfun}'s function {\sf eigs} to perform the eigenvalue decomposition of the resulting system of equations.
		
		\subsection{An illustrative example}
        \label{sec.example-integral-form}
			
			We first illustrate the procedure for converting a differential equation into its corresponding integral form using the 1D diffusion equation~(\ref{eq.heat-eq-tpbvp}),
			\begin{subequations}
			\label{eq.IO-heat-2}
				\begin{align}
					\label{eq.IO-heat-2-diffeq}
					\left( D^{(2)} - \mri \omega I \right) \phi(y) \,\, = \,\, & - d(y), \\[0.1cm]
					\label{eq.IO-heat-2-bc}
					\left(
					\left[
					\begin{array}{c}
						1 \\
						0
					\end{array}
					\right]
					E_{-1}
					+
					\left[
					\begin{array}{c}
						0 \\
						1
					\end{array}
					\right]
					E_{1} \right) \phi(y)
					\,\, = \,\, &
					\left[
					\begin{array}{c}
						0 \\
						0
					\end{array}
					\right].
				\end{align}
			\end{subequations}
System~(\ref{eq.IO-heat-2}) can be converted into an equivalent integral equation by introducing an auxiliary variable
			\begin{equation}
			\label{eq.integral-form-sol-u}
				\nu(y)
				\; = \;
				\left[ D^{(2)} \, \phi \right] (y).
			\end{equation}
			Integration of~(\ref{eq.integral-form-sol-u}) yields
			\begin{equation}
			\label{eq.phi-varphi-k}
				\begin{array}{rcl}
                \phi'(y)
				& = &
                \ds{\int_{-1}^{y}} \nu(\eta_{1}) \, \mrd \eta_{1}
                \; + \;
                k_1
                ~ = ~
                \left[
                J^{(1)} \, \nu
                \right] (y)
                \; + \; k_1,
                \\[0.35cm]
				\phi(y)
				& = &
				{\ds \int_{-1}^{y} \left( \int_{-1}^{\eta_{2}} \nu(\eta_{1}) \, \mrd \eta_{1} \right)
                \mrd\eta_{2}
                \; + \;
                k_{1} \left( y \, + \, 1 \right)
                \; + \;
                k_{2}}
				\\[0.4cm]
				& = &
				\left[ J^{(2)} \, \nu \right] (y)
				\; + \;
				K^{(2)} \, \mathbf{k},
				\end{array}
			\end{equation}
			where $J^{(1)}$ and $J^{(2)}$ denote the indefinite integration operators of degrees one and two, the vector
    $
    \mathbf{k}
    =
    \left[ \begin{array}{cc} k_{2} & k_{1} \end{array} \right]^T
    $
contains the constants of integration which are to be determined from the boundary conditions~(\ref{eq.IO-heat-2-bc}), and
			\begin{equation*}
				K^{(2)} \; = \; \left[ \begin{array}{cc} 1 & (y + 1) \end{array} \right].
			\end{equation*}
			The integral form of the 1D diffusion equation is obtained by substituting~(\ref{eq.phi-varphi-k}) into~(\ref{eq.IO-heat-2}),
			\begin{subequations}
			\label{eq.heat-varphi-bc}
				\begin{align}
					\label{eq.heat-varphi}
					\left( I - \mri \omega J^{(2)} \right) \nu(y) \, - \, \mri \omega \, K^{(2)} \, \mathbf{k} \,\, = \,\, & - d(y),
    \\[0.cm]
					\label{eq.heat-bc}
					\left[
					\begin{array}{cc}
						1 & 0 \\
						1 & 2
					\end{array}
					\right]
					\left[
					\begin{array}{c}
						k_2 \\
						k_1
					\end{array}
					\right]
					\; + \;
					\left(
					\left[
					\begin{array}{c}
						1 \\
						0
					\end{array}
					\right]
					E_{-1}
					\, + \,
					\left[
					\begin{array}{c}
						0 \\
						1
					\end{array}
					\right]
					E_{1} \right) \left[ J^{(2)} \nu \right] (y)
					\,\, = \,\, &
					\left[
					\begin{array}{c}
						0 \\
						0
					\end{array}
					\right].
				\end{align}
			\end{subequations}
Now, by observing that
			\[
				E_{-1} \left[ J^{(1)} \nu \right] (y)
				\; = \;
				{\ds \int_{-1}^{-1} \nu(\eta) \, \mrd \eta}
				\; = \;
				0,
			\]
			we can use~(\ref{eq.heat-bc}) to express the constants of integration $\bk$ in terms of $\nu$,
			\begin{equation}
			\label{eq.heat-bc-exp}
				\left[
				\begin{array}{c}
					k_{2} \\[0.1cm]
					k_{1}
				\end{array}
				\right]
				\; = \;
				-
				\cfrac{1}{2}
				\left[
				\begin{array}{rc}
					2 & 0 \\[0.1cm]
					-1 & 1
				\end{array}
				\right]
				\left[
				\begin{array}{c}
					0 \\[0.1cm]
					1
				\end{array}
				\right]
				E_{1}
				\,
				\left[
				J^{(2)}
				\,
				\nu
				\right] (y)
				\; = \;
				\left[
				\begin{array}{c}
					0 \\[0.1cm]
					-1/2
				\end{array}
				\right]
				E_{1}
				\,
				\left[ J^{(2)} \, \nu \right] (y)
				.
			\end{equation}
			Finally, substitution of~(\ref{eq.heat-bc-exp}) into~(\ref{eq.heat-varphi}) yields an equation for $\nu$,
			\begin{equation}
			\label{eq.heat-sol-u}
				\left(
				I \, - \, \mri \omega J^{(2)} \, + \, \cfrac{1}{2} \, \mri \omega \left( y \, + \, 1 \right) E_{1} J^{(2)}
				\right)
				\nu(y)
				\; = \;
				-d(y).
			\end{equation}
			
	\vspace*{0.15cm}
			
    Invertibility of the matrix that multiplies the integration constants
            $
            \mathbf{k}
            =
            \left[ \begin{array}{cc} k_{2} & k_{1} \end{array} \right]^T
            $
in~(\ref{eq.heat-bc}) facilitates derivation of an explicit expression for $\bk$ in terms of $\nu$. In situations where this invertibility condition fails to be satisfied, we next use the 1D reaction-diffusion equation with homogenous Neumann boundary conditions,
			\begin{subequations}
			\label{eq.heat-neumann}
				\begin{align}
					\label{eq.heat-neumann-diffeq}
					\left( D^{(2)} \, - \, c I \, - \, \mri \omega I \right) \phi(y) \,\, = \,\, & - d(y), \\[0.1cm]
					\label{eq.heat-neumann-bc}
					\left(
					\left[
					\begin{array}{c}
						1 \\
						0
					\end{array}
					\right]
					E_{-1}
					+
					\left[
					\begin{array}{c}
						0 \\
						1
					\end{array}
					\right]
					E_{1} \right) \left[ D^{(1)} \phi \right] (y)
					\,\, = \,\, &
					\left[
					\begin{array}{c}
						0 \\
						0
					\end{array}
					\right],
				\end{align}
			\end{subequations}
to illustrate a procedure for obtaining an input-output representation that only contains indefinite integration operators and point evaluation functionals. Substitution of~(\ref{eq.phi-varphi-k}) to~(\ref{eq.heat-neumann}) yields
			\begin{subequations}
			\label{eq.heat-NBC-v-bc}
				\begin{align}
					\label{eq.heat-NBS-v}
					\left( I \, - \, (\mri \omega \, + \, c ) \, J^{(2)} \right) \nu(y)
                \, - \,
                ( \mri \omega \, + \, c ) \, K^{(2)} \, \mathbf{k} \,\, = \,\, & - d(y),
    \\[0.cm]
					\label{eq.heat-NBC-bc}
					\left[
					\begin{array}{cc}
						0 & 1 \\
						0 & 1
					\end{array}
					\right]
					\left[
					\begin{array}{c}
						k_2 \\
						k_1
					\end{array}
					\right]
					\; + \;
					\left[
					\begin{array}{c}
						0 \\
						1
					\end{array}
					\right]
					E_{1} \, \left[ J^{(1)} \nu \right] (y)
					\,\, = \,\, &
					\left[
					\begin{array}{c}
						0 \\
						0
					\end{array}
					\right].
				\end{align}
			\end{subequations}		
			A positive reaction rate $c$ in~(\ref{eq.heat-neumann-diffeq}) ensures stability in the presence of Neumann boundary conditions.
			
			\vspace*{0.15cm}

Lack of invertibility of the matrix that multiplies the integration constants in~(\ref{eq.heat-NBC-bc}) is an obstacle to determining $\bk$ explicitly in terms of $\nu$. Instead, the dependence of $\nu$ on $\bk$ and $d$ can be obtained from~(\ref{eq.heat-NBS-v}),
			\begin{equation}
			\label{eq.heat-NBC-nu}
				\nu(y)
				\; = \;
				\left( I \, - \, (\mri \omega \, + \, c ) \, J^{(2)} \right)^{-1} \,
				\left( (\mri \omega \, + \, c ) \, K^{(2)} \, \bk \, - \, d(y)\right).
			\end{equation}
			Now, substitution of~(\ref{eq.heat-NBC-nu}) to~(\ref{eq.heat-NBC-bc}) yields
			\begin{equation}
			\label{eq.heat-NBC-k}
				\bk
				\; = \;
				\mathbf{G}^{-1}
				\left[
				\begin{array}{c}
					0 \\[0.1cm]
					1
				\end{array}
				\right]
				E_{1} \, J^{(1)} \,
				\left( I \, - \, (\mri \omega \, + \, c ) \, J^{(2)} \right)^{-1} \, d(y),
			\end{equation}
			where the matrix $\mathbf{G}$ is given by
			\begin{equation*}
				\mathbf{G}
				\; = \;
				\left[
				\begin{array}{cc}
					0 & 1 \\[0.1cm]
					0 & 1
				\end{array}
				\right]
				\, + \,
				\left[
				\begin{array}{c}
					0 \\[0.1cm]
					1
				\end{array}
				\right]
				E_{1} \, J^{(1)} \,
				\left( I \, - \, (\mri \omega \, + \, c ) \, J^{(2)} \right)^{-1}
                (\mri \omega \, + \, c ) \, K^{(2)}.
			\end{equation*}
			Finally, an equation for $\nu$ is obtained by substituting~(\ref{eq.heat-NBC-k}) into~(\ref{eq.heat-NBS-v}),
			\begin{equation}
			\label{eq.heat-NBC-sol}
				\left( I \, - \, (\mri \omega \, + \, c ) \, J^{(2)} \right) \nu(y)
				\; = \;
				\left(
				(\mri \omega \, + \, c ) \, K^{(2)} \,
				\mathbf{G}^{-1}
				\left[
				\begin{array}{c}
					0 \\[0.1cm]
					1
				\end{array}
				\right]
				E_{1} \, J^{(1)} \,
				\left( I \, - \, (\mri \omega \, + \, c ) \, J^{(2)} \right)^{-1}
				\, - \,
				I
				\right)
				d(y).
			\end{equation}
			
			\vspace*{0.15cm}
			
			Systems~(\ref{eq.heat-sol-u}) and~(\ref{eq.heat-NBC-sol}) only contain indefinite integration operators and point evaluation functionals which are known to be well-conditioned. This is a major advantage compared to their corresponding input-output differential equations~(\ref{eq.IO-heat-2}) and~(\ref{eq.heat-neumann}).

		\subsection{Integral form of a system of high-order differential equations}
		\label{sec.inteigs}
			
			We now present the procedure for converting a system of high-order differential equations~(\ref{eq.system-ode2-tts}),
    			\begin{equation}
			\label{eq.tts-diffeq-1}
				{\ct \ct^{\star}}:
				\left\{
				\begin{array}{rcl}
					\left[ \, \ca \, \bxi \, \right] (y)
					& \!\! = \!\! &
					\left[ \, \cb \, \bbf \, \right] (y),
					\\[0.1cm]
					\bvphi(y)
					& \!\! = \!\! &
					\left[ \, \cc \, \bxi \, \right] (y), \\[0.1cm]
					0
					& \!\! = \!\! &
					{\cal N} \, \bxi \, (y),
				\end{array}
				\right.
			\end{equation}
			to an equivalent integral form. The input and output vectors $\bbf (y)$ and $\bvphi (y)$ have $p$ elements, $\bxi (y)$ is a $2s$-vector, and the operators in~(\ref{eq.tts-diffeq-1}) are given by
			\begin{equation*}
				\ca
				\; = \;
				{\ds \sum_{i \, = \, 0}^{n} \mathbf{a}_{i}(y) \, \bD^{(i)}},
				\;\;
				\cb
				\; = \,
				{\ds \sum_{i \, = \, 0}^{k} \mathbf{b}_{i}(y) \, \bD^{(i)}},
				\;\;
				\cc
				\; = \;
				{\ds \sum_{i \, = \, 0}^{k} \mathbf{c}_{i}(y) \, \bD^{(i)}},
				\;\;
				{\cal N}
				\; = \;
				{\ds \sum_{i \, = \, 0}^{\ell}
				\left( \bYY_{a, i} \, \bE_{a} \,  +  \, \bYY_{b, i} \, \bE_{b} \right) \bD^{(i)}}.
			\end{equation*}
 			As illustrated in~Section~\ref{sec.example-integral-form}, instead of trying to find the solution $\bxi$ to~(\ref{eq.system-ode2-tts}) directly, we introduce two auxiliary variables, $\bnu$ and $\bk$. The $i$th component of the vector $\bnu (y) = \left[ \begin{array}{ccc} \nu_{1} (y) & \ldots & \nu_{2 s} (y) \end{array} \right]^T$ is determined by
			\begin{equation}
			\label{eq.u-xi-diffeq}
				\nu_i (y)
				\; = \;
				\left[
				\,
				D^{(n_i)} \, \xi_i
				\,
				\right]
				(y),
			\end{equation}
		    	where $n_{i}$ denotes the highest derivative of $\xi_{i}$ in
			\[
			\left[
			\,
			\ca
			\,
			\bxi
			\,
			\right]
			(y)
			\; = \;
			\left[
			\,	
			\cb
			\,
			\bbf
			\,
			\right]
			(y).
			\]
			Integration of~(\ref{eq.u-xi-diffeq}) yields
			\begin{equation}
			\label{eq.xi-u-c}
				\left[ D^{(j)} \xi_{i} \right] (y)
				\; = \;
				\left[ J^{(n_{i} - j)} \nu_{i} \right] (y)
				\, + \,
				K^{(n_{i} - j)} \, \bk_{i},
				\,\,\,
				j = 0, \ldots, n_{i},
			\end{equation}
			where $\bk_{i} \in \bbC^{n_{i}}$ is the vector of integration constants which are to be determined from the boundary conditions, $J^{(n_i)}$ is the indefinite integration operator of degree $n_i$ with $J^{(0)} = 0$, and $K^{(n_i)}$ is the matrix with columns that span the vector space of polynomials of degree less than $n_i$,
			\begin{equation*}
				\begin{array}{c}
					K^{(n_i)}
                    \; = \;
                    \left[
					\ba{cccc}
					K_{0}(y) & K_{1}(y) & \cdots & K_{n_i-1}(y)
					\ea
					\right],
                    ~~~
                    K^{(0)} \; = \; 0,
					\\[0.1cm]
					K_{0}(y) \; = \; 1,
					\,\,\,
					K_{j}(y) \; = \; \dfrac{1}{j!} \left( y - a \right)^{j},
                    ~
                    j \, \geq \, 1.
				\end{array}
			\end{equation*}
			Substitution of~(\ref{eq.xi-u-c}) into~(\ref{eq.tts-diffeq-1}) yields the integral representation of the operator $\ct \ct^{\star}$,
			\begin{equation}
				\label{eq.blocksys}
				\ct \ct^{\star}:
				\left\{
				\begin{array}{l}
					\left[
					\begin{array}{cc}
						{\cal L}_{11} & {\cal L}_{12} \\[0.1cm]
						{\cal L}_{21} & {\cal L}_{22} \\[0.1cm]
					\end{array}
					\right]
					\,
					\left[
					\begin{array}{c}
						\bnu \\[0.1cm]
						\bk
					\end{array}
					\right]
					\; = \;
					\left[
					\begin{array}{c}
						\cb \\[0.1cm]
						0
					\end{array}
					\right] \, \bbf,
	                \\[0.5cm]
					\bvphi
					\; = \;
					\left[
					\begin{array}{cc}
						{\cal P}_{1} & {\cal P}_{2}
					\end{array}
					\right]
					\left[
					\begin{array}{c}
						\bnu \\[0.1cm]
						\bk
					\end{array}
					\right],
				\end{array}
				\right.
			\end{equation}
			where
			\begin{equation*}
				\begin{array}{rclrcl}
					{\cal L}_{11}
					& \!\! = \!\! &
					{\ds \sum_{i \, = \, 0}^{n} \mathbf{a}_{i}(y) \, \bJ^{(n - i)}},
					&
					{\cal L}_{12}
					& \!\! = \!\! &
					{\ds \sum_{i \, = \, 0}^{n} \mathbf{a}_{i}(y) \, \mathbf{K}^{(n - i)}},
					\\[0.5cm]
					{\cal L}_{21}
					& \!\! = \!\! &
					{\ds \sum_{i \, = \, 0}^{\ell}
					{\bYY}_{b, i} \, \bE_{b} \, \bJ^{(n - i)}},
                    			&
					{\cal L}_{22}
					& \!\! = \!\! &
					{\ds \sum_{i \, = \, 0}^{\ell}
					\left( {\bYY}_{a, i} \, \bE_{a} \,  +  \, {\bYY}_{b, i} \, \bE_{b} \right) \mathbf{K}^{(n - i)}},
					\\[0.5cm]
					{\cal P}_{1}
					& \!\! = \!\! &
					{\ds \sum_{i \, = \, 0}^{k} \mathbf{c}_{i}(y) \, \bJ^{(n - i)}},
					&
					{\cal P}_{2}
					& \!\! = \!\! &
					{\ds \sum_{i \, = \, 0}^{k} \mathbf{c}_{i}(y) \, \bK^{(n - i)}},
					\end{array}
			        \end{equation*}
                    \begin{equation*}
				    \begin{array}{rclrcl}
					\bJ^{(n - i)}
					& \!\! = \!\! &
					\left[
					\begin{array}{ccc}
						J^{(n_{1} - i)} & \hspace*{-0.5cm} & \hspace*{-0.4cm}\\
						& \hspace*{-0.5cm} \ddots & \hspace*{-0.4cm}\\
						& \hspace*{-0.5cm} & \hspace*{-0.2cm} J^{(n_{2s} - i)}
					\end{array}
					\right],
					&
					\bK^{(n - i)}
					& \!\! = \!\! &
					\left[
					\begin{array}{ccc}
						K^{(n_{1} - i)} & \hspace*{-0.5cm} & \hspace*{-0.4cm}\\
						& \hspace*{-0.5cm} \ddots & \hspace*{-0.4cm}\\
						& \hspace*{-0.5cm} & \hspace*{-0.2cm} K^{(n_{2s} - i)}
					\end{array}
					\right],
                    \\[0.75cm]
                    J^{(i)} & \!\! = \!\! & 0,
                    ~~
                    K^{(i)} ~=~ 0,
                    ~~~
                    i \, \leq \, 0.
					\end{array}
			\end{equation*}
			Using~(\ref{eq.blocksys}) we can determine an expression for the integration constants,
			\begin{equation}
			\label{eq.L22-k-nu}
				{\cal L}_{22} \, \bk
	            		\; = \;
				- \left[ \, {\cal L}_{21} \, \bnu \, \right] (y).
			\end{equation}
			If the matrix ${\cal L}_{22}$ is invertible, equation~(\ref{eq.L22-k-nu}) in conjunction with~(\ref{eq.blocksys}) yields
			\begin{subequations}
				\label{eq.inteqs}
				\begin{align}
					\label{eq.inteqs-state}
					\bnu(y)
					\; = & \;
					\left[ \left( {\cal L}_{11} \, - \, {\cal L}_{12} \, {\cal L}_{22}^{-1} \, {\cal L}_{21} \, \right)^{-1} \, ({\cb} \, \bbf) \right] (y),
					\\[0.1cm]
                    \label{eq.inteqs-output}
					\bvphi(y)
					\; = & \; \left[ \left( {\cal P}_{1} \, - \, {\cal P}_{2} \, {\cal L}_{22}^{-1} \, {\cal L}_{21} \, \right) \bnu \right] (y),		
				\end{align}
			\end{subequations}
			and the representation of the operator $\ct \ct^{\star}$ is obtained by substituting~(\ref{eq.inteqs-state}) into~(\ref{eq.inteqs-output}). Thus, determination of the left singular functions $\{ \bu_n \}$ of the operator $\ct$ amounts to solving the following eigenvalue problem
			\begin{equation}
			\label{eq.eig-intbvp}
				\left[ \left( {\cal P}_{1} \, - \, {\cal P}_{2} \, {\cal L}_{22}^{-1} \, {\cal L}_{21} \, \right) \, \left( {\cal L}_{11} \, - \, {\cal L}_{12} \, {\cal L}_{22}^{-1} \, {\cal L}_{21} \, \right)^{-1} ( {\cb} \, \bu_n ) \right] (y)
				\; = \;
				\sigma_n^2 \, \bu_n (y),
			\end{equation}
			where $\sigma_n$ denotes the corresponding singular value of $\ct$.
			
	\vspace*{0.15cm}
			
	On the other hand, if ${\cal L}_{22}$ is singular, we can determine an expression for $\bnu$ in terms of $\bk$ and $\bbf$ from~(\ref{eq.blocksys}),
			\begin{equation}
			\label{eq.nu-f-k}
				\bnu(y)
				\; = \;
				\left[ {\cal L}_{11}^{-1}
				\,
				\cb \, \bbf \right] (y)
				\, - \,
				{\cal L}_{11}^{-1}
				\,
				{\cal L}_{12} \, \bk.
			\end{equation}
			Furthermore, substitution of~(\ref{eq.nu-f-k}) into~(\ref{eq.L22-k-nu}) yields
			\begin{equation}
			\label{eq.k-f-G}
				\bk
				\; = \;
                -
				\mathbf{G}^{-1}
				{\cal L}_{21}
				\,
				\left[ {\cal L}_{11}^{-1}
				\,
				\cb \, \bbf \right](y),
			\end{equation}
			where the matrix $\mathbf{G}$ is given by
			\begin{equation*}
				\mathbf{G}
				\; = \;
				{\cal L}_{22}
				-
				{\cal L}_{21}
				\,
				{\cal L}_{11}^{-1}
				\,
				{\cal L}_{12}.
			\end{equation*}
			This expression for $\bk$ in conjunction with~(\ref{eq.blocksys}) yields
			\begin{subequations}
				\label{eq.inteqs-newform}
				\begin{align}
					\label{eq.inteqs-new-state}
					\bnu(y)
					\; = & \;
					\left[ {\cal L}_{11}^{-1}
					\left(
					\cb
					\, + \,
					{\cal L}_{12} \, \mathbf{G}^{-1} \, {\cal L}_{21} \, {\cal L}_{11}^{-1} \cb \right) \bbf \right] (y),
					\\[0.1cm]
                    			\label{eq.inteqs-new-output}
					\bvphi(y)
					\; = & \;
					\left[ {\cal P}_{1} \, \bnu \right] (y)
					\, - \,
					\left[ {\cal P}_{2} \, \mathbf{G}^{-1} \, {\cal L}_{21} \, {\cal L}_{11}^{-1} \, \cb \, \bbf \right] (y).		
				\end{align}
			\end{subequations}
			The integral representation of the operator $\ct \ct^{\star}$ can be obtained by substituting~(\ref{eq.inteqs-new-state}) into~(\ref{eq.inteqs-new-output}), and the left singular pair $(\sigma_{n}, \bu_{n})$ of the operator $\ct$ is determined from the solution to the following eigenvalue problem
			\begin{equation}
			\label{eq.eig-intbvp-new}
				\left[
				\left(
				{\cal P}_{1} \, {\cal L}_{11}^{-1}
				\, + \,
				{\cal P}_{1} \, {\cal L}_{11}^{-1} \, {\cal L}_{12} \, \mathbf{G}^{-1} \, {\cal L}_{21} \, {\cal L}_{11}^{-1}
				\, - \,
				{\cal P}_{2} \, \mathbf{G}^{-1} \, {\cal L}_{21} \, {\cal L}_{11}^{-1}
				\right)
				( {\cb} \, \bu_n ) \right] (y)
				\; = \;
				\sigma_n^2 \, \bu_n (y).
			\end{equation}

	\subsection{Integral form of a spatial state-space representation}
    	\label{sec.ss-solve}
		
		We next describe a procedure for transforming a spatial state-space representation~(\ref{eq.HHstar-ss}),
		\beq
			\label{eq.HHstar-ss-1}
			{\cal T} {\cal T}^\star:
			\left\{
			\begin{array}{rcl}
				\bq'(y)
				& \!\! = \!\! &
				\bA(y) \, \bq(y)
                \; + \;
                \bB(y) \, \bbf(y),
				\\[.1cm]
				\bvphi(y)
                & \!\! = \!\! &
                \bC(y) \, \bq(y),
				\\[.1cm]
				0
                & \!\! = \!\! &
                \bL_a \, \bq(a) \; + \; \bL_b \, \bq(b),
			\end{array}
			\right.
		\eeq
into a system of first-order integral equations. In a similar manner as in Section~\ref{sec.inteigs}, we introduce two auxiliary variables $\bnu$ and $\bk$ so that
		\begin{equation}
		\label{eq.qu-sol}
			\bnu(y) \; = \; \bq'(y)
            ~~ \Rightarrow ~~
            \bq(y) = \left[ \, \bJ \, \bnu \, \right](y) \, + \, \bk,
		\end{equation}
		where $\bJ$ is a block diagonal matrix of the first order indefinite integration operators $J^{(1)}$,
		\begin{equation*}
			\bJ
			\; = \;
			\left[
			\begin{array}{ccc}
				J^{(1)} &  & \\[-0.1cm]
				& \ddots & \\
				&  &  J^{(1)}
			\end{array}
			\right].
		\end{equation*}
		Substitution of~(\ref{eq.qu-sol}) into~(\ref{eq.HHstar-ss-1}) yields a system of first order integral equations for the operator $\ct \ct^{\star}$,
		\begin{subequations}
		\label{eq.ss-integral}
			\begin{align}
				\label{eq.ss-integral-states}
				\bnu(y)
				\; = & \;\;
				\bA(y) \,
				\left[ \, \bJ \, \bnu \, \right](y)
				\, + \,
				\bA(y) \, \bk
				\, + \,
				\bB(y) \, \bbf(y),
				\\[0.15cm]
				\label{eq.ss-integral-output}
				\bvphi(y)
				\; = & \;\;
				\bC(y)
				\left[ \, \bJ \, \bnu \, \right] (y)
				\, + \,
				\bC(y) \, \bk,
				\\[0.15cm]
				\label{eq.ss-integral-bc}
				0
				\; = & \;\; \left( \bL_a \, \bE_{a}
				\, + \,
				\bL_b \, \bE_{b} \right)
				\left[ \,
				\bJ \, \bnu
				\, \right](y)
				\, + \,
				\left(
				\bL_a
				\, + \,
				\bL_b
				\right) \bk.
			\end{align}
		\end{subequations}
		An expression for $\bnu$ in terms of the forcing $\bbf$ and the integration constants $\bk$ can be obtained from~(\ref{eq.ss-integral-states}),
		\begin{equation}
		\label{eq.ss-integral-u-sol}
			\bnu(y)
			\; = \;
			\left[
			\left(
			\bI - \bA \, \bJ
			\right)^{-1}
			\left(
			\bB \, \bbf
			\right)
			\right] (y)
			\, + \,
			\left[
			\left(
			\bI - \bA \, \bJ
			\right)^{-1}
			\bA
			\right] (y)
			\,\,
			\bk.
		\end{equation}
		Furthermore, substitution of~(\ref{eq.ss-integral-u-sol}) into~(\ref{eq.ss-integral-bc}) yields
		\begin{equation}
		\label{eq.ss-integral-k-sol}
			\bk \; = \;
			- \bH^{-1} \, \bL_{b} \, \bE_{b}
			\left[
			\bJ
			\left(
			\bI - \bA \, \bJ
			\right)^{-1}
			\bB \, \bbf
			\right]
			(y),
		\end{equation}
		where $\bH$ is a matrix given by
		\begin{equation*}
			\bH
			\; = \;
			\bL_{b} \,
			\bE_{b}
			\left[
			\bJ
			\left(
			\bI - \bA \, \bJ
			\right)^{-1}
			\bA
			\right](y)
			\, + \,
			\bL_{a}
			\, + \,
			\bL_{b}.
		\end{equation*}
		Finally, substitution of~(\ref{eq.ss-integral-u-sol}) and~(\ref{eq.ss-integral-k-sol}) into~(\ref{eq.ss-integral-output}) yields
		\begin{equation}
		\label{eq.ss-intbvp}
			\begin{array}{rcl}
			\bvphi(y)
			& \!\! = \!\! &
			\left[
			\bC
			\,
			\bJ
			\left(
			\bI \, - \bA \, \bJ
			\right)^{-1}
			\bB \,
			\bbf
			\right](y)
			\, - \,
			\left[
			\bC
			\,
			\bH^{-1} \, \bL_{b} \, \bE_{b} \, \bJ
			\left(
			\bI \, - \bA \, \bJ
			\right)^{-1}
			\bB \,
			\bbf
			\right](y)
			\\[0.2cm]
			&&
			\, - \,
			\left[
			\bC
			\,
			\bJ
			\left(
			\bI \, - \bA \, \bJ
			\right)^{-1}
			\bA \, \bH^{-1} \, \bL_{b} \, \bE_{b} \, \bJ
			\left(
			\bI \, - \bA \, \bJ
			\right)^{-1}
			\bB \,
			\bbf
			\right](y),
			\end{array}
		\end{equation}
		where invertibility of the matrix $\bH$ follows from the well-posedness of the two-point boundary value problem~(\ref{eq.HHstar-ss-1}). Thus, the singular values $\sigma_{n}$ and the associated left singular functions ${\bu_{n}}$ of $\ct$ can be obtained by solving the following eigenvalue problem
		\begin{equation}
		\label{eq.ss-inteigs}
			\begin{array}{l}
			\left[
			\bC
			\,
			\bJ
			\left(
			\bI \, - \bA \, \bJ
			\right)^{-1}
			\bB \,
			\bu_{n}
			\right](y)
			\, - \,
			\left[
			\bC
			\,
			\bH^{-1} \, \bL_{b} \, \bE_{b} \, \bJ
			\left(
			\bI \, - \bA \, \bJ
			\right)^{-1}
			\bB \,
			\bu_{n}
			\right](y)
			\\[0.2cm]
			\, - \,
			\left[
			\bC
			\,
			\bJ
			\left(
			\bI \, - \bA \, \bJ
			\right)^{-1}
			\bA \, \bH^{-1} \, \bL_{b} \, \bE_{b} \, \bJ
			\left(
			\bI \, - \bA \, \bJ
			\right)^{-1}
			\bB \,
			\bu_{n}
			\right](y)
			\; = \;
			\sigma_{n}^{2}
			\,
			\bu_{n}(y).
			\end{array}
		\end{equation}
		
		\vspace*{0.15cm}
		
		In summary, the principal left singular pair of the operator $\ct$ can be determined by rewriting either the input-output differential equation~(\ref{eq.system-ode2-tts}) or the system of first-order differential equations~(\ref{eq.HHstar-ss}) representing $\ct \ct^{\star}$ into their respective integral forms~(\ref{eq.blocksys}) and~(\ref{eq.ss-integral}). The resulting eigenvalue problems~(\ref{eq.eig-intbvp}) and~(\ref{eq.ss-inteigs}) are solved using {\sf Chebfun}~\cite{chebfunv4}. The detailed discussion on how {\sf Chebfun} can be used to solve the eigenvalue problems~(\ref{eq.eig-intbvp}) and~(\ref{eq.ss-inteigs}) is relegated to~\ref{sec.app-chebfun-implementation}.
		
\section{Examples}
\label{sec.example}

	We next use our method to study frequency responses of two systems from fluid mechanics: three-dimensional incompressible channel flow of Newtonian fluids, and two-dimensional inertialess channel flow of viscoelastic fluids. In the latter example, we show how numerical instabilities encountered when using finite dimensional approximation techniques can be alleviated. The utility of theoretical and computational tools of this paper goes beyond fluids; they can be used to examine dynamics of a broad class of physical systems with normal or non-normal dynamical generators, and spatially constant or varying coefficients.
			
	\subsection{Three-dimensional incompressible channel flows of Newtonian fluids}
	\label{sec.examples-LNS}
	
		\begin{figure}
			\begin{center}
		            	{
		            	\includegraphics[width=0.35\columnwidth]
				{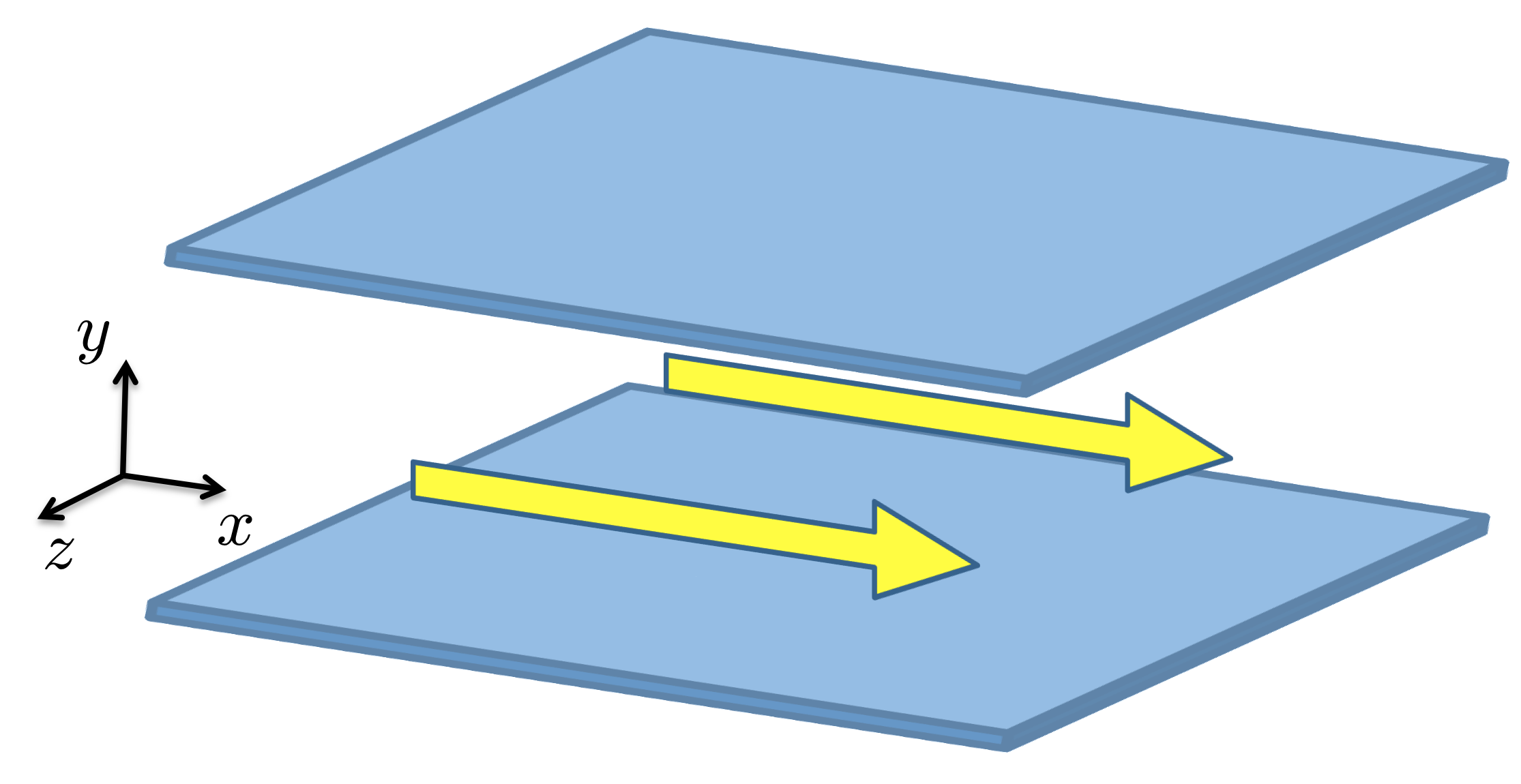}
				}
		  	\end{center}
			\vspace{-0.5cm}
			\caption{Channel flow geometry.}
			\label{fig.3D-channel}
		\end{figure}
	
		We first study the dynamics of infinitesimal three-dimensional fluctuations in a pressure-driven channel flow with base velocity $U(y) = 1 - y^2$; see figure~\ref{fig.3D-channel} for geometry. As shown in~\cite{jovbamJFM05}, the linearized Navier-Stokes (NS) equations can be brought to the evolution form~(\ref{eq.system}) with state
        $
		\bphi
		=
		\left[
		\begin{array}{cc}
			\phi_1 & \phi_2
		\end{array}
		\right]^{T}$,
		where $\phi_1$ and $\phi_2$ are the normal velocity and vorticity fluctuations. Furthermore,
		$
		\bd
		=
		\left[
		\begin{array}{ccc}
			d_1 & d_2 & d_3
		\end{array}
		\right]^{T}
		$
        and
        $
		\bvphi
		=
		\left[
		\begin{array}{ccc}
			u & v & w
		\end{array}
		\right]^{T}
		$
are the input and output fields whose components represent the body forcing and velocity fluctuations in the three spatial directions, $x$, $y$, and $z$. Owing to translational invariance in $x$ and $z$,~(\ref{eq.system}) is parameterized by the corresponding wave numbers $k_x$ and $k_z$ with the boundary conditions on the normal velocity and vorticity,
		\[
		\ba{rcl}
		\phi_1 (k_x, \pm1, k_z, t)
		& \!\! = \!\! &
		D^{(1)} {\phi_1} (k_x, \pm1, k_z, t)
		~ = ~ 0,
		\\[0.1cm]
		\phi_2(k_x, \pm1, k_z, t)
		& \!\! = \!\! &
		0,
		~~~~~~~
		k_x, k_z \in \mathbb{R}, \;\;\; t \geq 0.
		\ea
		\]
The operators in~(\ref{eq.system}) are given in~\ref{sec.app-LNS} and, for any pair of $k_x$ and $k_z$, they are matrices of differential operators in $y \in [-1, 1]$.

	\vspace*{0.15cm}
		
\begin{figure}
			\begin{center}
				\begin{tabular}{c}
					{
			            	\includegraphics[width=0.45\columnwidth]
					{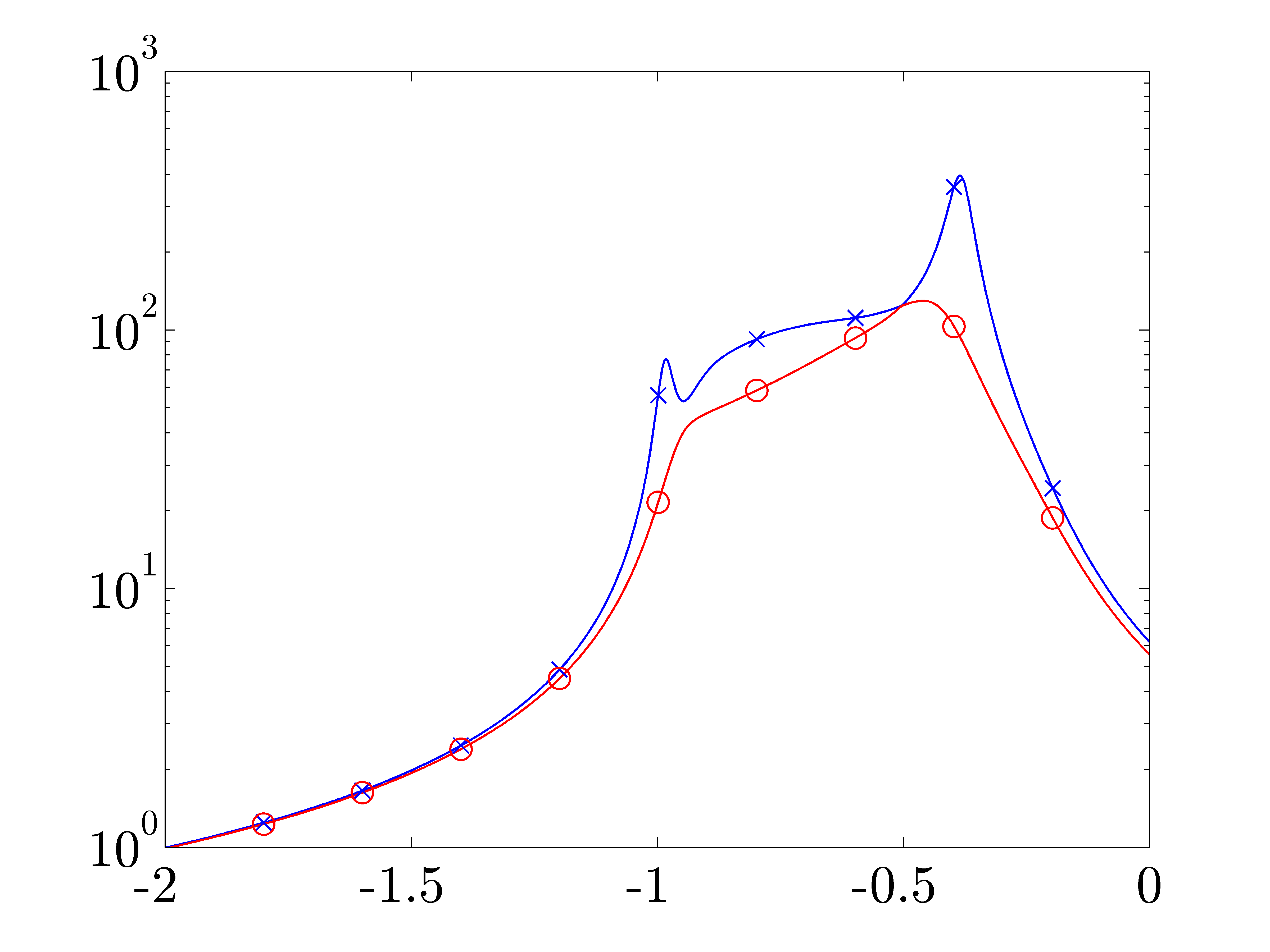}
					}
					\\[-0.1cm]
					\hspace{0.4cm}
					{\Large $\omega$}
				\end{tabular}
		  	\end{center}
			\vspace{-0.5cm}
			\caption{Two largest singular values of the frequency response operator for the linearized Navier-Stokes equations as a function of the temporal frequency $\omega$ in a channel flow with $R = 2000$, $k_x = 1$, and $k_z = 1$: blue $\times$, $\sigma_{1}(\ct)$; and red $\circ$, $\sigma_{2}(\ct)$.}
			\label{fig.LNS_S1_S2}
	 	\end{figure}

		In what follows, we set the Reynolds number to $R = 2000$, $k_x = k_z = 1$ and compute the singular values of $\ct$ using the method developed in Section~\ref{sec.inteigs}. Figure~\ref{fig.LNS_S1_S2} shows two largest singular values, $\sigma_1$ and $\sigma_2$, of the frequency response operator $\ct$ for the linearized NS equations as a function of the temporal frequency $\omega$. The largest singular value $\sigma_{1}$ exhibits two distinct peaks at $\omega \approx -1$ and $\omega \approx -0.4$. Our results have been verified against predictions resulting from earlier studies~\cite{schhen01,mj-phd04}; cf.\ figure~\ref{fig.LNS_S1_S2} with~figure 4.10b in~\cite{schhen01}. We also note that these peaks are caused by different physical mechanisms which can be uncovered by investigating responses from individual forcing to individual velocity components~\cite{jovbamJFM05}. The discussion of these mechanisms is beyond the scope of this paper.

		\begin{figure}
			\begin{center}
				\begin{tabular}{cc}
				\subfigure[]{
				\includegraphics[width=0.45\columnwidth]
				{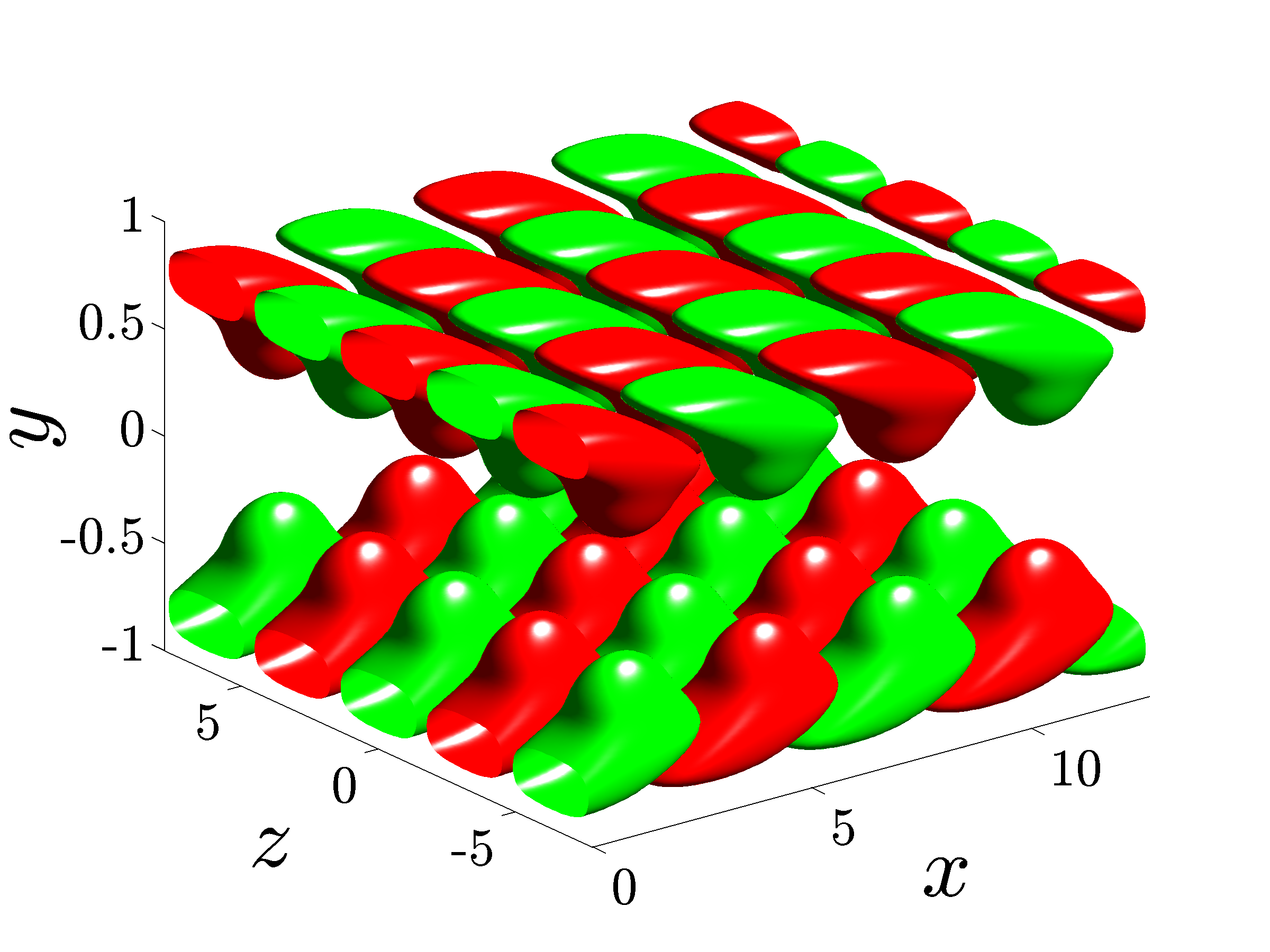}
				\label{fig.u_isosurface_a}
				}
				&
		            	\subfigure[]{
		            	\includegraphics[width=0.45\columnwidth]
				{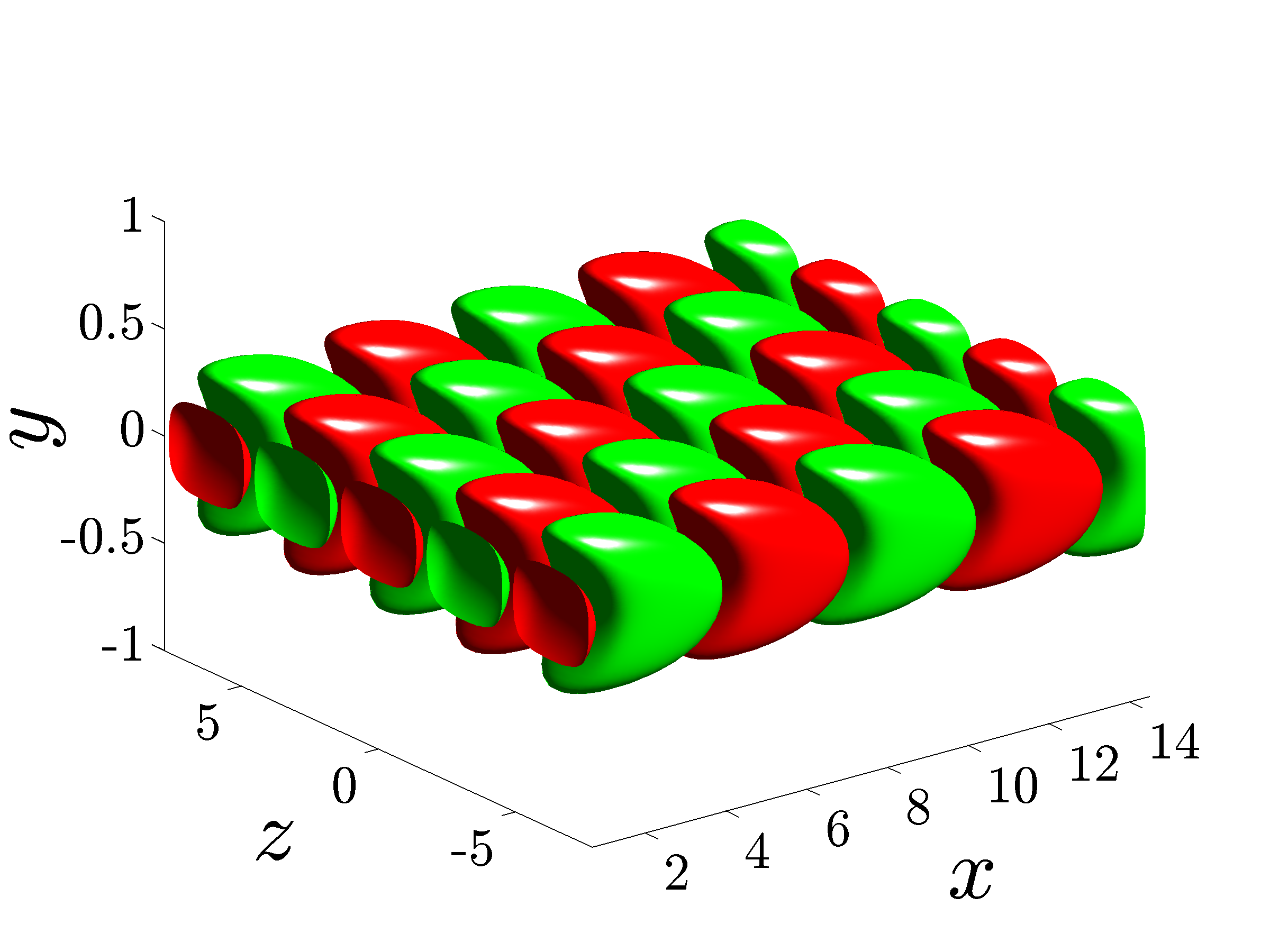}
				\label{fig.u_isosurface_b}
				}
				\end{tabular}
		  	\end{center}
			\vspace{-0.5cm}
			\caption{Spatial structure of the streamwise velocity fluctuations for largest singular value of the frequency response operator in a pressure-driven channel flow with R = 2000, $k_x = k_z = 1$, (a) $\omega = -0.385$, and (b) $\omega = -0.982$. High and low velocity regions are represented by red and green colors. Isosurfaces of $u$ are taken at $\pm 0.55$.}
			\label{fig.u_isosurface}
	 	\end{figure}
		
		\begin{figure}
			\begin{center}
				\begin{tabular}{m{6cm}m{6cm}}
					\hspace*{-2cm}
					\begin{sideways}
						\hspace{0.6cm}
						{\Large $y$}
					\end{sideways}
					\hspace*{-0.4cm}
					\begin{tabular}{c}
						{
						\includegraphics[width=0.45\columnwidth]
						{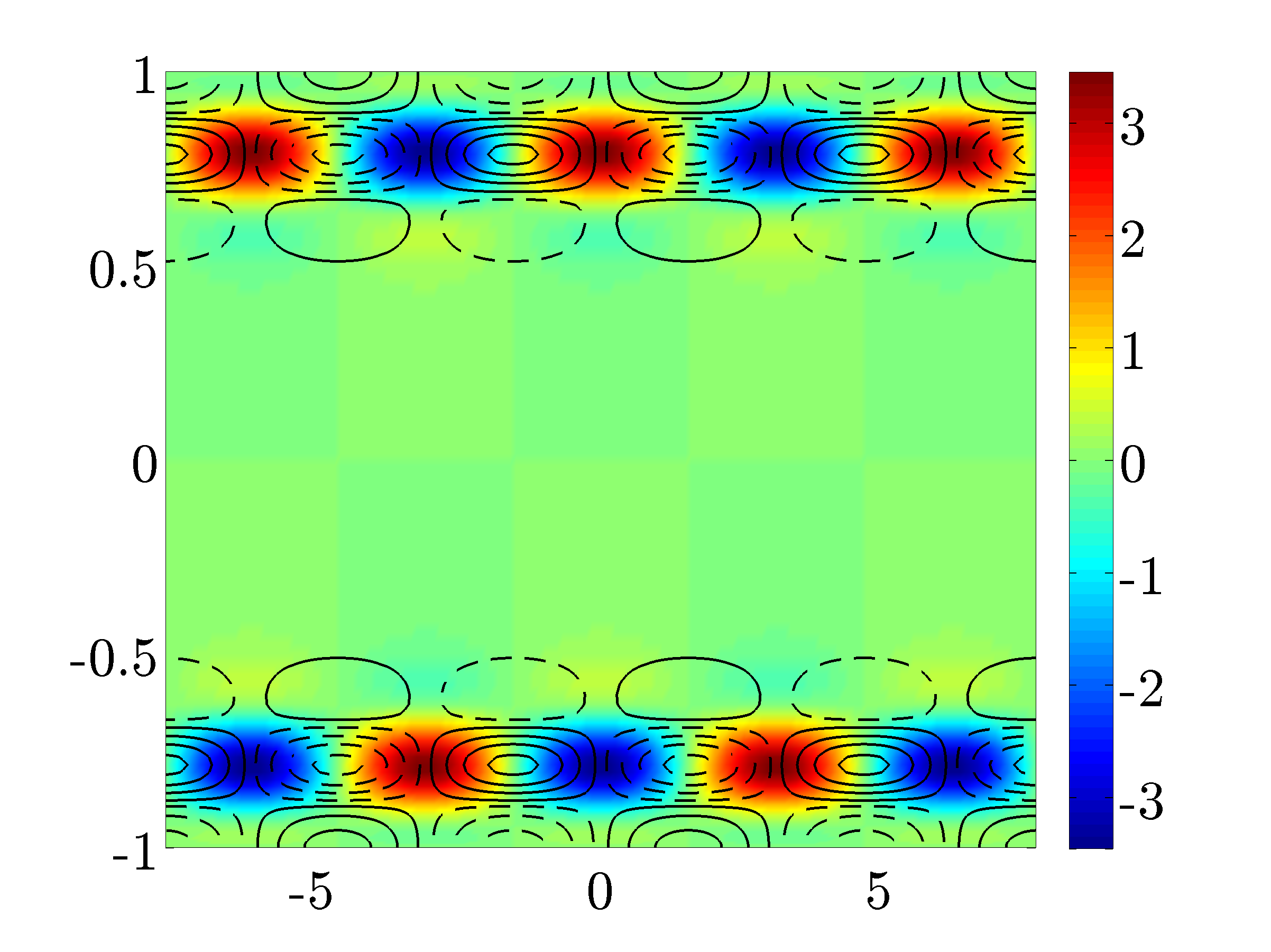}
						}
						\\[-0.2cm]
						{\Large $z$}
						\\
						\subfigure[]{
						\label{fig.u_wx_a}
						}
					\end{tabular}
					&
					\hspace*{-0.3cm}
					\begin{tabular}{c}
				            	{
				            	\includegraphics[width=0.45\columnwidth]
						{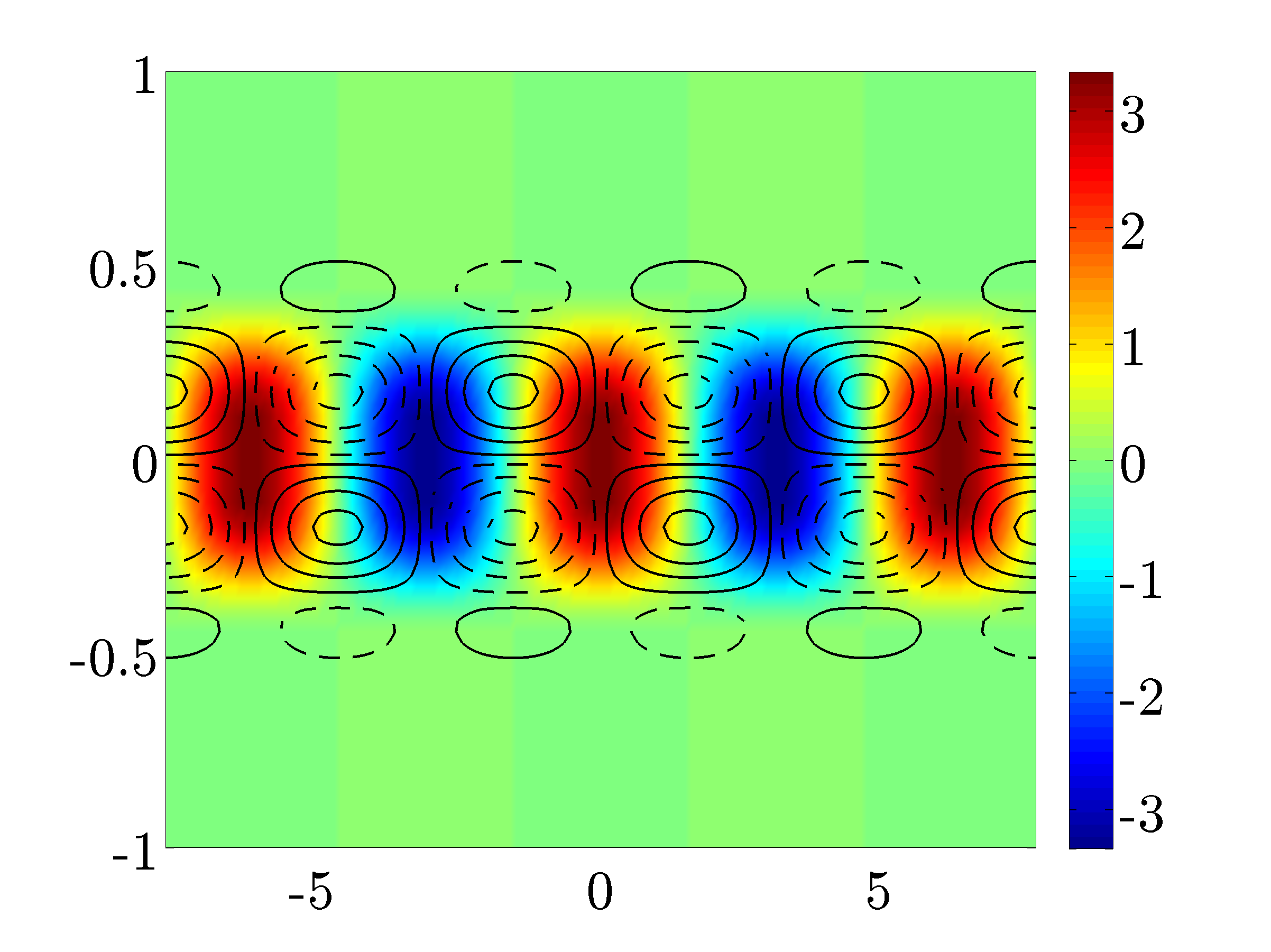}
						}
						\\[-0.2cm]
						{\Large $z$}
						\\
						\subfigure[]{
						\label{fig.u_wx_b}
						}
					\end{tabular}
				\end{tabular}
		  	\end{center}
			\vspace{-0.5cm}
			\caption{Spatial structure of the streamwise velocity (color plots) and vorticity, $w_y - v_z$, (contour lines) fluctuations for largest singular value of the frequency response operator in the cross section of a pressure-driven channel flow with $R = 2000$, $k_x = k_z = 1$, (a) $\omega = -0.385$, and (b) $\omega = -0.982$. Red color represents high speed and blue color represents low speed streaks.}
			\label{fig.u_wx}
	 	\end{figure}
		
		\vspace*{0.15cm}
		
		Figure~\ref{fig.u_isosurface} shows isosurfaces of the most amplified streamwise velocity fluctuations corresponding to the two peaks shown in figure~\ref{fig.LNS_S1_S2}. These output structures are purely harmonic in $x$, $z$, and $t$, and their profiles in $y$ are determined by the left principal singular functions of the frequency response operator at $\omega = -0.385$ and $\omega = -0.982$. For $\omega = -0.385$, $u$ is localized in the near-wall region. On the other hand, for $\omega = -0.982$ the fluctuations occupy the center of the channel. The development of the streamwise velocity (color plots), and streamwise vorticity $w_y - v_z$ (contour lines) fluctuations in the channel's cross-section is shown in figure~\ref{fig.u_wx}. For $\omega = -0.385$, the most amplified set of fluctuations results in pairs of counter rotating streamwise vortices that generate high and low velocity in the vicinity of the lower and upper walls. In contrast, for $\omega = -0.982$ there is a large concentration of arrays of counter rotating streamwise vortices in the center of the channel. Even though the spatial patterns identified by our analysis represent an idealized view of the flow, their utility in understanding the early stages of transition to turbulence has been well-documented~\cite{sch07}. The spatial structure of input forcing that triggers largest response of velocity fluctuations is determined by the right principal singular function of the frequency response operator ${\cal T}$ (i.e., the principal eigenfunction of the operator ${\cal T}^\star {\cal T}$). For brevity, we do not report these forcing structures here.
			
	\subsection{Inertialess channel flow of viscoelastic fluids}
	
	We next compute the frequency responses of the inertialess flow of viscoelastic fluids presented in Section~\ref{sec.motivating-ex-SOB}. This example illustrates the utility of our method in situations where standard finite dimensional approximations may fail to produce accurate results. For this example, the input-output and spatial state-space representations of the frequency response operator are given in~\ref{sec.app-SOB-kz0-TPBVSR}. We compute the largest singular value using the procedure described in Section~\ref{sec.computation} and provide comparison of our results with those obtained using a pseudo-spectral collocation method~\cite{weired00}.
 	
	\begin{figure}
		\begin{center}
		\begin{tabular}{cc}
			$\sigma_{\max} \left( \ct \right)$
			&
			$\sigma_{\max} \left( \ct \right)$
			\\
			{
			\includegraphics[width=0.45\columnwidth]
			{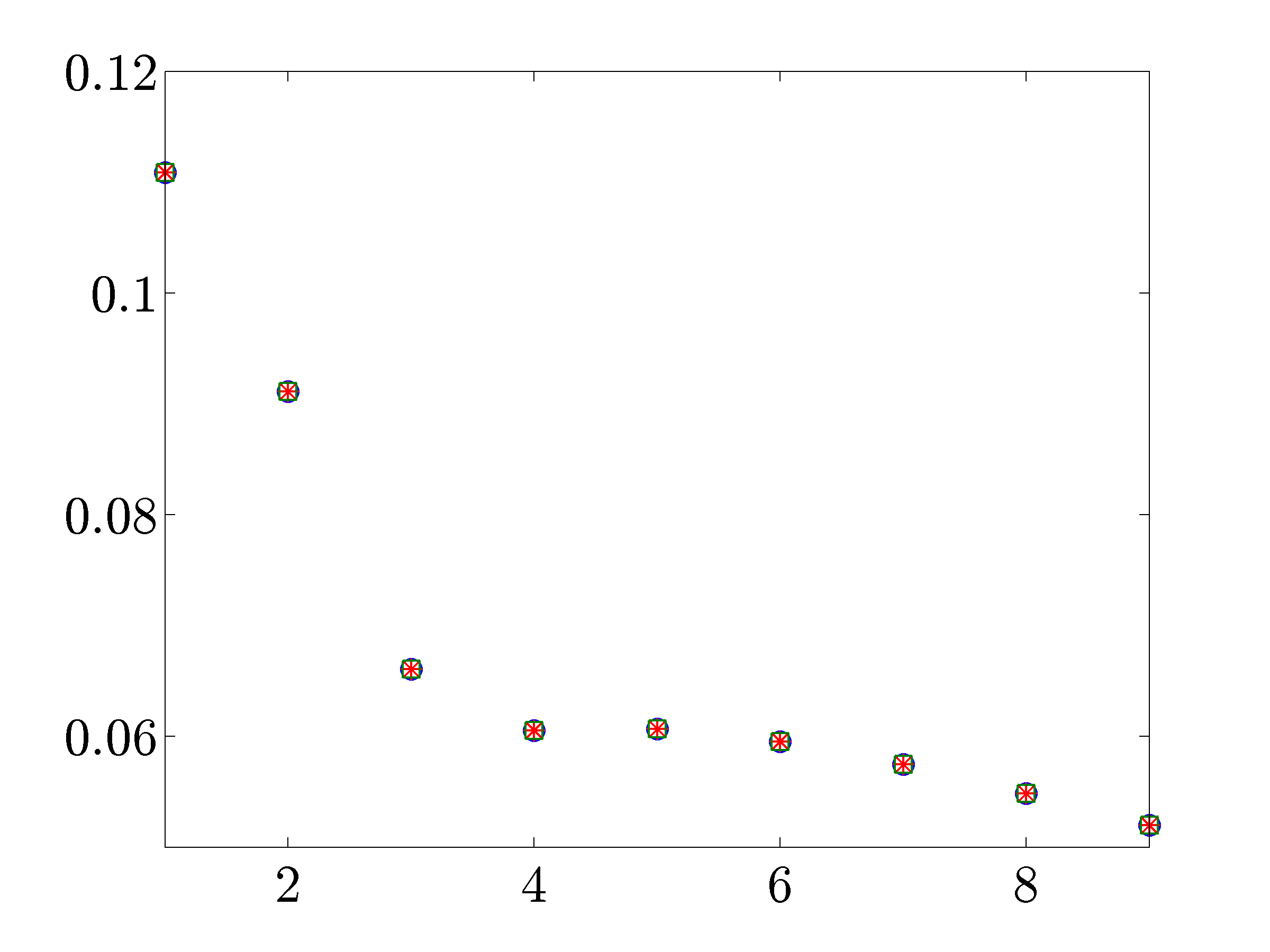}
			}
			&
	            	{
	            	\includegraphics[width=0.45\columnwidth]
			{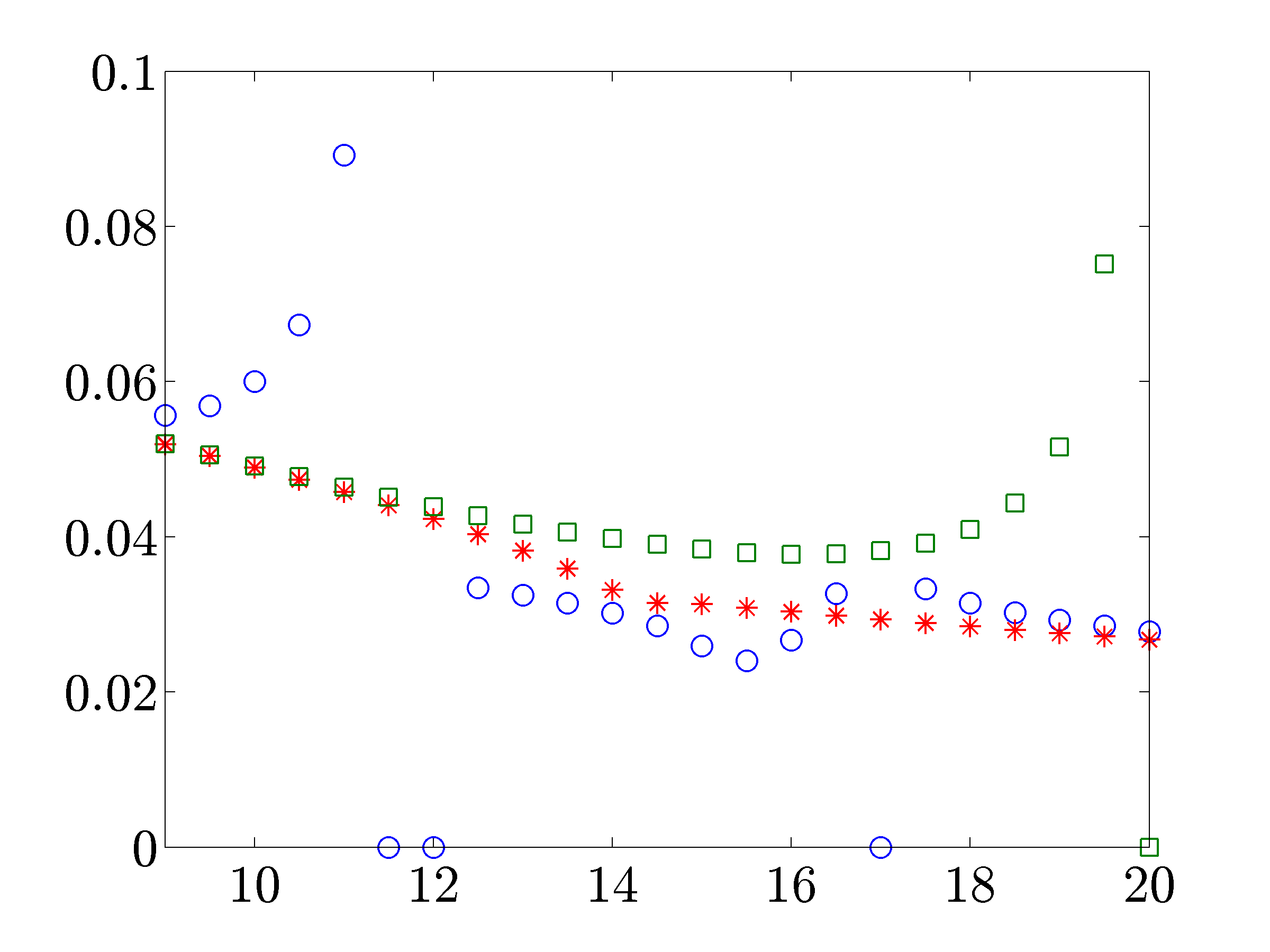}
			}
			\vspace{-0.2cm}
			\\
			{\large $\We$}
			&
			{\large $\We$}
			\\
			\subfigure[]{
			\label{fig.Smax-SOB-chebdif-We1-9}
			}
			&
			\subfigure[]{
			\label{fig.Smax-SOB-chebdif-We10-100}
			}
            \\
			{
			\includegraphics[width=0.45\columnwidth]
			{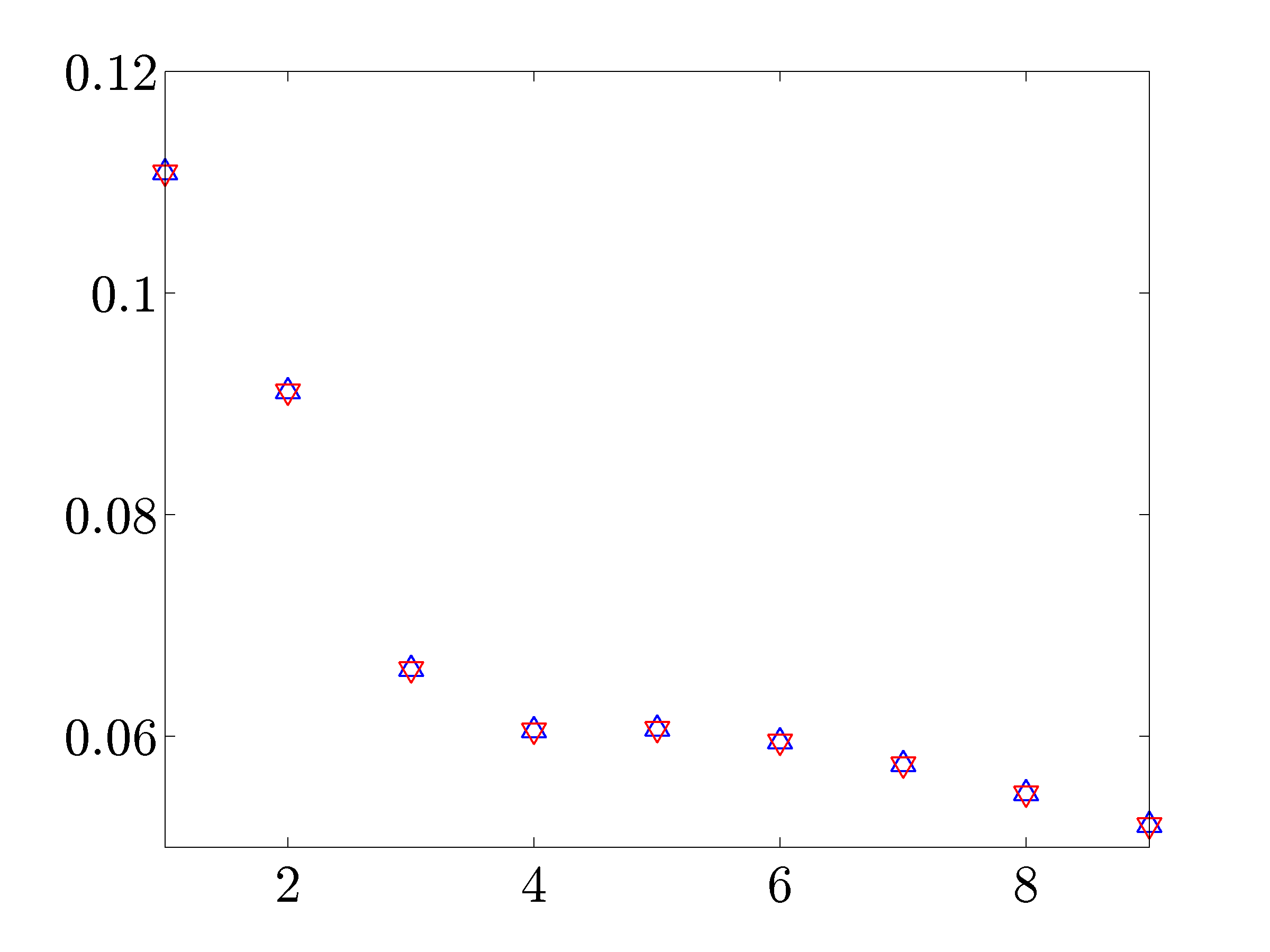}
			}
			&
	            	{
	            	\includegraphics[width=0.45\columnwidth]
			{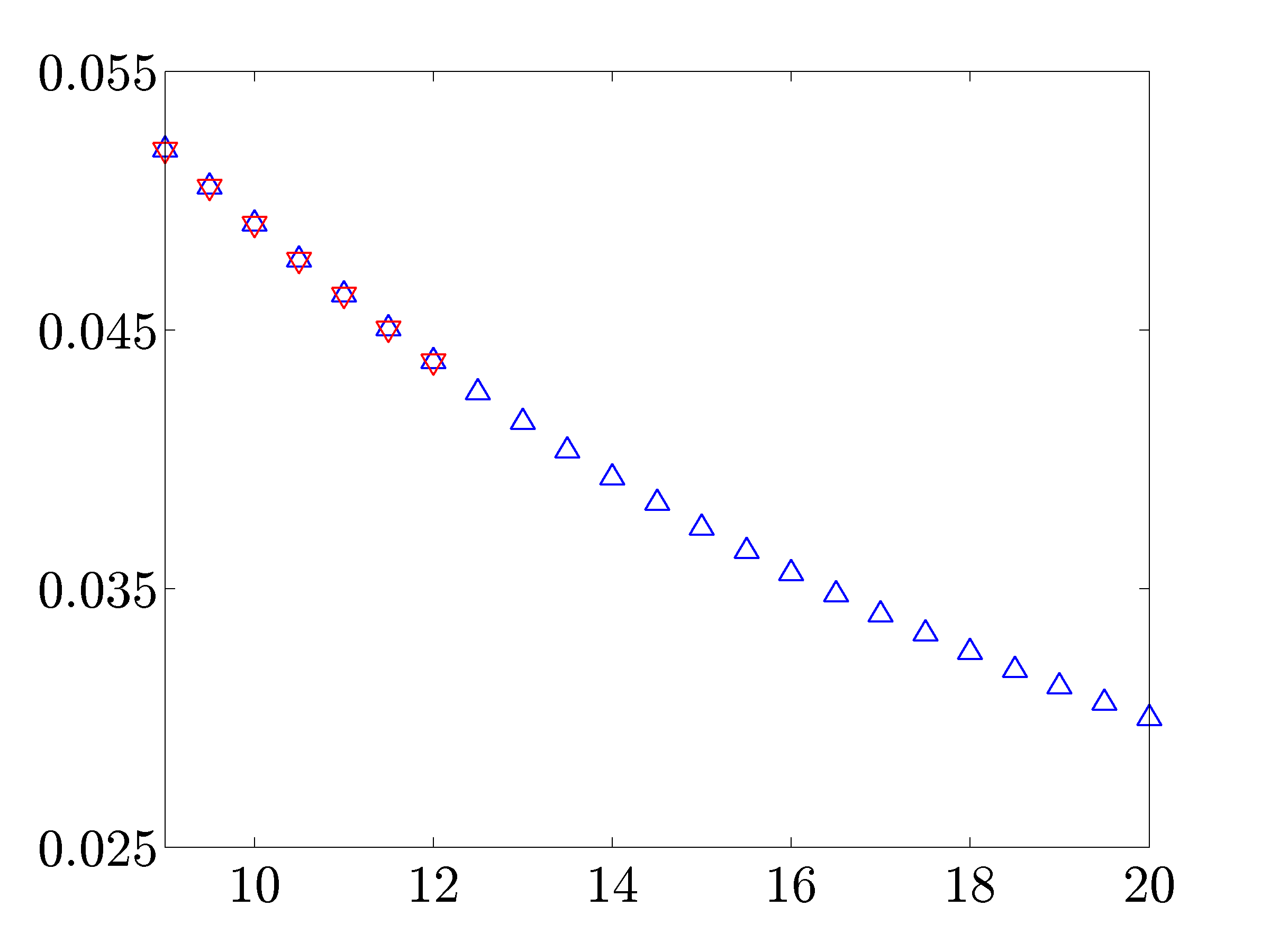}
			}
			\vspace{-0.2cm}
			\\
			{\large $\We$}
			&
			{\large $\We$}
			\\
			\subfigure[]{
			\label{fig.Smax-SOB-inteigs-We1-9}
			}
			&
			\subfigure[]{
			\label{fig.Smax-SOB-inteigs-We9-20}
			}
			\end{tabular}
	  	\end{center}
		\vspace{-0.5cm}
		\caption{The largest singular values of the frequency response operator for an inertialess shear-driven channel flow of viscoelastic fluids as a function of $\We$ at $k_x = 1$, $\beta = 0.5$, and $\omega = 0$. Results are obtained using: (a) and (b) Pseudo-spectral method with $N = 100$, blue $\circ$; $N = 150$, red *; and $N = 200$, green $\square$; (c) and (d) {\sf Chebfun} with integral forms of input-output differential equations, blue $\vartriangle$; and spatial state-space representations, red $\triangledown$.}
		\label{fig.Smax-SOB-kz0-kx1-om0}
 	\end{figure}
	
	\vspace*{0.15cm}
	
	It is well-known that inertialess flows of viscoelastic fluids exhibit spurious numerical instabilities at high-Weissenberg numbers~\cite{kup05,gra98}. In view of this, we fix $k_x = 1$, $\beta = 0.5$, and $\omega = 0$ and examine the effects of the Weissenberg number, $\We$, on the frequency response. We first compute the largest singular value of $\ct$ using a pseudo-spectral collocation method~\cite{weired00}. This is achieved by approximating the operators in the input-output representation~(\ref{eq.system-ode2-tts}) of $\ct \ct^{\star}$ with differentiation matrices of different sizes. Figure~\ref{fig.Smax-SOB-chebdif-We1-9} shows that $\sigma_{\max}$ converges as the number of collocation points, $N$, increases from 50 to 200 for $1 \leq \We \leq 9$. However, for $\We > 9$ the increased number of collocation points in $y$ does not necessarily produce convergent results; see figure~\ref{fig.Smax-SOB-chebdif-We10-100}. Furthermore, in certain cases, the eigenvalues of the operator $\ct \ct^{\star}$ computed using pseudo-spectral method have large negative values. This is clearly at odds with the fact that $\ct \ct^{\star}$ is a non-negative self-adjoint operator, which indicates that the negative eigenvalues arise from numerical artifacts.
	
	\vspace*{0.15cm}
	
	Figures~\ref{fig.Smax-SOB-inteigs-We1-9} and~\ref{fig.Smax-SOB-inteigs-We9-20} show the largest singular value of the operator $\ct$ computed using the method of Section~\ref{sec.computation}. For $1 \leq \We \leq 9$, the largest singular values obtained in {\sf Chebfun} for both input-output and spatial state-space integral representations of $\ct \ct^{\star}$ are equal to each other and they agree with the results of pseudo-spectral method; see figure~\ref{fig.Smax-SOB-inteigs-We1-9}. For $\We > 9$ we see that the largest singular value computed using {\sf Chebfun} exhibits nice trends as $\We$ increases. Furthermore, the automatic Chebyshev spectral collocation method employed by {\sf Chebfun} makes sure that grid point convergence of the singular values is satisfied. We note that the singular values computed using the input-output and spatial state-space integral representations of $\ct \ct^{\star}$ are equal to each other for $\We \leq 12$. On the laptop used for computations, {\sc Matlab} has experienced memory issues when solving the eigenvalue problem in the state-space formulation~(\ref{eq.ss-inteigs}) for $\We > 12$. These memory issues may arise from solving a large system of linear equations internally in {\sf Chebfun}. While internal memory issues can be alleviated using a platform with larger memory capacity, we show these limitations in order to illustrate the trade-off arising from the use of the state-space and the input-output formulations in Chebfun. In Chebfun, the input-output formulation appears to be better suited for efficient computations than the state-space formulation. We further note that the singular values can be computed accurately using the input-output integral representation at much higher Weissenberg numbers.
	
	\vspace*{0.15cm}
	
\begin{figure}
		\begin{center}
		\begin{tabular}{cc}
			$\text{Re} \left( u(y) \right)$
			&
			$\text{Im} \left( v(y) \right)$
			\\
	            	{
	            	\includegraphics[width=0.45\columnwidth]
			{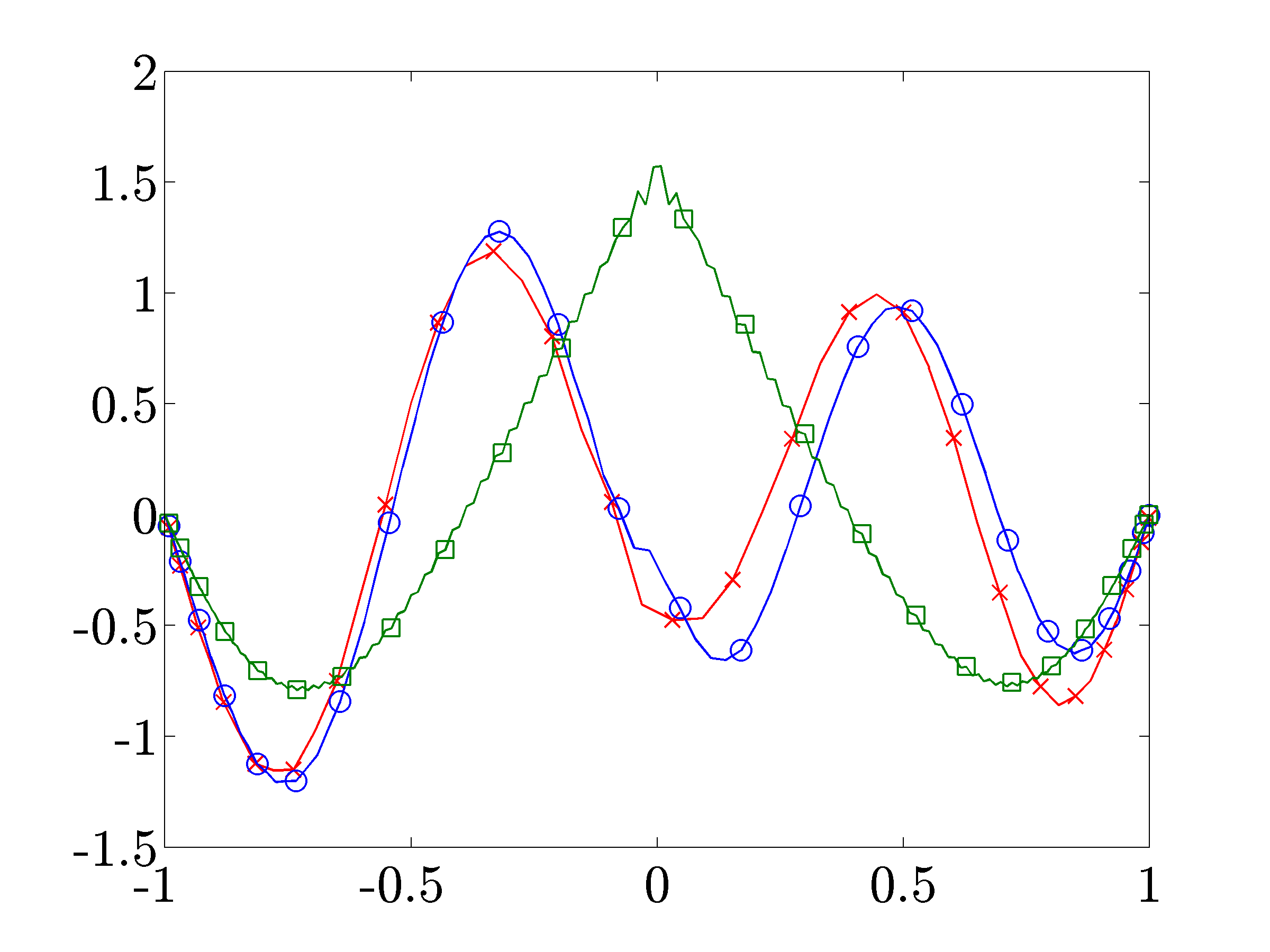}
			}
			&
	            	{
			\includegraphics[width=0.45\columnwidth]
			{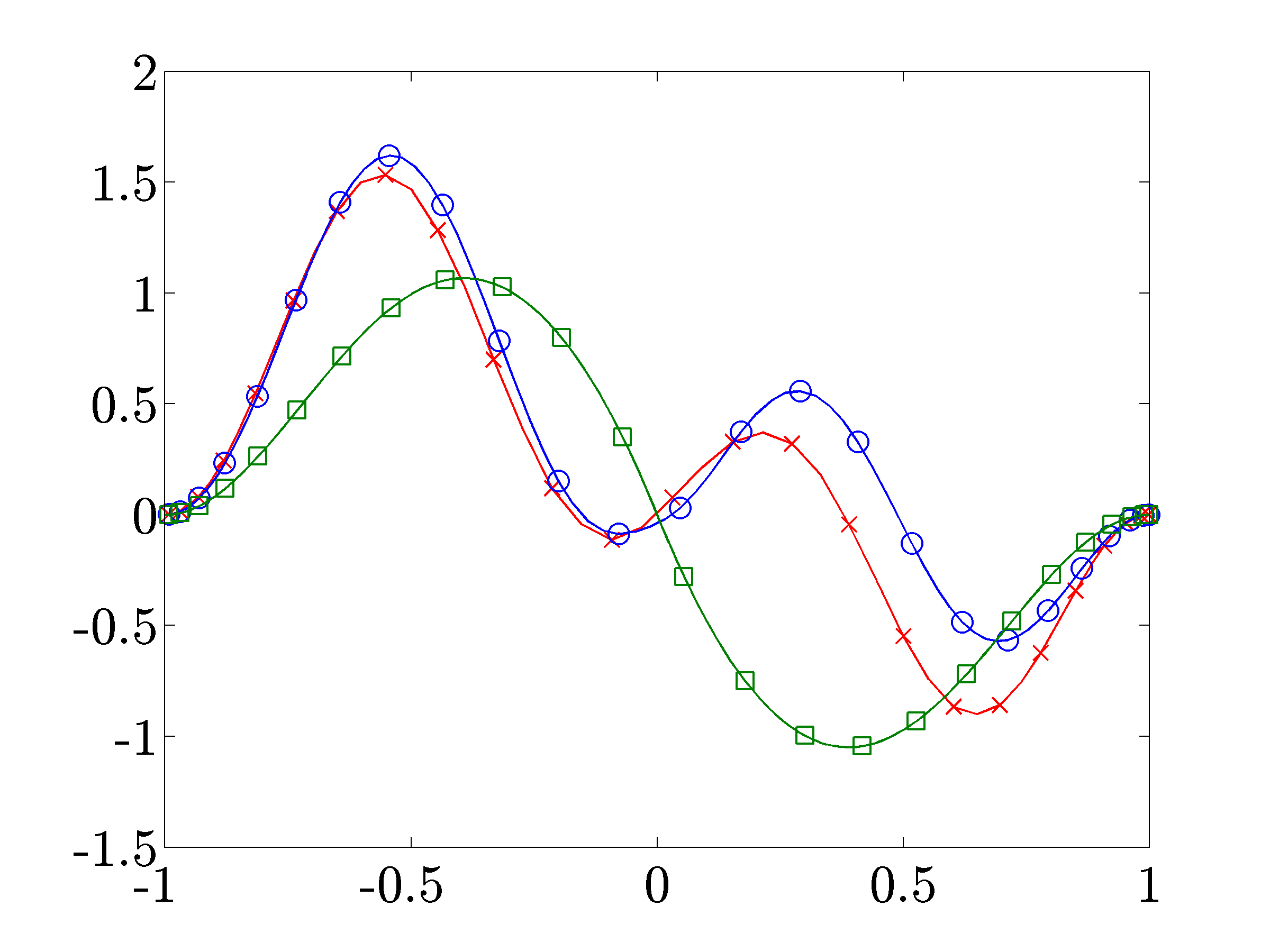}
			}
			\vspace{-0.2cm}
			\\
			{\Large $y$}
			&
			{\Large $y$}
			\\
			\subfigure[]{
			\label{fig.umax-SOB-chebdif}
			}
			&
			\subfigure[]{
			\label{fig.vmax-SOB-chebdif}
			}
			\\
	            	{
	            	\includegraphics[width=0.45\columnwidth]
			{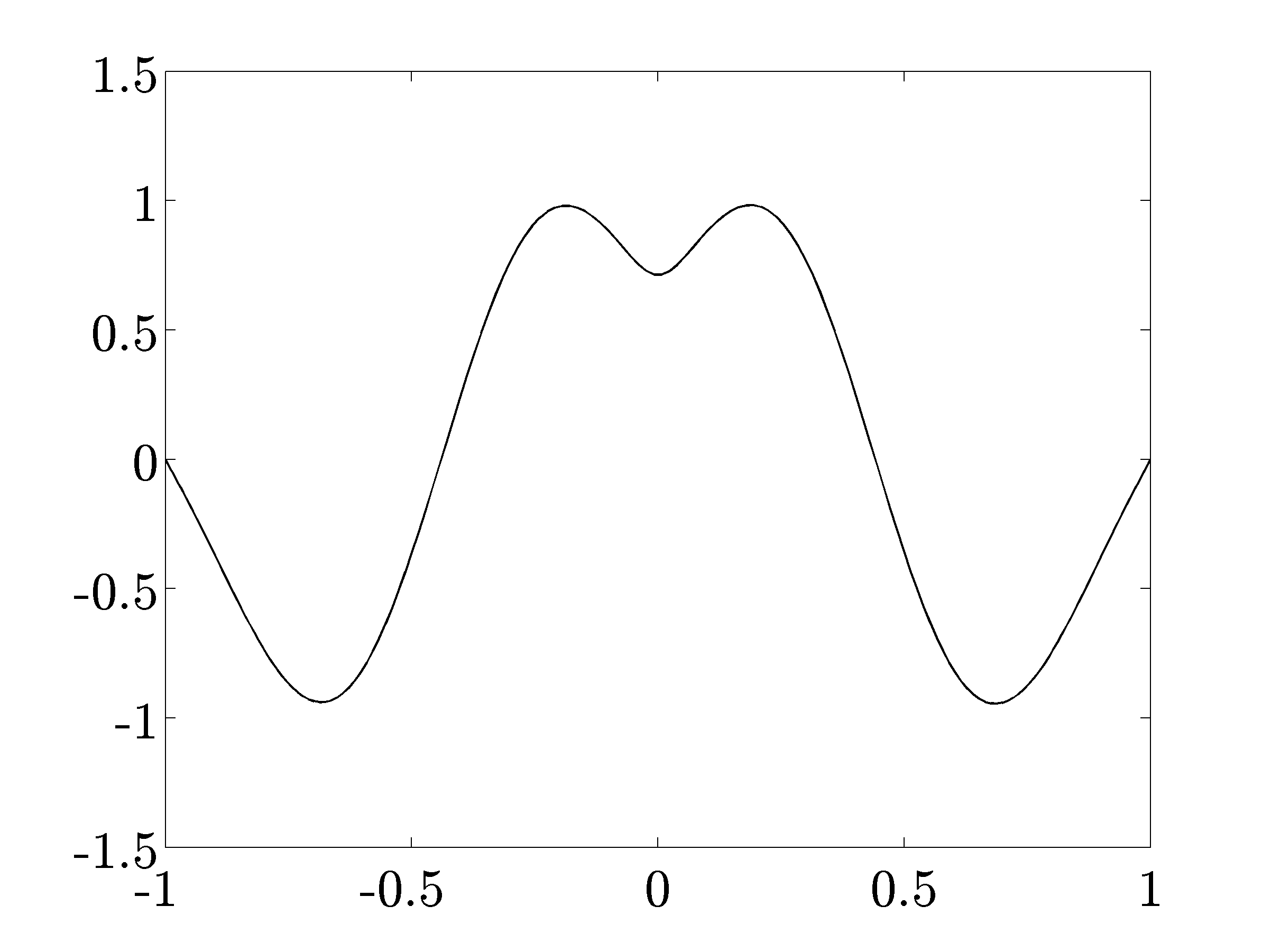}
			}
			&
	            	{
	            	\includegraphics[width=0.45\columnwidth]
			{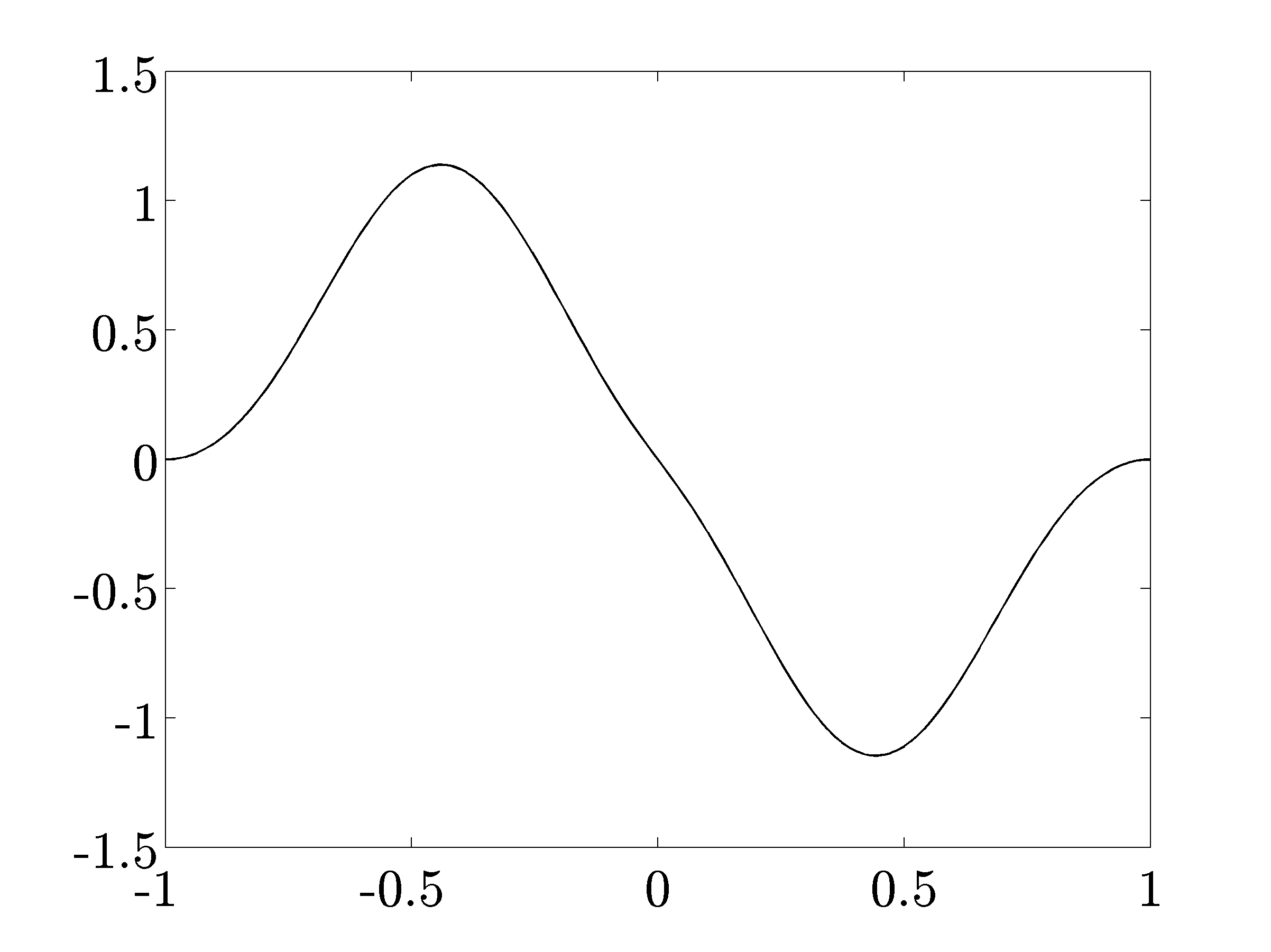}
			}
			\vspace{-0.2cm}
			\\
			{\Large $y$}
			&
			{\Large $y$}
			\\
			\subfigure[]{
			\label{fig.umax-SOB-bvp4c}
			}
			&
			\subfigure[]{
			\label{fig.vmax-SOB-bvp4c}
			}
			\end{tabular}
	  	\end{center}
		\vspace{-0.5cm}
		\caption{Wall-normal shapes of the streamwise ($u$) and wall-normal velocity ($v$) fluctuations for the largest singular value of the frequency response operator in an inertialess shear-driven flow of viscoelastic fluids with $\We = 19.5$, $k_x = 1$, $\beta = 0.5$, and $\omega = 0$. First column: real part of $u$; second column: imaginary part of $v$. Results are obtained using: (a) and (b) Pseudo-spectral method with $N = 50$, red $\times$; $N = 100$, blue $\circ$; $N = 200$, green $\square$; (c) and (d) {\sf Chebfun} with integral form of input-output differential equations.}
		\label{fig.umax-vmax-SOB}
 	\end{figure}

	We next present the wall-normal shapes of the principal singular functions corresponding to the streamwise ($u$) and wall-normal ($v$) velocity fluctuations in a flow with $\We = 19.5$. These are obtained using pseudo-spectral method and {\sf Chebfun} with the input-output integral representation. Figures~\ref{fig.umax-SOB-chebdif} and~\ref{fig.vmax-SOB-chebdif} show the spatial profiles of velocity fluctuations that experience the largest amplification in the presence of disturbances. These profiles are obtained using pseudo-spectral method with different number of collocation points. Note the lack of convergence as the number of collocation points is increased. On the other hand, {\sf Chebfun} does not suffer from numerical instabilities, and the corresponding principal singular functions exhibit the expected symmetry with respect to the center of the channel; see figures~\ref{fig.umax-SOB-bvp4c} and~\ref{fig.vmax-SOB-bvp4c}. Similar trends are observed for larger values of $\We$.
					
\section{Concluding remarks}
\label{sec.conclusion}

We have developed a method for computing the principal singular value and the corresponding singular functions of the frequency response operator for distributed systems with a spatial variable that belongs to a compact interval. Our method avoids the need for numerical approximation of differential operators in the evolution equation. This is achieved by recasting the frequency response operator as a two point boundary value problem; the resulting system of differential equations is then brought into an equivalent integral form which alleviates ill-conditioning and is well-suited for employing {\sf Chebfun} computing environment. When dealing with spatial differential operators of high order our method exhibits two advantages over conventional techniques: numerical ill-conditioning associated with high-order differential matrices is overcome; and boundary conditions are easily implemented and satisfied. We have provided examples from Newtonian and viscoelastic fluid dynamics to illustrate the utility of our developments.

	\vspace*{0.15cm}
	
	Our method has been enhanced by the development of easy-to-use {\sc Matlab} functions which take the system's coefficients and boundary condition matrices as inputs and yield the desired number of left (or right) singular pairs as the output. The coefficients and boundary conditions of the adjoint systems are automatically implemented within the code using the method described in this paper. The burden of finding the adjoint operators and boundary conditions is thus removed from the user who can instead focus on interpreting results and understanding the essential physics.
	
	\vspace*{0.15cm}

	Even though we have confined our attention to computation of the frequency responses for PDEs, the developed framework allows users to employ {\sf Chebfun} as a tool for determining singular value decomposition of compact operators that admit two point boundary value representations. In particular, our approach paves the way for overloading {\sc Matlab}'s command {\sf svds}, from matrices to compact operators.
	
	\vspace*{0.15cm}
	
	While the body of the paper focuses on PDEs with distributed input and output fields, by considering an Euler-Bernoulli beam with boundary actuation in~\ref{sec.app-beam}, we illustrate how {\sf Chebfun} can be used to compute frequency responses of systems with boundary inputs. This problem turns out to be much simpler than the problems with distributed inputs, and it can be implemented with only few lines of code in {\sf Chebfun}. We also use this example to demonstrate the utility of integral formulation in producing accurate results even for systems with poorly scaled coefficients.
	
	\vspace*{0.15cm}

	In all examples that we considered, it is much more efficient to compute the eigenvalue pairs for a system of high-order integral equations~(\ref{eq.eig-intbvp}) than for a system of first-order integral equations~(\ref{eq.ss-inteigs}). We believe that larger number of dependent variables is reducing efficiency of computations that rely on spatial state-space representation. We note that {\sf Chebfun} automatically adjust the number of collocation points in order to obtain solutions with an {\em a priori\/} specified tolerance. The computational speed can be increased by lowering this tolerance using the following command in {\sc Matlab}
	\begin{verbatim}
	   chebfunpref('res', tolerance).
	\end{verbatim}
	
	Our ongoing efforts are focused on employing {\sf Chebfun} as a tool for computing the peak (over temporal frequency) of the largest singular value of the frequency response operator. In systems and controls literature, $\sup_\omega \sigma_{\max} \, (\ct (\omega))$ is known as the $H_\infty$ norm and its computation requires identification of purely imaginary eigenvalues of a Hamiltonian operator in conjunction with bisection~\cite{boybalkab89}. In addition to quantifying the worst-case amplification of purely harmonic (in time) deterministic (in space) disturbances, the inverse of the $H_\infty$ norm determines the size of an unstructured modeling uncertainty that can destabilize the nominal system. Thus, large frequency response peaks indicate small stability margins (i.e., poor robustness properties to modeling imperfections), and they are a reliable predictor of systems in which small modeling imperfections can introduce instability. This interpretation of the $H_\infty$ norm is closely related to the notion of pseudospectra of linear operators~\cite{treemb05} and it has been used to provide useful insight into dynamics of systems with non-normal generators~\cite{tretrereddri93,sch07,liejovkumJFM13,mj-phd04}. 
	
	\vspace*{0.15cm}
	
We finally note that the frequency response analysis can also be used to study the dynamics of systems with two or three spatial variables that belong to a compact interval. However, in 2D and 3D the two-point boundary value structure of the frequency response operator that we exploit in this paper is lost. Furthermore, for 2D, and especially for 3D problems, one would have to develop iterative solvers for the corresponding eigenvalue problems. This would necessitate determination of finite dimensional approximations of both the frequency response operator $\cal T$ and its adjoint ${\cal T}^\star$. Once these are available, standard power-iteration-based methods (e.g., Lanczos algorithm) can be utilized to determine spatial structures of the principal input and output directions.  We note that a recent extension of {\sf Chebfun} to two-dimensional problems -- {\sf Chebfun2} -- may be used to address this challenge in 2D.	
		
\section*{Acknowledgments}

We would like to thank Tobin A.\ Driscoll for useful discussions and for his help with {\sf Chebfun} computing environment. This work was supported in part by the National Science Foundation under CAREER Award CMMI-06-44793 and by the University of Minnesota Doctoral Dissertation Fellowship.

\section*{Supplementary material}

All {\sc Matlab} codes for computing frequency responses are publicly available at
    \begin{center}
{\sf www.umn.edu/$\sim$mihailo/software/chebfun-svd/}
    \end{center}

\appendix

\section{Conversion to a spatial state-space realization}
\label{sec.app-realization}

	We next describe how a high-order ODE with spatially varying coefficients can be converted to a family of first-order ODEs~(\ref{eq.T-ssfirst}). We consider the following ordinary differential equation with boundary conditions:
	\begin{subequations}
		\label{eq.system-forma}
		\begin{align}
			\label{eq.system-forma-states}
			\phi^{(n)}(y) & \; = \;
            -\sum_{i \, = \, 0}^{n \, - \, 1} \alpha_{i}(y) \, \phi^{(i)}(y)
            \, + \,
            \sum_{i \, = \, 0}^{m} \beta_{i}(y) \, d^{(i)}(y), \,\, m \, < \, n \, - \, \ell, \\[0.1cm]
			\label{eq.system-forma-output}
			\varphi(y) & \; = \; \sum_{i \, = \, 0}^{k} \gamma_{i}(y) \, \phi^{(i)}(y), \,\, k \, < \, n \, - \, m, \\[0.1cm]
			\label{eq.system-forma-bc}
			0 & \; = \; \sum_{i \, = \, 0}^{\ell} N_{i, a} \, \phi^{(i)}(a) \, + \, N_{i, b} \, \phi^{(i)}(b), \,\, \ell \, < \, n,
		\end{align}
	\end{subequations}
	where $\phi^{(i)} = \mrd^{i} \phi / \mrd y^{i}$. Since coefficients $\{ \beta_{i}(y)\}$ in~(\ref{eq.system-forma-states}) are spatially varying, the standard observer and controller canonical forms cannot be used to obtain a system of first-order ODEs~(\ref{eq.T-ssfirst}). Instead, we introduce a new variable $w(y)$,
	\begin{equation}
	\label{eq.w-lhs}
		w(y) = \sum_{i \, = \, 0}^{m} \beta_{i}(y) \, d^{(i)}(y),
	\end{equation}
	and substitute~(\ref{eq.w-lhs}) into~(\ref{eq.system-forma-states}) to obtain
	\begin{equation}
		\label{eq.system-formb-states}
		\phi^{(n)}(y) = -\sum_{i \, = \, 0}^{n \, - \, 1} \alpha_{i}(y) \, \phi^{(i)}(y) \, + \, w(y),
	\end{equation}
	Here, a state-space realization of~(\ref{eq.system-formb-states}) is given by the controller canonical form,
	\begin{subequations}
		\label{eq.system-formb-ss}
		\begin{align}
			\bz'(y) \; = & \;\; \bA_{1}(y) \, \bz(y) \, + \, {\bf e}_{n} \, w(y), \\[0.1cm]
			\phi(y) \; = & \;\; {\bf e}_{1}^T \, \bz(y),
		\end{align}
	\end{subequations}
	where
	\begin{equation*}
		\bA_{1}(y) \; = \;
		\left[
		\begin{array}{c|c}
			\mathbf{0}_{\, (n-1) \times 1} & \bI_{\, (n-1) \times (n-1)} \\[0.2cm]
			\hline
			&
			\\[-0.2cm]
			-\alpha_{0}(y) &
			\begin{array}{ccc}
			-\alpha_{1}(y) & \cdots & -\alpha_{n - 1}(y)
			\end{array}
		\end{array}
		\right],
	\end{equation*}
	and ${\bf e}_{i}$ is the $i$th unit vector. It is a standard fact that the solution to~(\ref{eq.system-formb-ss}) is given by
	\begin{equation}
		\label{eq.sol-system-formb}
		\bz(y)
		\; = \;
		\bPhi_{1}(y, a) \, \bz(a)
		\, + \,
		\int_{a}^{y} \bPhi_{1}(y, \eta) \, {\bf e}_{n} \, w(\eta) \, \mrd \eta,
	\end{equation}
	where $\bPhi_{1}(y,\eta)$ is the state-transition matrix of $\bA_{1} (y)$. Substituting~(\ref{eq.w-lhs}) into~(\ref{eq.sol-system-formb}) yields
	\begin{equation}
		\label{eq.sol-system-formb2}
		\bz(y)
		\; = \;
		\bPhi_{1}(y, a) \, \bz(a)
		\, + \,
		{\ds \int_{a}^{y}
		\left(
		\bPhi_{1}(y, \eta) \, {\bf e}_{n}
		\left(
		\sum_{i \, = \, 0}^{m} \, \beta_{i}(\eta) \, d^{(i)}(\eta)
		\right)
		\right)
		\mrd \eta}.
	\end{equation}
	Application of integration by parts to the integral in~(\ref{eq.sol-system-formb2}) along with a change of variables leads to the following two point boundary value state-space representation of~(\ref{eq.system-forma})
	\begin{subequations}
		\label{eq.system-formc-ss}
		\begin{align}
			\bx'(y)
			\; = & \;\;
			\bA_{0}(y) \, \bx(y)
			\, + \,
			\bB_{0} \, d(y), \\[0.1cm]
			\varphi(y)
			\; = & \;\;
			\bC_{0} \, \bx(y), \\[0.1cm]
			0
			\; = & \;\;
			\bN_{a} \, \bx(a) \, + \, \bN_{b} \, \bx(b),	
		\end{align}
	\end{subequations}
	where
	\begin{equation*}
		\begin{array}{rcl}
			\bx(y)
			& \!\! = \!\! &
			\ds{
			\bz(y)
			\, - \,
			\sum_{i \, = \, 0}^{m \, - \, 1}
			\left(
			\sum_{j \, = \, 1}^{m \, - \, i} \bQ_{j - 1}
			( \beta_{i + j}(y) )
			\right) d^{(i)}(y)},
			\\[0.7cm]
			\bA_{0}(y)
			& \!\! = \!\! &
			\bA_{1}(y),
			\;\;
			\bB_{0}(y)
			\; = \;
			\ds{ \sum_{i \, = \, 0}^{m} \bQ_{i}( \beta_{i}(y) )},
			\\[0.3cm]
			\bC_{0}(y)
			& \!\! = \!\! &
			\underset{k \, + \, 1}{
			\left[
			\underbrace{
			\begin{array}{ccc}
				\gamma_{0}(y) & \cdots & \gamma_{k}(y)
			\end{array}
			}
			\right.
			}
			\underset{n \, - \, k \, - \, 1}{
			\left.
			\underbrace{
			\begin{array}{ccc}
				0 & \cdots & 0
			\end{array}
			}
			\right]
			},
			\\[1.0cm]
			\bN_{a}
			& \!\! = \!\! &
			\left[
			\begin{array}{ccc}
				N_{0, a} & & \\
				& \ddots & \\
				& & N_{\ell,a}
			\end{array}
			\right],
			\,\,
			\bN_{b} \; = \;
			\left[
			\begin{array}{ccc}
				N_{0, b} & & \\
				& \ddots & \\
				& & N_{\ell,b}
			\end{array}
			\right].	
		\end{array}
	\end{equation*}
	We note that, for a given function $\beta$, $\bQ_{i}$ can be recursively determined from
	\begin{equation*}
		\begin{array}{rcl}
			\bQ_{i} ( \beta(y) )
			& \!\! = \!\! &
			\bA_{1}(y) \, \bQ_{i \, - \, 1} ( \beta(y) )
            \, - \,
            \cfrac{\mrd}{\mrd y} \, \bQ_{i \, - \, 1} ( \beta(y) ),
            \;\;\;
            i = 1, \, \ldots, \, m, \\[0.2cm]
			\bQ_{0} ( \beta(y) )
			& \!\! = \!\! &
			{\bf e}_{n} \, \beta(y).
		\end{array}
	\end{equation*}

\section{Implementation of eigenvalue problems in integral formulation using {\sf Chebfun}}
\label{sec.app-chebfun-implementation}	

	The eigenvalue problems~(\ref{eq.eig-intbvp}) and~(\ref{eq.ss-inteigs}) derived in Sections~\ref{sec.inteigs} and~\ref{sec.ss-solve} are solved using {\sf Chebfun}. Here, we show how to implement the functions and operators in {\sf Chebfun} to solve~(\ref{eq.eig-intbvp}); a similar procedure can be used to solve~(\ref{eq.ss-inteigs}). The eigenvalue problem~(\ref{eq.eig-intbvp}) requires the construction of a number of operators and quasimatrices (terminology used by the authors of {\sf Chebfun} to denote vectors of functions). The operator $\ca$ in~(\ref{eq.tts-diffeq-1}) is represented by the coefficients $\mathbf{a}_{i}(y)$ which are functions determining columns of a quasimatrix. For example, consider the differential equations representing the operator $\ct \ct^{\star}$ for the 1D diffusion equation
	\begin{equation}
	\label{eq.heat-eq-tts}
	\hspace{-0.2cm}
		\begin{array}{rcl}
			\left[
			\begin{array}{cc}
				D^{(2)} - \mri \omega I & -I \\[0.1cm]
				0 & D^{(2)} + \mri \omega I
			\end{array}
			\right]
			\left[
			\begin{array}{c}
				\xi_{1}(y) \\[0.1cm]
				\xi_{2}(y)
			\end{array}
			\right]
			& \!\!\! = \!\!\! &
			\left[
			\begin{array}{c}
				0 \\[0.1cm]
				I
			\end{array}
			\right]
			f(y),
			\\[0.4cm]
			\phi(y)
			& \!\!\! = \!\!\! &
			\left[
			\begin{array}{cc}
				I & 0
			\end{array}
			\right]
			\left[
			\begin{array}{c}
				\xi_{1}(y) \\[0.1cm]
				\xi_{2}(y)
			\end{array}
			\right],
			\\[0.4cm]
			\left[
			\begin{array}{cc}
				1 & 0\\[0.1cm]
				0 & 0
			\end{array}
			\right]
			\left[
			\begin{array}{c}
				\xi_{1}(-1) \\[0.1cm]
				\xi_{1}'(-1)
			\end{array}
			\right]
			+
			\left[
			\begin{array}{cc}
				0 & 0\\[0.1cm]
				1 & 0
			\end{array}
			\right]
			\left[
			\begin{array}{c}
				\xi_{1}(+1) \\[0.1cm]
				\xi_{1}'(+1)
			\end{array}
			\right]
			& \!\!\! = \!\!\! &
			\left[
			\begin{array}{c}
				0 \\[0.1cm]
				0 \\[0.1cm]
			\end{array}
			\right],
			\\[0.5cm]
			\left[
			\begin{array}{cc}
				1 & 0\\[0.1cm]
				0 & 0
			\end{array}
			\right]
			\left[
			\begin{array}{c}
				\xi_{2}(-1) \\[0.1cm]
				\xi_{2}'(-1)
			\end{array}
			\right]
			+
			\left[
			\begin{array}{cc}
				0 & 0\\[0.1cm]
				1 & 0
			\end{array}
			\right]
			\left[
			\begin{array}{c}
				\xi_{2}(+1) \\[0.1cm]
				\xi_{2}'(+1)
			\end{array}
			\right]
			& \!\!\! = \!\!\! &
			\left[
			\begin{array}{c}
				0 \\[0.1cm]
				0 \\[0.1cm]
			\end{array}
			\right].
		\end{array}
	\end{equation}
	The code used to generate operator $\ca$ for the 1D diffusion equation is given by
	\begin{verbatim}
	   %% Operator A for the 1D diffusion equation
	   dom = domain(-1,1);     % domain of functions				
	   fone = chebfun(1,dom);  % fone(y) = 1
	   fzero = chebfun(0,dom); % fzero(y) = 0
	
	   % w is the temporal frequency and 1i is the imaginary unit
	   % (1,1) element of operator A
	   A11 = [-1i*w*fone, fzero, fone];  % -i*w*xi_1 + 0*D^{(1)}*xi_1 + 1*D^{(2)}*xi_1
	   % (1,2) element of operator A
	   A12 = [-fone, fzero, fzero];   % -1*xi_2 + 0*D^{(1)}*xi_2 + 0*D^{(2)}*xi_2
	   % (2,1) element of operator A
	   A21 = [fzero, fzero, fzero];   % 0*xi_1 + 0*D^{(1)}*xi_1 + 0*D^{(2)}*xi_1
	   % (2,2) element of operator A
	   A22 = [1i*w*fone, fzero, fone];   % i*w*xi_1 + 0*D^{(1)}*xi_1 + 1*D^{(2)}*xi_1
	
	   % form operator A using cell-array construction
	   A = {A11, A12; A21, A22};
	\end{verbatim}
	The variable {\sf dom} denotes the domain of the functions, and {\sf fone} and {\sf fzero} represent unit and zero functions. The dimension of each {\sf Chebfun}'s function in {\sc Matlab} is $\infty \times 1$, where the first index represents the continuous variable $y$. Hence, the quasimatrices {\sf A11}, {\sf A12}, {\sf A21}, and {\sf A22} have dimensions $\infty \times 3$. Since the dimension of quasimatrices prohibits the construction of matrix of functions, we  instead utilize {\sc Matlab}'s cell arrays (using curly brackets) to represent the operator $\ca$. The boundary condition matrices are given by
	\begin{verbatim}
	   Ya1 = [1, 0; 0, 0];  Ya2 = [1, 0; 0, 0];
	   Yb1 = [0, 0; 1, 0];  Yb2 = [0, 0; 1, 0];
	   Ya  = {Ya1; Ya2};    Yb  = {Yb1; Yb2};
	\end{verbatim}
	The code used to generate the quasimatrix $\mathbf{K}^{(n)}$ is given by
	\begin{verbatim}
	   n = size(A,1); % number of states in your system of ODEs
	   % determine the highest differential order of each component of \xi in the equations
	   ni = zeros(n,1);
	   for j = 1:n
	      ni(j) = size( A{j,j}, 2) - 1;
	   end
	   % indefinite integration operator
	   J = cumsum(dd);
	   %% Construct each component of K
	   Ki = chebfun(1,dd);
	   for j = 2 : max(ni)
	      Ki(:,j) = J*Ki(:, j-1);
	   end
	   % construct quasimatrix K using cell-array
	   for j = 1:n
	      K{j} = Ki(:, 1:ni(j));
	   end
	\end{verbatim}
	The indefinite integration operator is obtained using {\sf Chebfun}'s command {\sf cumsum}. The variable {\sf ni} contains the highest differential order of each state $\xi_{i}$ in the system. We next determine the matrix ${\cal L}_{22}$ appearing in~(\ref{eq.blocksys}) by applying the boundary condition operator ${\cal N}$ to $\bK$. The following code is used to generate ${\cal L}_{22}$
	\begin{verbatim}
	   %% Determine the matrix L_{22}
	   % loop through each component of \xi
	   for j = 1:n
	      % quasimatrix K associated with \xi_{j}
	      Kj = K{j};
	      L22{j} = Ya{j} + Yb{j}*toeplitz([1 zeros(1, ni(j)-1)], Kj( b, : ));
	   end
	\end{verbatim}
	The qausimatrix ${\cal L}_{12}$ is obtained by multiplying coefficients of the operator $\ca$ with the quasimatrix $\bK$,
	\begin{verbatim}
	   %% Determine the functional operator L_{12}
	   % loop through each component of L_{12}, which has size n x n
	   for i = 1:n
	      for j = 1:n
	         % initialize the (i,j) component of L_{12} and
	         % get the quasimatrix K associated with \xi_{j}
	         L12ij = 0; Kj = K{j};
	         % get the (i,j) component of operator A
	         Aij = A{i,j};
	         for indni = 1 : ni(j)
	            L12ij = L12ij + diag( Aij(:, ind) )*Kj;
	            Kj = [  chebfun(0,dd), Kj(:, 1:ni(j) - 1) ];
	         end
	         L12{i, j} = L12ij;
	      end
	   end
	\end{verbatim}
	The operator ${\cal L}_{11}$ in~(\ref{eq.blocksys}) is realized using the following {\sc Matlab}'s commands
	\begin{verbatim}
	   %% Determine the operator L_{11}
	   % loop through each component of L_{11}, which has size n x n
	   for i = 1 : n
	      for j = 1 : n
	      	% get the (i,j) component of A
	         Aij = A{i,j};
	         % initialize (i,j) component of L11 with Aij_0
	         L11ij = diag( Aij(:,1) );
	         for indni = 1 : ni(j) - 1
	            L11ij = L11ij*J + diag( Aij(:, indni + 1) );
	         end
	         L11ij = L11ij*J + diag( Aij(:, ni(j) + 1) );
	         L11{i,j} = L11ij;
	      end
	   end			
	\end{verbatim}
	The boundary point evaluation functional $E_{b}$ is easily constructed by
	\begin{verbatim}
	   Eb = linop(@(n) [zeros(1,n-1) 1], @(u) feval(u,b), dd);
	\end{verbatim}
	In a similar manner, the operator ${\cal L}_{21}$ is realized by
	\begin{verbatim}
	   %% Determine the operator L_{21}
	   % loop through each component of L_{21} which has size of n x 1
	   for j = 1:n
	      % get the j component of the boundary condition matrix Yb
	      Ybj = Yb{j};
	      L21j = Ybj(:,1)*Eb;
	      for indni = 1 : ni(j) - 1
	         L21j = L21j*J + Ybj(:, ind+1)*Eb;
	      end
	      L21{j} = L21j*J;
	   end
	\end{verbatim}
	We note that the operators ${\cal P}_{1}$ and ${\cal P}_{2}$ in~(\ref{eq.blocksys}) can be constructed using similar procedure. We have shown how to construct all operators and quasimatrices appearing in~(\ref{eq.blocksys}). However, the eigenvalue problem~(\ref{eq.eig-intbvp}) requires the operator ${\cal L}_{12} \, {\cal L}_{22}^{-1} \, {\cal L}_{21}$. This operator can only be realized using explicit construction~\cite{dri10} because {\sf Chebfun} syntax does not allow this expression to be formed directly.
	\begin{verbatim}
	   %% determining the operator H = L_{12} L_{22}^{-1} L_{21}
	   % looping through each component of H which has size of n x n
	   for i = 1:n
	      for j = 1:n		
	         L12ij = L12{i,j};
	         L22j = L22{j};
	         L21j = L21{j};
	
	         % m-by-m discretization of H (discretized form)
	         mat = @(m) L12ij( chebpts(m,dom), : )*( L22j \ L21j(m) );
	         % functional expression of H (functional form)
	         op = @(v) L12ij*( L22j \ (L21j*v) );
	         % explicit construction of a linear operator in Chebfun
	         H{i,j} = linop(mat,op,dom);
	      end
	   end
	\end{verbatim}
	
	A similar procedure is used to construct the operator ${\cal P}_{2} \, {\cal L}_{22}^{-1} \, {\cal L}_{21}$. Finally, {\sf Chebfun}'s eigenvalue solver ({\sf eigs}) is used to compute the eigenvalues and eigenfunctions. We note that we use similar method to construct the operators for the spatial state-space representation of the eigenvalue problem discussed in Section~\ref{sec.ss-solve}. For brevity, they are not presented here. All codes for solving the eigenvalue problems in the integral formulation using {\sf Chebfun} are available at~{\sf www.umn.edu/$\sim$mihailo/software/chebfun-svd/}.
	
\section{Representations of the frequency response operator for the linearized Navier-Stokes equations}
\label{sec.app-LNS}

	In this section, we provide the input-output and spatial state-space representations of the frequency response operator for the linearized NS equations. The input-output differential equations for the three-dimensional incompressible channel flow are given by
	\begin{equation}
	\label{eq.LNS-ode}
		\ct:
		\left\{
		\begin{array}{l}
			\left( \mathbf{a}_{4} \, \bD^{(4)} \, + \, \mathbf{a}_{2}(y) \, \bD^{(2)} \, + \, \mathbf{a}_{0}(y) \right) \bphi(y) \; = \; \left( \mathbf{b}_{1} \, \bD^{(1)} \, + \, \mathbf{b}_{0} \right) \bd(y), \\[0.2cm]
			\left[
			\begin{array}{c}
				u \\
				v \\
				w
			\end{array}
			\right]
			\; = \;
			\left( \mathbf{c}_{1} \, \bD^{(1)} \, + \, \mathbf{c}_{0} \right) \bphi(y), \\[0.6cm]
			0 = \left( (\bW_{-1,1} \, \bE_{-1} \, + \, \bW_{1,1} \, \bE_{1}) \bD^{(1)} \, + \, (\bW_{-1,0} \, \bE_{-1} \, + \, \bW_{1,0} \, \bE_{1}) \right) \bphi(y),
		\end{array}
		\right.
	\end{equation}
	where
	\begin{equation*}
		\begin{array}{rcl}
			\mathbf{a}_{4}(y)
			& \!\! = \!\! &
			\left[
			\begin{array}{cc}
				1 & 0 \\[0.1cm]
				0 & 0
			\end{array}
			\right],
			\;\;
			\mathbf{a}_{2}(y)
			\; = \;
			\left[
			\begin{array}{cc}
				a_{2, 1}(y) & 0 \\[0.1cm]
				0 & 1
			\end{array}
			\right],
			\;\;
			\mathbf{a}_{0}(y)
			\; = \;
			\left[
			\begin{array}{cc}
				a_{0, 1}(y) & 0 \\[0.1cm]
				-\mri k_z \, U'(y) & a_{0, 2}(y)
			\end{array}
			\right],
			\\[0.6cm]
			a_{2, 1}(y)
			& \!\! = \!\! &
			-\left( 2 \kappa^{2} \, + \, \mri k_x R \, U(y) \, + \, \mri \omega R \right),
			\\[0.3cm]
			a_{0, 1}(y)
			& \!\! = \!\! &
			\kappa^{4} \, + \, \mri k_x \kappa^{2} R \, U(y) \, + \, \mri k_x R \, U''(y) \, + \, \mri \omega \kappa^{2} R,
			\\[0.3cm]
			a_{0, 2}(y)
			& \!\! = \!\! &
			-\left( \kappa^{2} \, + \, \mri k_x R \, U(y) \, + \, \mri \omega R \right),
			\;\;
			\kappa^{2} \; = \; k_x^{2} + k_{z}^2,
			\\[0.3cm]
			\mathbf{b}_{1}
			& \!\! = \!\! &
			\left[
			\begin{array}{ccc}
				\mri k_x R & 0 & \mri k_z R \\[0.1cm]
				0 & 0 & 0
			\end{array}
			\right],
			\;\;
			\mathbf{b}_{0}
			\; = \;
			\left[
			\begin{array}{ccc}
				0 & \kappa^{2} R & 0 \\[0.1cm]
				-\mri k_z R & 0 & \mri k_x R
			\end{array}
			\right],
			\\[0.5cm]
			\mathbf{c}_{1}^{T}
			& \!\! = \!\! &
			\cfrac{1}{\kappa^{2}}
			\left[
			\begin{array}{ccc}
				\mri k_x & 0 & \mri k_z \\[0.1cm]
				0 & 0 & 0
			\end{array}
			\right],
			\;\;
			\mathbf{c}_{0}^{T}
			\; = \;
			\cfrac{1}{\kappa^{2}}
			\left[
			\begin{array}{ccc}
				0 & \kappa^{2} & 0 \\[0.1cm]
				-\mri k_z & 0 & \mri k_x
			\end{array}
			\right],
			\\[0.5cm]
			\bW_{-1,0}
			& \!\! = \!\! &
			\left[
			\begin{array}{cccccc}
				1 & 0 & 0 & 0 & 0 & 0 \\[0.1cm]
				0 & 0 & 0 & 0 & 1 & 0
			\end{array}
			\right]^T,
			\;\;
			\bW_{1,0}
			\; = \;
			\left[
			\begin{array}{cccccc}
				0 & 1 & 0 & 0 & 0 & 0 \\[0.1cm]
				0 & 0 & 0 & 0 & 0 & 1
			\end{array}
			\right]^T,
			\\[0.5cm]
			\bW_{-1,1}
			& \!\! = \!\! &
			\left[
			\begin{array}{cccccc}
				0 & 0 & 1 & 0 & 0 & 0 \\[0.1cm]
				0 & 0 & 0 & 0 & 0 & 0
			\end{array}
			\right]^T,
			\;\;
			\bW_{1,1}
			\; = \;
			\left[
			\begin{array}{cccccc}
				0 & 0 & 0 & 1 & 0 & 0 \\[0.1cm]
				0 & 0 & 0 & 0 & 0 & 0
			\end{array}
			\right]^T.
		\end{array}
	\end{equation*}
	The spatial state-space representation of $\ct$ is obtained by rewriting~(\ref{eq.LNS-ode}) into a system of first-order differential equations given by~(\ref{eq.T-ssfirst}) with the following matrices
	\begin{equation*}
		\begin{array}{rl}
			\bA_{0} \, = \!\!\! &
			\left[
			\begin{array}{cccccc}
				0 & 1 & 0 & 0 & 0 & 0\\[0.1cm]
				0 & 0 & 1 & 0 & 0 & 0\\[0.1cm]
				0 & 0 & 0 & 1 & 0 & 0\\[0.1cm]
				-a_{0, 1}(y) & 0 & -a_{2, 1}(y) & 0 & 0 & 0 \\[0.1cm]
				0 & 0 & 0 & 0 & 0 & 1\\[0.1cm]
				\mri k_z R \, U'(y) & 0 & 0 & 0 & -a_{0, 2}(y) & 0
			\end{array}
			\right],
			\;\;
			\bB_{0} \; = \;
			\left[
			\begin{array}{ccc}
				0 & 0 & 0\\[0.1cm]
				0 & 0  & 0\\[0.1cm]
				\mri k_x R & 0 & \mri k_z R \\[0.1cm]
				0 & \kappa^{2} R & 0 \\[0.1cm]
				0 & 0 & 0 \\[0.1cm]
				-\mri k_z R & 0 & \mri k_x R
			\end{array}
			\right],
			\\[1.5cm]
			\bC_{0}
			\, = \!\!\! &
			\cfrac{1}{\kappa^{2}}
			\left[
			\begin{array}{cccccc}
				0 & \mri k_x & 0 & 0 & -\mri k_z & 0 \\[0.1cm]
				\kappa^{2} & 0 & 0 & 0 & 0 & 0\\[0.1cm]
				0 & \mri k_z & 0 & 0 & \mri k_x & 0
			\end{array}
			\right],
			\\[0.8cm]
			\bN_{-1}
			\, = \!\!\! &
			\left[
			\begin{array}{cccc}
				I_{2 \times 2} & 0_{2 \times 2} & 0_{2 \times 1} & 0_{2 \times 1} \\[0.1cm]
				0_{2 \times 2} & 0_{2 \times 2} & 0_{2 \times 1} & 0_{2 \times 1} \\[0.1cm]
				0_{1 \times 2} & 0_{1 \times 2} & 1 & 0 \\[0.1cm]
				0_{1 \times 2} & 0_{1 \times 2} & 0 & 0 \\[0.1cm]
			\end{array}
			\right],
			\,\,\,
			\bN_{1} \,\, = \,\,
			\left[
			\begin{array}{cccc}
				0_{2 \times 2} & 0_{2 \times 2} & 0_{2 \times 1} & 0_{2 \times 1} \\[0.1cm]
				I_{2 \times 2} & 0_{2 \times 2} & 0_{2 \times 1} & 0_{2 \times 1} \\[0.1cm]
				0_{1 \times 2} & 0_{1 \times 2} & 0 & 0 \\[0.1cm]
				0_{1 \times 2} & 0_{1 \times 2} & 1 & 0 \\[0.1cm]
			\end{array}
			\right].
		\end{array}
	\end{equation*}
	The input-output and state-space representations of the adjoint of the operator $\ct$ can be determined using the procedure presented in Section~\ref{sec.representations-ctstar}.

\section{Representations of the frequency response operator for the inertialess channel flow of viscoelastic fluids}
\label{sec.app-SOB-kz0-TPBVSR}
	
	We next show how to formulate the input-output and spatial state-space representations of the frequency response operator for the inertialess flow of viscoelastic fluids. We begin by rewriting~(\ref{eq.ex-psi-ode}) into the input-output representation~(\ref{eq.system-ode2}),
	\begin{equation}
	\label{eq.psi-ode}
		\ct:
		\left\{
		\begin{array}{l}
			\left( D^{(4)} \, + \, a_{3}(y) \, D^{(3)} \, + \, a_{2}(y) \, D^{(2)} \, + \, a_{1}(y) \, D^{(1)} \, + \, a_{0}(y) \right) \psi(y) = \left( \mathbf{b}_{1}(y) \, \bD^{(1)} \, + \, \mathbf{b}_{0}(y) \right) \bd(y),
			\\[0.3cm]
			\left[
			\begin{array}{c}
				u \\
				v
			\end{array}
			\right]
			=
			\left( \mathbf{c}_{1} \, D^{(1)} \, + \, \mathbf{c}_{0} \right) \psi(y), \\[0.5cm]
			0 = \left( (\bW_{-1,1} \, E_{-1} \, + \, \bW_{1,1} \, E_{1}) D^{(1)} \, + \, (\bW_{-1,0} \, E_{-1} \, + \, \bW_{1,0} \, E_{1}) \right) \psi(y),
		\end{array}
		\right.
	\end{equation}
	where
	\begin{subequations}
		\begin{align}
			\non
			a_{0}(y)
			\; = & \;\;
			\cfrac{k_{x}^4}{a_{4}(y)}
			\left(
			\beta
			\, - \,
			\frac{
			2 \, \We^2 \, (\beta \, - \, 1)
			\left(
			2 \We^2 \, + \, 1
			\right)
			}
			{
			(\mri k_{x} \We \, y \, + \, \mri \omega \, + \, 1)^3
			}
			\, - \,
			\frac{
			(\beta \, - \, 1)
			\left(
			2 \We^2 \, + \, 1
			\right)
			}
			{
			\mri k_{x} \We \, y \, + \, \mri \omega \, + \, 1
			}
			\right),
			\\[0.2cm]
			\non
			a_{1}(y)
			\; = & \;\;
			\cfrac{1}{a_{4}(y)}
			\frac{
			2 \, \mri k_{x}^3 \, \We \, (\beta \, - \, 1) \, (\mri \omega \, + \, \mri k_{x} \We \, y)
			\left(
			\mri k_{x} \We \, y \, + \, \mri \omega \, - \, 2 \We^2 \, + \, 1
			\right)
			}
			{
			(\mri k_{x} \We \, y \, + \, \mri \omega \, + \, 1)^3
			},
			\\[0.2cm]
			\non
			a_{2}(y)
			\; = & \;\;
			\cfrac{1}{a_{4}(y)}
			\left(
			- \, 2 \, \beta \, k_{x}^2
			\, + \,
			\cfrac{
			2 \, k_{x}^2
			\left( \beta \, - \, 1 \right)
			\left( \We^2 \, + \, 1 \right)
			}
			{
			\mri k_{x} \We \, y \, + \, \mri \omega \, + \, 1
			}
			\, - \,
			\cfrac{
			4 \left( \beta \, - \, 1 \right) k_{x}^2 \, \We^2
			}
			{
			(\mri k_{x} \We \, y \, + \, \mri \omega \, + \, 1)^2
			}
			\, + \,
			\cfrac{
			2 \, \left( \beta \, - \, 1 \right) k_{x}^2 \, \We^2
			}
			{
			(\mri k_{x} \We \, y \, + \, \mri \omega \, + \, 1)^3
			}
			\right),
			\\[0.2cm]
			\non
			a_{3}(y)
			\; = & \;\;
			-\cfrac{1}{a_{4}(y)}
			\frac{
			2 \, \mri k_{x} \We \, (\beta \, - \, 1) \, (\mri k_{x} \We \, y \, + \, \mri \omega)
			}
			{
			(\mri k_{x} \We \, y \, + \, \mri \omega \, + \, 1)^2
			},
			\;\;\;
			a_{4}(y)
			\; = \;
			\frac{
			\beta \, \mri  k_{x} \We \, y \, + \, \beta \, \mri \omega \, + \, 1
			}
			{
			\mri k_{x} \We \, y \, + \, \mri \omega \, + \, 1
			},
			\\[0.2cm]
			\non
			b_{1}(y)
			\; = & \;\;
			 -\cfrac{1}{\beta \, a_{4}(y)},
			\;\;\;
			b_{0}(y)
			\; = \;
			\cfrac{\mri k_x}{\beta \, a_{4}(y)},
			\;\;\;
			\mathbf{b}_{1}(y)
			\; = \;
			\left[
			\begin{array}{cc}
				b_{1}(y) & 0
			\end{array}
			\right],
			\;\;\;
			\mathbf{b}_{0}(y)
			\; = \;
			\left[
			\begin{array}{cc}
				0 & b_{0}(y)
			\end{array}
			\right],
			\\[0.2cm]
			\non
			\mathbf{c}_{1} \; = & \;\;
			\left[ \begin{array}{cc} 1 & 0 \end{array} \right]^{T},
			\;\;
			\mathbf{c}_{0} \; = \; \left[ \begin{array}{cc} 0 & -\mri k_x \end{array} \right]^{T},
			\;\;
			\left[ \begin{array}{cccc} \bW_{-1,1} & \bW_{1,1} & \bW_{-1,0} & \bW_{1,0} \end{array} \right] \; = \; \bI_{4 \times 4}.
		\end{align}
	\end{subequations}
	The spatial state-space representation of $\ct$ is obtained by rewriting~(\ref{eq.psi-ode}) into a system of first-order differential equations. Using the procedure described in~\ref{sec.app-realization} yields
	\begin{equation*}
		\begin{array}{rl}
			\bA_{0} \, = \!\!\! &
			\left[
			\begin{array}{cccc}
				0 & 1 & 0 & 0 \\
				0 & 0 & 1 & 0 \\
				0 & 0 & 0 & 1 \\
				-a_{0}(y) & -a_{1}(y) & -a_{2}(y) & -a_{3}(y)
			\end{array}
			\right],
			\;\;
			\bB_{0} \; = \;
			\left[
			\begin{array}{cc}
				0 & 0 \\
				0 & 0 \\
				b_{1}(y) & 0 \\
				-b'_{1}(y) \, - \, a_{3}(y) \, b_{1}(y) & b_{0}(y)
			\end{array}
			\right],
			\\[0.8cm]
			\bC_{0} \, = \!\!\! &
			\left[
			\begin{array}{cccc}
				0 & 1 & 0 & 0 \\
				-\mri k_x & 0 & 0 & 0
			\end{array}
			\right],
			\;\;
			\bN_{-1} \; = \;
			\left[
			\begin{array}{cc}
				I_{2 \times 2} & 0_{2 \times 2} \\
				0_{2 \times 2} & 0_{2 \times 2}
			\end{array}
			\right],
			\,\,\,
			\bN_{1} \,\, = \,\,
			\left[
			\begin{array}{cc}
				0_{2 \times 2} & 0_{2 \times 2} \\
				I_{2 \times 2} & 0_{2 \times 2}
			\end{array}
			\right].
		\end{array}
	\end{equation*}
	The input-output and state-space representations of the adjoint of the operator $\ct$ can be determined using the procedure described in Section~\ref{sec.representations-ctstar}.

\section{Frequency response of an Euler-Bernoulli beam}
\label{sec.app-beam}

	\begin{figure}
		\begin{center}
	            	{
	            	\includegraphics[width=0.4\columnwidth]
			{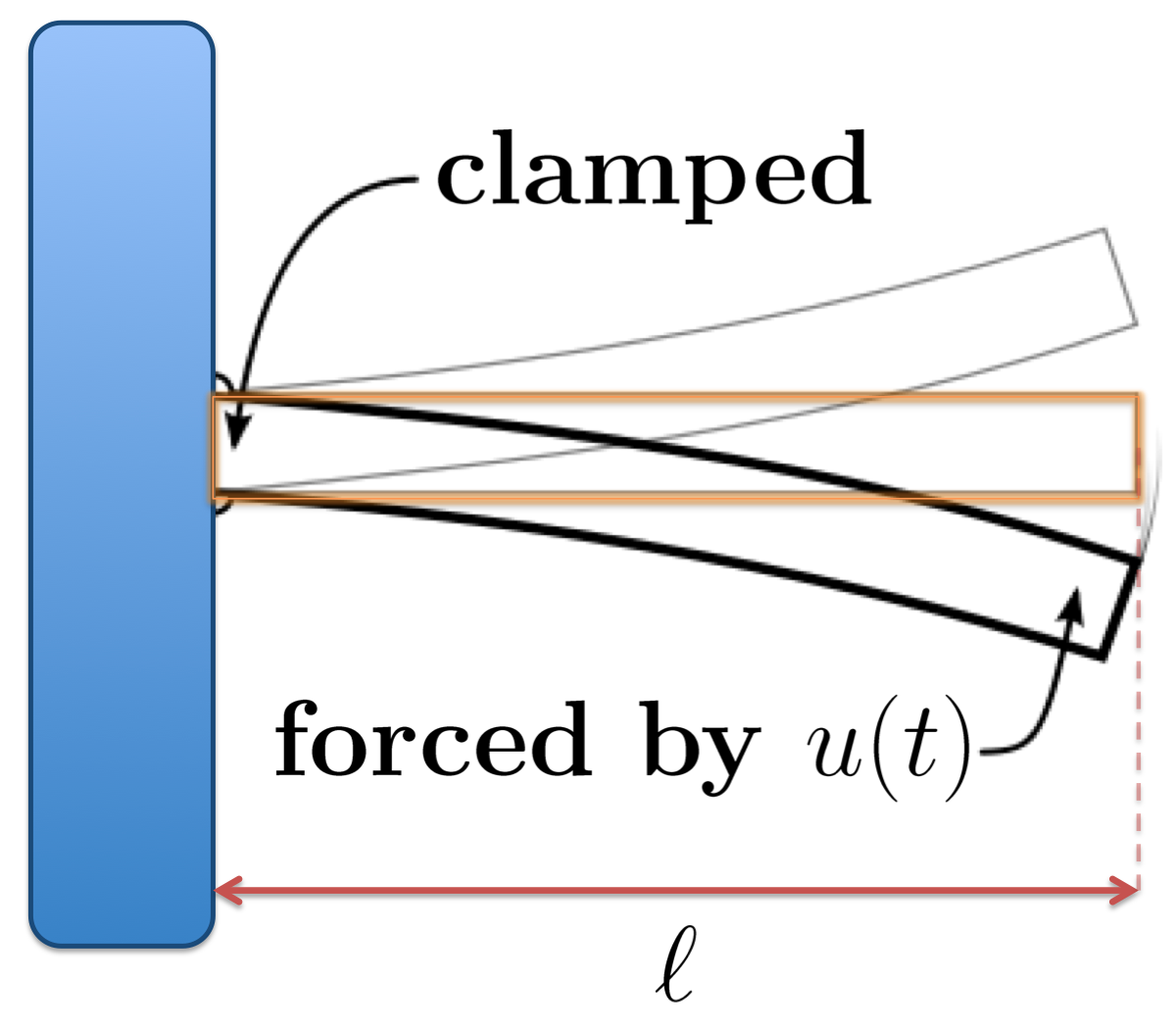}
			}
	  	\end{center}
		\vspace{-0.5cm}
		\caption{An Euler-Bernoulli beam that is clamped at the left end and subject to a boundary actuation at the other end.}
		\label{fig.beam}
	\end{figure}

	In this section, we consider an Euler-Bernoulli beam that is clamped at the left end and subject to a boundary actuation $u(t)$ at the other end; see figure~\ref{fig.beam} for an illustration. The vertical displacement of the beam $\phi(y, t)$ is governed by~\cite{clopen93},
	\begin{subequations}
	\label{eq.beam-governingeq}
		\begin{align}
			\mu \, \phi_{tt}(y,t) \, + \, \cfrac{\alpha \, E_{I}}{\ell^{4}} \, \phi_{tyyyy}(y,t) \, + \, \cfrac{E_{I}}{\ell^{4}} \, \phi_{yyyy}(y,t)
			\; = & \;\;
			0,
			\;\;
			y \in \left[ \, 0, 1 \, \right],
			\\[0.1cm]
			\label{eq.beam-bc-right}
			\phi(0, t) \; = \; \phi_{y}(0, t)
			\; = & \;\; 0,
			\\[0.1cm]
			\label{eq.beam-bc-left}
			\phi_{yy}( 1, t ) \; = \; 0, \;\;\;
			\cfrac{\alpha \, E_{I}}{\ell^{3}} \, \phi_{tyyy}(1, t) \, + \, \cfrac{E_{I}}{\ell^{3}} \, \phi_{yyy}(1, t)
			\; = & \;\; u(t).
		\end{align}
	\end{subequations}
	Here, the input $u(t)$ denotes the force acting on the tip of the beam, $\ell$ is the length of the beam, $\mu$ is the mass per unit length of the beam, $E_{I}$ is the flexural stiffness, and $\alpha$ is the Voigt damping factor.
	
	\vspace*{0.15cm}

	Equation~(\ref{eq.beam-governingeq}) can be used to model the movement of a micro-cantilever in atomic force microscopy applications~\cite{farjovsalACC09} with	
    \beq
	\begin{array}{rclrcl}
		\ell
		& \!\! = \!\! &
		240 \times 10^{-6} \, \text{m},
		&
		\mu
		& \!\! = \!\! &
		1.88 \times 10^{-7} \, \text{kg/m},
		\\[0.1cm]
		E_{I}
		& \!\! = \!\! &
		7.55 \times 10^{-12} \, \text{Nm}^{2},
		&
		\alpha
		& \!\! = \!\! &
		5 \times 10^{-8} \, \text{s}.
	\end{array}
    \label{eq.beam-parameters}
	\eeq
	In contrast to the body of the paper, the forcing $u(t)$ does not enter to the equation as an additive input but as a boundary condition. We next show how easily frequency response in this case can be computed using {\sf Chebfun}.
	
	\vspace*{0.15cm}

\begin{figure}
		\begin{center}
		\begin{tabular}{cc}
			$\left| \ct(\omega) \right|$
			&
			$\angle \, \ct(\omega)$
			\\
			{
			\includegraphics[width=0.45\columnwidth]
			{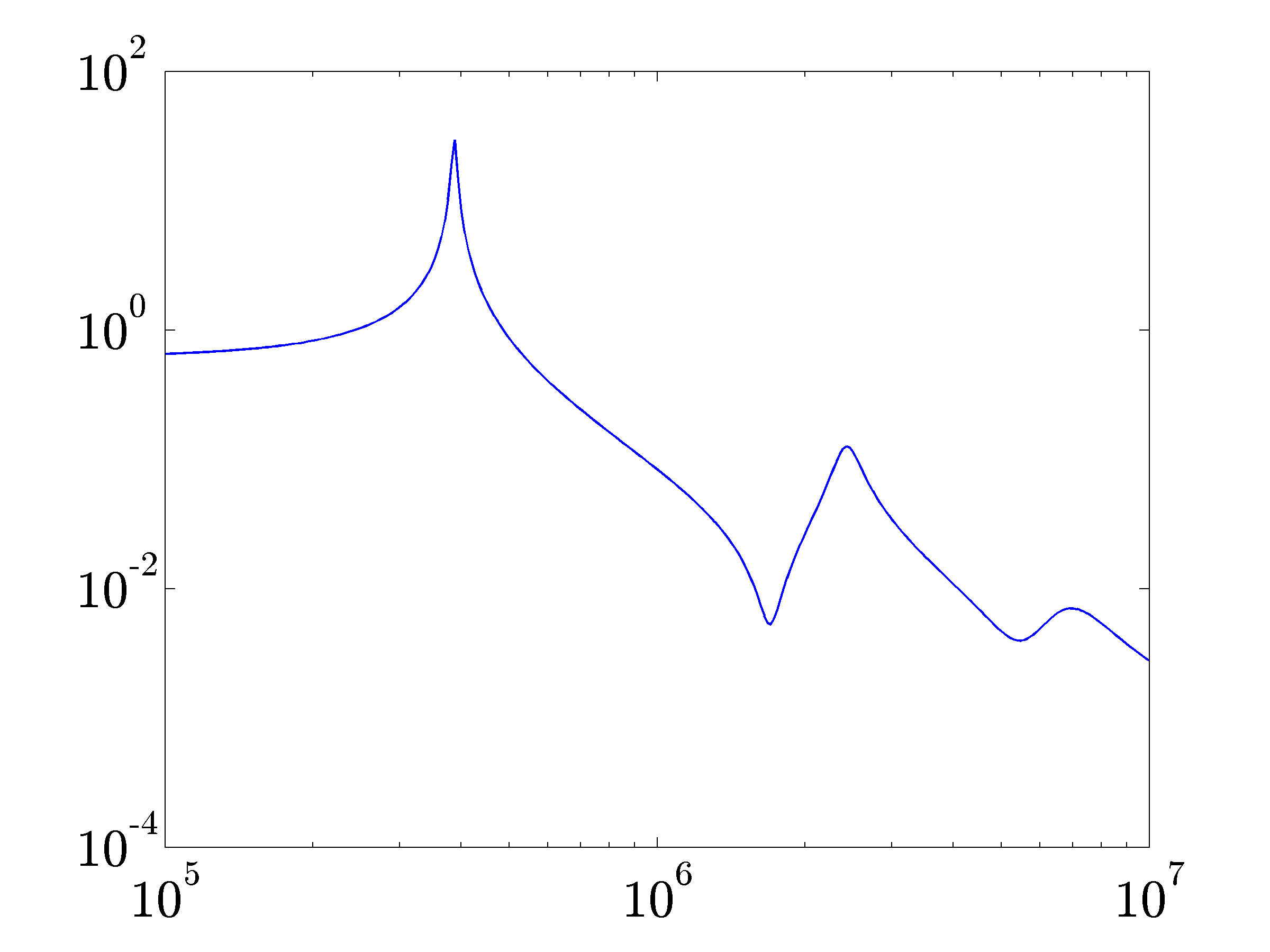}
			}
			&
	            	{
	            	\includegraphics[width=0.45\columnwidth]
			{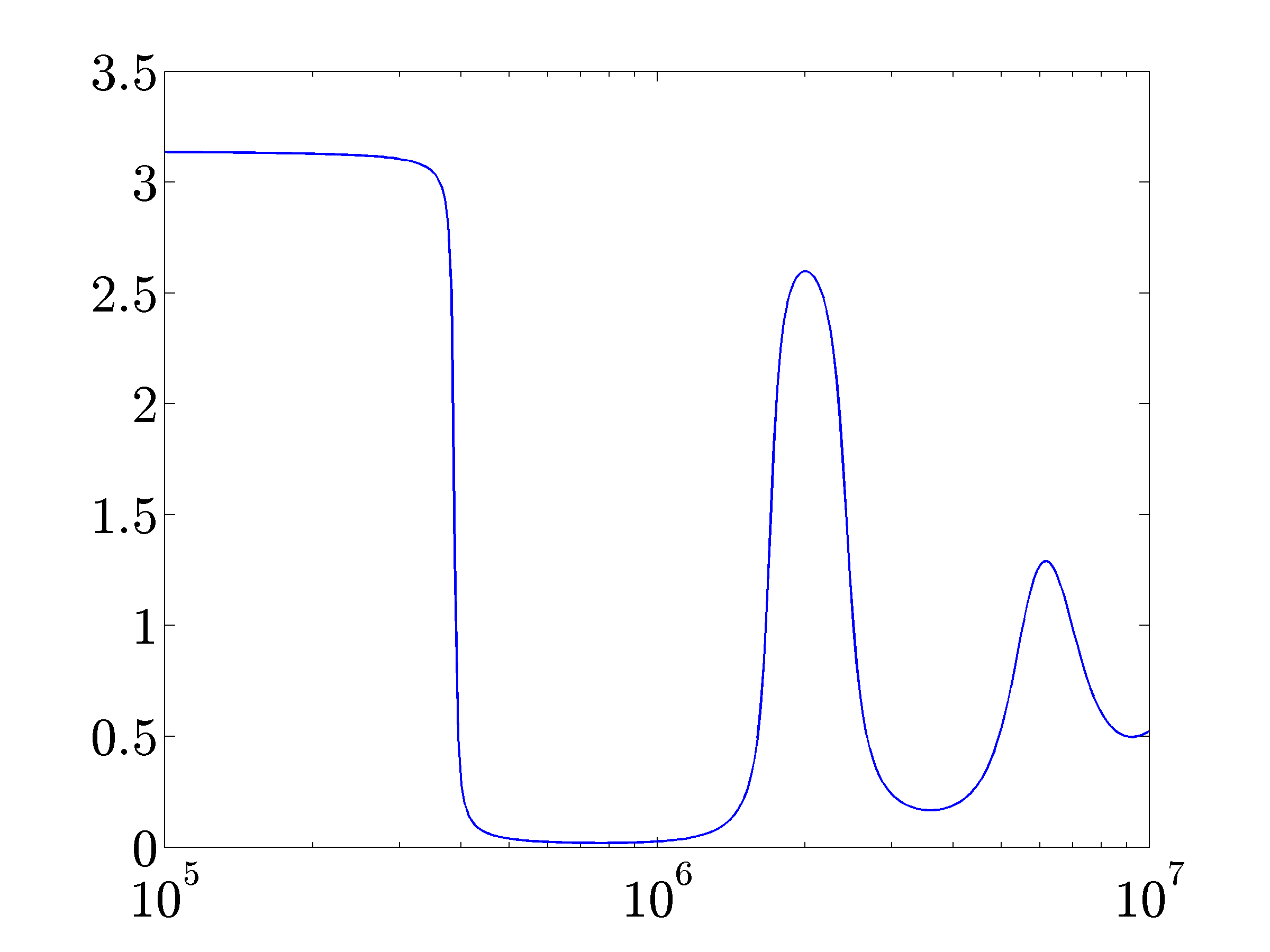}
			}
			\vspace{-0.2cm}
			\\
			{\large $\omega$}
			&
			{\large $\omega$}
			\\
			\subfigure[]{
			\label{fig.beam-magH}
			}
			&
			\subfigure[]{
			\label{fig.beam_phaseH}
			}
			\end{tabular}
	  	\end{center}
		\vspace{-0.5cm}
		\caption{Frequency response of the Euler-Bernoulli beam~(\ref{eq.beam-governingeq})-(\ref{eq.beam-parameters}) with the output determined by the vertical displacement of the beam at the right end. (a) magnitude of the frequency response $| \ct (\omega) |$;  (b) phase of the frequency response $\angle \, \ct (\omega)$.}
		\label{fig.beam-freqresp}
 	\end{figure}

	Application of the temporal Fourier transform to~(\ref{eq.beam-governingeq}) yields
	\begin{equation}
	\label{eq.beam-freqresp}
		\ct(\omega):
		\left\{
		\begin{array}{l}
			\cfrac{E_{I}}{\ell^{4}} \left( 1 \, + \, \mri \, \omega \, \alpha \right) \,
            \phi''''(y,\omega)
			\, - \,
			\mu \, \omega^{2} \, \phi(y,\omega)
			\; = \;
			0,
			\\[0.35cm]
			\phi(0,\omega) \; = \; \phi'(0,\omega)
			\; = \; 0,
			\\[0.15cm]
			\phi''( 1,\omega ) \; = \; 0, \;\;\;
			\cfrac{E_{I}}{\ell^{3}} \left( 1 \, + \, \mri \, \omega \, \alpha \right) \,
            \phi'''(1,\omega) \; = \; u(\omega).
		\end{array}
		\right.
	\end{equation}
At each $\omega$, the mapping from $u (\omega)$ to $\phi (y,\omega)$ can be obtained by computing the solution to~(\ref{eq.beam-freqresp}) with $u(\omega) = 1$ using {\sf Chebfun}. The energy of the beam is determined by
    \[
    E (\omega)
    ~=~
    \dfrac{1}{2}
    \left(
    \inner{\phi''(\cdot,\omega)}{\phi''(\cdot,\omega)}
    ~+~
    \omega^2 \inner{\phi(\cdot,\omega)}{\phi(\cdot,\omega)}
    \right),
    \]
and it can be simply computed with the aid of {\sf Chebun}'s functions {\sf diff} and {\sf cumsum}. On the other hand, if the output is given by the vertical displacement at the right end of the beam, the frequency response is simply determined by the magnitude and phase of the complex number $\phi(1,\omega)$; see figure~\ref{fig.beam-freqresp}.
	
	\vspace*{0.15cm}
	
For parameters given by~(\ref{eq.beam-parameters}), even the use of {\sf Chebfun}'s differential operators to construct
	\[
		\ca_0
		\; = \;
		\cfrac{E_{I}}{\ell^{4}} \left( 1 \, + \, \mri \, \omega \, \alpha \right) D^{(4)}
		\, - \,
		\mu \, \omega^{2} \, I,
	\]
with appropriate boundary conditions may lead to unfavorable conditioning of differentiation matrices. This can be alleviated by determining and solving instead the integral form of~(\ref{eq.beam-freqresp}). The procedure for achieving this closely follows the method presented in Section~\ref{sec.inteigs}. The {\sc Matlab} code used for computing the frequency response with integral formulation can be found at~{\sf www.umn.edu/$\sim$mihailo/software/chebfun-svd/}.
	
\newpage

\end{document}